\documentclass[preprint,12pt]{elsarticle}
\usepackage[utf8]{inputenc}
\usepackage[margin=0.5in]{geometry}
\usepackage{graphicx}
\usepackage{enumitem}
\usepackage{xcolor}
\usepackage{cancel}
\usepackage{hyperref}
\usepackage{mathtools}
\usepackage{multirow}
\usepackage{nicefrac}
\usepackage{comment}
\DeclareMathOperator*{\argmax}{arg\,max}
\DeclareMathOperator*{\argmin}{arg\,min}
\usepackage{hhline}
\usepackage{placeins}
\numberwithin{figure}{section}
\numberwithin{table}{section}
\usepackage{subcaption}
\usepackage{caption}
\usepackage{float}
\usepackage{bm}
\usepackage{physics}
\renewcommand*{\arraystretch}{1.5}
\allowdisplaybreaks

\usepackage{amsfonts}
\usepackage{amssymb}
\usepackage{amsmath}
\usepackage{color}
\journal{Journal of Computational Physics}
\begin{document}

\begin{frontmatter}

\title{Bayesian finite element regression for vascular flow reconstruction with quantified uncertainty}

\author[Berkeley]{Cem Gormezano\corref{cor1}}
\ead{cem.gormezano@berkeley.edu}
\author[Berkeley]{Shawn Shadden}
\address[Berkeley]{Mechanical Engineering, University of California, Berkeley, California, USA}
\cortext[cor1]{Corresponding author}

\begin{abstract}
Reconstructing accurate velocity and pressure fields from under-resolved noisy measurements of blood flow is an ill-posed inverse problem due to unknown inlet and outlet boundary conditions. We present a Bayesian finite element regression framework that reconstructs steady three-dimensional velocity and pressure fields, with quantified uncertainty, from noisy velocity observations without offline training data. We represent velocity and pressure fields in Taylor–Hood finite element basis functions, and construct physics-informed priors on the nodal degrees of freedom from maximum-entropy principles. Combined with a likelihood specified by a noise-model, this yields a posterior whose maximum-a-posteriori estimate (MAP) gives velocity and pressure reconstructions. The MAP estimate is computed by solving a large-scale sparse nonlinear least-squares problem where pressure is eliminated analytically, no-slip walls are enforced exactly, and gradient is computed without forward/adjoint solves or automatic differentiation. A Laplace approximation of the posterior quantifies the uncertainties in our reconstructions and propagates them to clinically relevant quantities of interest including, pressure drop, flow rates, and wall shear stress. On patient-specific cerebral aneurysm, aortic aneurysm, and aortic coarctation geometries, the method reconstructs velocity and pressure more accurately than tricubic interpolation and comparably to a PINN, while recovering region-of-interest wall shear stress more accurately than both.
\end{abstract}


\end{frontmatter}

\section{Introduction} 

High-resolution imaging of blood flow is critical in the diagnosis and treatment of cardiovascular disease. Many cardiovascular conditions are driven by local flow features like narrow jets, recirculation zones, and near-wall boundary layers. The accuracy of unobserved actionable diagnostic and prognostic hemodynamic quantities, such as pressure gradients and wall shear stress (WSS) are very sensitive to sufficient resolution of these flow structures. However, flow images obtained through the state of the art noninvasive techniques such as Doppler ultrasound or phase-contrast magnetic resonance imaging (PC-MRI), tend to be low-resolution and noisy. For example, 4D flow MRI, which is one of the most clinically advanced techniques for acquiring unsteady 3D cardiovascular velocity fields, has spatial resolution in the range $0.8 \ \text{mm}^2 - 3 \ \text{mm}^2$ depending on the region of interest \cite{Bissell2023, Markl2012}. Scanning times vary in the range $8 \ \text{min} - 25 \ \text{min}$  with higher scan times giving more accurate velocity measurements \cite{Bissell2023, Markl2012}. Consequently, successful super-resolution and denoising also plays an important role in reducing scanning times. Although, 2D-PC-MRI \cite{Bollache2016} and ultrasound \cite{Jensen2016p1, Jensen2016p2} may achieve better resolution, they are restricted by 2D field of view and 2D directional information.

A number of methods within the field of flow reconstruction can be utilized to denoise and enhance the spatial resolution of cardiovascular flow fields. Methods based on proper orthogonal decomposition (POD) \cite{Lee2010, Brindise2017, Fathi2018}, robust proper orthogonal decomposition (RPOD) \cite{Chatpattanasiri2023}, spectral proper orthogonal decomposition (SPOD) \cite{Nekkanti2021}, dynamic mode decomposition (DMD) \cite{Arzani2021, Pialot2026} make use of the low rank structure and spatiotemporal coherence of fluid flow to filter measurement noise and reconstruct missing data. These methods are easy to implement and computationally efficient. Nevertheless, they are purely measurement driven and do not enforce governing equations along with boundary conditions, yielding physically inconsistent results.

To integrate smoothness and incompressibility to their reconstructions several approaches project the flow field to predefined basis functions (polynomials \cite{Lekien2005}, radial basis functions \cite{Busch2013}, wavelets \cite{Ong2015}) which are smooth or divergence-free by construction. Such methods give physically plausible flow fields, yet do not make use of the momentum balance and require additional post-processing, like solving the pressure Poisson equation (PPE), to obtain physically consistent pressure fields.

Computational fluid dynamics (CFD) offers a pipeline which completely exploits the Navier-Stokes equations to obtain velocity and pressure fields with arbitrary resolution \cite{Updegrove2017, Arthurs2021}. Variational data assimilation, formulates the reconstruction task as an inverse problem and solves for the boundary conditions, segmentation, and model parameters, which minimize the difference between the under-resolved noisy measurements and CFD predictions \cite{Koltukluoglu2018, Koltukluoglu2019v2, Koltukluoglu2019v1, Funke2019}. Once the unknown boundary conditions and parameters are recovered, a forward solution to Navier-Stokes equations produces fully-resolved velocity and pressure fields. Such methods give excellent results on 4D-PC-MRI data synthesized from CFD generated fields. However, computing the derivatives of the loss function with respect to unknowns (e.g., boundary conditions, model parameters, segmentation field) requires the solution of the forward and adjoint equations at each iteration of the optimizer, making this approach computationally intensive. Such approach further requires the corresponding forward problem to be well posed for adjoint and forward problems to be solvable. Additionally, modeling errors arising from uncertainties in blood rheology are not explicitly accounted for when the optimization is hard-constrained by the Navier–Stokes equations. Some of the recent work by Kontogiannis et al. \cite{Kontogiannis2022, Kontogiannis2025} has focused on integrating uncertainties in model parameters and segmentation to inversion through Bayesian inference. They sample from the posterior distribution of model parameters by solving a forward Navier-Stokes problem for each sample, which can become expensive.

Deep learning based techniques have also been explored for flow reconstruction. Purely data-driven neural network (NN) architectures which use CFD simulations as the training data have been successfully employed to denoise and super-resolve cardiovascular flow fields. Although, convolutional neural network architectures are commonly used in this effort \cite{Rutkowski2021, Shit2022}, less traditional architectures like graph neural networks have also been employed \cite{Barwey2025}. These methods assume the presence of exact ground truth data and do not ensure physical consistency of the model’s output. Alternatively, physics-informed neural networks (PINNs) which are guided only by real-time observations, governing equations and boundary conditions have successfully been used to reconstruct flows from noisy and sparse measurements \cite{Raissi2019, Karniadakis2021, Fathi2020}. PINNs, integrate governing equations softly via regularization terms in the loss function, making these approaches less sensitive to errors associated with modeling and representation of geometry. Sun and Wang use Bayesian PINNs to propagate modeling error to their reconstructions in 2D aneurysms and stenoses \cite{Sun2020}.  PINNs have potential limitations as well. Evaluating the Laplacian in Navier-Stokes equation loss term requires nested automatic differentiation at every collocation point which is memory heavy and computationally expensive. Several authors remedy this using finite difference stencils \cite{Han2021} or the weak form of Navier-Stokes with the finite element method (FEM) \cite{Han2022}. Still, without a loss term based on a data set, these types of PINNs exhibit slow convergence in training and require re-training across different geometries and boundary conditions. Addition of a loss term based on low-resolution or ground truth data set significantly improves training stability, convergence speed, and reconstruction accuracy \cite{Han2021, Sautory2024}. These networks do not need re-training on different geometries allowing instantaneous velocity reconstruction pressure inference. Nonetheless, these datasets tend to contain similar geometries and make these networks less generalizable to different geometries.

Karnakov et al \cite{Karnakov2023, Karnakov2024}, introduced the optimization of discrete loss (ODIL) framework to solve forward and inverse problems in physics. ODIL minimizes an objective function composed of a data loss term and a regularizer formed from the discretization of governing equations using traditional gradient-based optimization methods (Gauss-Newton and limited memory quasi-Newton methods). Gradients with respect to discretized field variables are obtained via automatic differentiation. They show that, ODIL with finite volume discretization and multi-resolution decomposition can successfully reconstruct flow fields from noisy sparse measurements at a three to five orders of magnitude lower computational cost than PINNs on benchmark problems like Poisson equation, Burgers equation, and lid-driven cavity problems. Recently, \cite{Amoudruz2025} et al. extended this framework to a Bayesian setting. A similar reconstruction approach is based on the least squares finite element method (LSFEM) \cite{Bochev2009}, which was shown to achieve accurate reconstruction of laminar flow over a 2D backward-facing step \cite{Schwarz2018} and through a 3D pipe \cite{Heys2010, Dwight2010}. LSFEM differs from ODIL in that it uses a first-order formulation of the Navier-Stokes equations and it minimizes an objective functional rather than an objective function. It relies on solving the optimality conditions obtained from taking the Frechet derivative of the objective functional using the finite element method (FEM). It does not rely on automatic differentiation as Frechet derivatives are evaluated with finite element assembly. In \cite{Schwarz2018}, authors also connect LSFEM to Kalman filters and Bayesian inference to perform data assimilation.


Our approach formulates the reconstruction problem as a Bayesian regression problem with finite element basis functions. We encode the governing equations into the prior distribution on the coefficients of the velocity and pressure basis functions, which are commonly called degrees of freedom in finite element community. We let this prior distribution be the least informative prior distribution such that expectations of discretized governing equations are small. We infer the velocity and pressure degrees of freedom through maximum a posteriori (MAP) estimate of the posterior distribution. Once the most likely degrees of freedom are obtained, we can represent velocity and pressure fields to arbitrary resolution everywhere with finite element basis functions. The objective function in MAP estimation resembles the objective functions in ODIL and LSFEM frameworks with a couple of caveats. Unlike ODIL, which uses the finite volume and finite difference discretizations, we use FEM to discretize the governing equations. The use of finite element basis functions allows us to impose desirable properties on the reconstruction, such as continuity and smoothness, which are important in accurate prediction of WSS. Similar to LSFEM, we do not use automatic differentiation. The derivatives of the objective function with respect to degrees of freedom are obtained through finite-element assembly. Unlike LSFEM, we use the weak form of the governing equations in our regularization terms eliminating the need for auxiliary variables like stress or vorticity appearing in the first order formulation of Navier-Stokes equations, hence reducing the dimensionality of the MAP estimation. We solve the resulting optimization problem via matrix-free quasi-Newton method with line-search. 
 
In summary, the main contributions of this paper are : (1) We develop a methodology for reconstructing steady 3D cardiovascular flow images from noisy under-resolved observations, that does not require a training data-set besides the observations, does not require automatic differentiation, and does not require forward or adjoint solves. (2) We demonstrate the ability of our method to jointly infer the unobserved pressure field. (3) We estimate the uncertainties in our velocity and pressure reconstructions using a Laplace approximation and propagate them to hemodynamic quantities like pressure drop, flow-rate, and wall-shear stress. (4) We compare our method to a baseline tricubic vector-interpolation method \cite{Lekien2005} and a basic multilayer perceptron (MLP) PINN in terms of velocity, pressure, and wall shear stress relative reconstruction errors over a range of signal to noise ratios (SNR), resolutions, and vascular geometries. (5) We develop a weakly divergence-free variant of the proposed method.


\section{Methods}\label{sec:Methods} 

Before describing our methodology, we first fix some notations which will be used throughout the paper. We use bold face letters to indicate vector and matrix quantities. Given a sequence of vectors or matrices $\bm{T}_i \in \mathbb{R}^{m\times n}$ for $i = 1, \dots, N$,  the notation, $\left[\bm{T}_i\right]_{i=1}^N \in \mathbb{R}^{N*m\times n}$ refers to vertical concatenation of $\bm{T}_i$ from $1$ to $N$.  All the field variables (velocity $\bm{v}$, pressure $\bm{p}$, shear-tress $\bm{\tau}$), in this section are nondimensionalized such that $\bm{v}  = \bm{v}/V, \ \bm{x} = \bm{x}/L, \ \bm{p}  = \bm{v}/(\rho V^2)$, and $\bm{\tau} = \bm{\tau}/(\rho V^2)$ where $V$ is the maximum speed over all observations, $L$ is the maximum distance between any two points on the rim of the inlet, and $\rho$ is the density.

\subsection{Finite element regression (FER)}

Given a mesh, the finite element regression relies on representing velocity $\bm{v} \colon \mathbb{R}^3 \mapsto \mathbb{R}^3$ and pressure fields $\bm{p} \colon \mathbb{R}^3 \mapsto \mathbb{R}$ with finite element basis functions $\{\phi_i(\bm{x})\}_{i = 1}^{\text{NVB}}$ and $\{\psi_i(\bm{x})\}_{i = 1}^{\text{NPB}}$ of choice such that,

\begin{align}
    \begin{bmatrix}
       \bm{v}^h(\bm{x}) \\
       p^h(\bm{x})
    \end{bmatrix} = 
    \begin{bmatrix}
        \sum\limits_{i = 1}^{\text{NVB}} \hat{\bm{v}}_i\phi_i(\bm{x})\\[1.0em]
        \sum\limits_{i = 1}^{\text{NPB}} \hat{p}_i\psi_i(\bm{x})
    \end{bmatrix} = 
    \underbrace{\begin{bmatrix}
    \left(\left[\phi_i(\bm{x})\bm{I}_{3\times 3}\right]_{i = 1}^{\text{NVB}}\right)^T & \bm{0}\\[1.0em]
    \bm{0} & \left(\left[\psi_i(\bm{x})\right]_{i=1}^{\text{NPB}}\right)^T 
    \end{bmatrix}}_{\displaystyle \bm{N}(\bm{x})}
    \begin{bmatrix}
    \left[\hat{\bm{v}}_i\right]_{i = 1}^{\text{NVB}} \\[1.0em]
    \left[\hat{p}_i\right]_{i=1}^{\text{NPB}},
    \end{bmatrix} 
    \label{eq:fe_expansion}
\end{align}

\noindent where $[\hat{\bm{v}}_i]_{i=1}^{\text{NVB}} \in \mathbb{R}^{3\text{NVB}}$ is the vector of velocity degrees of freedom and $[\hat{p}_i]_{i=1}^{\text{NPB}} \in \mathbb{R}^{\text{NPB}}$ is the vector of pressure degrees of freedom. The above representation assumes nodal basis functions, however it can easily be extended to basis functions associated other mesh entities like edges or faces. The goal of FER is to treat the velocity and pressure degrees of freedom uncertain and learn them from the flow field observations via Bayesian inference while incorporating the governing equations. Let $\Omega$ be a 3D vascular region whose boundary can be expressed as the union of inlets, outlets and vessel walls such that $\partial \Omega = \Gamma_{\text{in}} \cup \Gamma_{\text{wall}} \cup \Gamma_{\text{out}}$. Moreover, let $\Omega^h \coloneq \bigcup_{i = 1}^{\text{NE}} K_i$ be a triangulation of $\Omega$ in to tetrahedral subdomains $K_i$. We choose the velocity and pressure finite element function spaces to be,

\begin{align}
\mathcal{V}^h
&\coloneq \left\{\bm{v}^h \in H^1(\Omega^h)^3 \
 \Big\vert \  \bm{v}^h|_{K_i} \in [\mathbb{P}_k(K_i)]^3 \ \ \forall K_i \in \Omega^h, \ \bm{v}^h = \bm{0} \ \text{on} \ \Gamma_{\text{wall}}^h \right\} \\
\mathcal{P}^h
&\coloneq
\left\{p^h \in L^2(\Omega^h) \
 \Big\vert \ p^h|_{K_i} \in \mathbb{P}_l(K_i) \ \ \forall K_i \in \Omega^h \right\},
 \label{eq:spaces}
\end{align}

\noindent forming continuous Lagrange elements of order $k$ and $l$ respectively. We set $k=2$ and $l=1$ which gives the well known Taylor-Hood element pair. For this pair, $\text{NVB} = \text{NV} + \text{NED}$ and $\text{NVB} = \text{NV}$ where NV is the number of vertices and NED is the number of edges of the mesh. Note that, the velocity finite element space does not have any constraint on the boundary conditions at the inlet/outlet representing our ignorance about the flow variables at these locations. From Bayes theorem we have,

\begin{align}
     \underbrace{\mathbb{P}\left(\left[\hat{\bm{v}}_i\right]_{i=1}^{\text{NVB}}, \left[\hat{p}_i\right]_{i=1}^{\text{NPB}} \Bigm| \left\{(\bm{x}_i, \tilde{\bm{v}}_i)\right\}_{i=1}^{\text{NOBS}}\right)}_{\textbf{Posterior}} \propto \underbrace{\mathbb{P}\left(\left\{(\bm{x}_i, \tilde{\bm{v}}_i)\right\}_{i=1}^{\text{NOBS}} \Bigm| \left[\hat{\bm{v}}_i\right]_{i=1}^{\text{NVB}}\right)}_{\textbf{Likelihood}} \cdot \underbrace{\mathbb{P}\left(\left[\hat{\bm{v}}_i\right]_{i=1}^{\text{NVB}}, \left[\hat{p}_i\right]_{i=1}^{\text{NPB}}\right)}_{\textbf{Prior}}.
     \label{eq:Bayes_theorem}
\end{align}

\noindent Our goal is to obtain the posterior distribution on the left, that is the probability of realizing a set of velocity and pressure nodal values given the velocity observations. Once the posterior is obtained, its mode, that is the most probable nodal values given the observations, can be combined with the basis functions via \eqref{eq:fe_expansion} to reconstruct velocity and pressure fields anywhere in the domain to arbitrary resolution. Additionally, the posterior variance provides a measure of the uncertainty associated with these reconstructions. The remainder of this section will focus on constructing each of the terms in \eqref{eq:Bayes_theorem} and using the resulting posterior to obtain reconstructions and quantify their uncertainty.

\subsubsection{Prior distribution}

We derive a physics-based prior that encodes known characteristics of physiological blood flow, such as smoothness, boundedness, mass conservation, and momentum conservation. We quantify the boundedness and smoothness of the velocity and pressure fields by introducing the quadratic forms,

\begin{align*}
    \mathcal{L}_v\left(\left[\hat{\bm{v}}\right]\right) &\coloneqq \frac{1}{|\Omega^h|}\int_{\Omega^h} \norm{\bm{v}^h}_2^2 + \alpha_v\norm{\nabla\bm{v}^h}_2^2 \ d\Omega^h = \frac{1}{|\Omega^h|}\left[\hat{\bm{v}}\right]^T \left(\bm{M}_v + \alpha_v\bm{K}_v\right) \left[\hat{\bm{v}}\right], \\
     \mathcal{L}_p\left(\left[\hat{p}\right]\right) &\coloneqq\frac{1}{|\Omega^h|}\int_{\Omega^h} (\bm{p}^h)^2 + \alpha_p \norm{\nabla p^h}_2^2 \ d\Omega^h = \frac{1}{|\Omega^h|}\left[\hat{p}\right]^T \left(\bm{M}_p + \alpha_p\bm{K}_p\right) \left[\hat{p}\right],
\end{align*}

\noindent where we dropped the enumeration up to NVB and NPB from $\left[\hat{\bm{v}}_i\right]_{i = 1}^{\text{NVB}}$ and $\left[\hat{p}_i\right]_{i = 1}^{\text{NPB}}$, respectively, for notational simplicity. Here, $\bm{M}_v$ and $\bm{M}_p$ are the finite element mass matrices associated with velocity and pressure, respectively, while $\bm{K}_v$ and $\bm{K}_p$ are the corresponding stiffness matrices. $\alpha_v$ and $\alpha_p$ are design parameters that characterize boundedness of the field relative to its smoothness. To represent our ignorance on the degrees of freedom before acquiring observations, we seek the least informative (maximum entropy) probability distribution on $\left[\hat{\bm{v}}\right]$ and $\left[\hat{p}\right]$ under the expectation constraints,

\begin{align*}
\mathbb{E}\left[\mathcal{L}_v\left(\left[\hat{\bm{v}}\right]\right)\right] \leq \delta_v \quad \text{and} \quad \mathbb{E}\left[\mathcal{L}_p\left(\left[\hat{p}\right]\right)\right] \leq \delta_p,
\end{align*}

\noindent respectively. Such distribution can be obtained by solving the variational optimization problem,

\begin{align}
    \max_{\mathbb{Q}} \ -\int \mathbb{Q}\left(\left[\hat{\bm{v}}\right], \left[\hat{p}\right]\right) \log \mathbb{Q}\left(\left[\hat{\bm{v}}\right], \left[\hat{p}\right]\right) \ d\left[\hat{\bm{v}}\right] d\left[\hat{p}\right], \quad \text{s.t.} \quad
    \mathbb{E}\left[\mathcal{L}_v\right] \leq \delta_v, \quad \mathbb{E}\left[\mathcal{L}_p\right] \leq \delta_p.
    \label{eq:max_entropy_formulation}
\end{align}

\noindent The solution to \eqref{eq:max_entropy_formulation} is given by, 

\begin{align}
    \mathbb{Q}\left([\hat{\bm{v}}], \left[\hat{p}\right]\right) \propto \exp\left\{- \frac{\lambda_v}{2|\Omega^h|}[\hat{\bm{v}}]^T\left(\bm{M}_v + \alpha_v\bm{K}_v\right)[\hat{\bm{v}}] - \frac{\lambda_p}{2|\Omega^h|}[\hat{p}]^T\left(\bm{M}_p + \alpha_p\bm{K}_p\right)[\hat{p}]\right\},
    \label{eq:smoothness_prior}
\end{align}

\noindent where $\lambda_v = 3\text{NVB}/\delta_v$ and $\lambda_p = \text{NPB}/\delta_p$. $\delta_v$ and $\delta_p$ can be chosen by setting  the desired level of boundedness and smoothness in the reconstructions. They can also be estimated by computing $\mathcal{L}_v$ and $\mathcal{L}_p$ on CFD generated flow fields or flow-fields coming from experimental datasets. The distribution $\mathbb{Q}\left([\hat{\bm{v}}], \left[\hat{p}\right]\right)$ allocates more probability mass on of velocity and pressure fields which have bounded $L^2$ and $H^1$ norms. It also defines two independent mean zero Gaussian processes over the admissible velocity and pressure fields such that,

\begin{align*}
\bm{v}^h(\bm{x}) \sim& \ \mathcal{GP}\left(0, \left(\left[\phi_i(\bm{x})\bm{I}_{3\times 3}\right]_{i = 1}^{\text{NVB}}\right)^T \lambda_v\left(\bm{M}_v + \alpha_v\bm{K}_v\right) \left(\left[\phi_i(\bm{y})\bm{I}_{3\times 3}\right]_{i = 1}^{\text{NVB}}\right)\right), \\
 p^h(\bm{x}) \sim& \ \mathcal{GP}\left(0, \left(\left[\psi_i(\bm{x})\right]_{i = 1}^{\text{NVP}}\right)^T \lambda_p\left(\bm{M}_p + \alpha_p\bm{K}_p\right) \left(\left[\psi_i(\bm{y})\right]_{i = 1}^{\text{NVP}}\right)\right).
\end{align*}

\noindent The constants $(\alpha_v, \lambda_v, \alpha_p, \lambda_p) > 0$ determine the variance and correlation length of the Gaussian processes. 

Steady incompressible fluid flow is governed by the Navier-Stokes equations,

\begin{align}
\begin{aligned}
    R_c(\bm{x}) \coloneqq \nabla \cdot \bm{v} &= 0, \quad \bm{x}\in \Omega, \\
  \bm{R}_m(\bm{x}) \coloneqq (\nabla \bm{v})\bm{v} -2\text{Re}^{-1} \nabla \cdot \bm{\varepsilon}\left(\bm{v}\right) + \nabla p &= \bm{0}, \quad \bm{x}\in \Omega, \\
  \label{eq:governing_equations}
\end{aligned}
\end{align}

\noindent where $\bm{\varepsilon}\left(\bm{v}\right) \coloneqq \frac{1}{2}\left(\nabla \bm{v} + \nabla \bm{v}^T\right)$. To quantify how much velocity and pressure fields deviate from Navier-Stokes equations, we project the residuals $R_c(\bm{x})$ and $\bm{R}_m(\bm{x})$ to the space of velocity and pressure basis functions, respectively, and compute their norms. We represent the norm of the projected residuals in terms of velocity and pressure coefficients,

\begin{align}
\begin{aligned}
    \mathcal{L}_c\left(\left[\hat{p}\right]\right) \coloneqq \frac{1}{|\Omega^h|}\int_{\Omega^h} \left(\Pi_pR_c(\bm{x})\right)^2 \quad d\Omega^h &= \frac{1}{|\Omega^h|}\norm{\bm{M}_{p}^{-\frac{1}{2}}\bm{B}[\hat{\bm{v}}]}_2^2, \\
    \mathcal{L}_m\left(\left[\hat{\bm{v}}\right], \left[\hat{p}\right]\right) \coloneqq \frac{1}{|\Omega^h|}\int_{\Omega^h} \norm{\Pi_v\bm{R}_m(\bm{x})}_2^2 \quad d\Omega^h &= \frac{1}{|\Omega^h|}\norm{\bm{M}_{v}^{-\frac{1}{2}}\left(\text{Re}^{-1}\tilde{\bm{K}}_v[\hat{\bm{v}}] + \bm{A}\left([\hat{\bm{v}}]\right) -\tilde{\bm{B}}^T[\hat{p}]\right)}_2^2.
    \label{eq:projected_residual}
\end{aligned}
\end{align}

\noindent It is worthwhile to mention that, in the LSFEM approach \cite{Bochev2009}, $\mathcal{L}_c$ and $\mathcal{L}_m$ are defined directly as $L^2$-norms of the strong-form residuals without projection. Because the strong momentum residual contains second derivatives of velocity, which do not lie in $L^2(\Omega)$ for
$C^0$ Taylor–Hood spaces, this requires recasting Navier–Stokes as a first-order system in auxiliary variables (e.g., velocity gradient, vorticity, stress), leaving only first derivatives. In contrast, we perform projection and integration by parts to reduce the differentiation order of the viscous term, which gives a Galerkin residual that vanishes exactly when the discrete weak form is satisfied The matrices in \eqref{eq:projected_residual} are defined by the bilinear forms,

\begin{align*}
\left(\left[\hat{\bm{w}}_i\right]_{i=1}^{\text{NVB}}\right)^T\tilde{\bm{K}} \left[\hat{\bm{v}}_i\right]_{i=1}^{\text{NVB}} &= 2\int_{\Omega^h} \bm{\varepsilon}(\bm{v}^h)\colon\bm{\varepsilon}(\bm{w}^h) \ d\Omega - 2 \int_{\Gamma_{\text{in}}^h \cup \Gamma_{\text{out}}^h} \bm{\varepsilon}(\bm{v}^h)\bm{w}^h\cdot\bm{n} \ d\Gamma,  \\
\left(\left[\hat{\bm{q}}_i\right]_{i=1}^{\text{NPB}}\right)^T \bm{B} \left[\hat{\bm{v}}_i\right]_{i=1}^{\text{NVB}} &= \int_{\Omega} q^h\nabla\cdot\bm{v}^h \ d\Omega, \\
\left(\left[\hat{\bm{w}}_i\right]_{i=1}^{\text{NVB}}\right)^T \tilde{\bm{B}}^T \left[\hat{\bm{p}}_i\right]_{i=1}^{\text{NPB}} &= \int_{\Omega^h} p^h\nabla\cdot\bm{w}^h \ d\Omega - \int_{\Gamma_{\text{in}}^h \cup \Gamma_{\text{out}}^h} p^h\bm{w}^h\cdot\bm{n} \ d\Gamma, \\
\left(\left[\hat{\bm{w}}_i\right]_{i=1}^{\text{NVB}}\right)^T\bm{A}\left( \left[\hat{\bm{v}}_i\right]_{i=1}^{\text{NVB}}\right) &= \int_{\Omega^h} \bm{w}^h\cdot(\nabla\bm{v}^h)\bm{v}^h \ d\Omega,\\
\end{align*}

\noindent where $\left[\hat{\bm{w}}_i\right]_{i=1}^{\text{NVB}}$ and $\left[\hat{q}_i\right]_{i=1}^{\text{NPB}}$ are the nodal values of arbitrary velocity and pressure test functions, respectively, belonging to \eqref{eq:spaces}. Observe that the definitions of $\tilde{\bm{K}}$ and $\tilde{\bm{B}}^T$ contain boundary integrals over the inlet and outlets. In a forward problem, these would be eliminated by a prescribed velocity or traction. Since the inlet and outlet conditions are unknown, the integrals are retained, and the inlet/outlet behavior is determined by the data and the physics prior rather than being imposed a priori. We integrate information contained in the governing equations \eqref{eq:governing_equations} to our prior, by seeking the distribution that adds the least information to the smoothness prior, subject to constraints on the expected projected norms of the discretized continuity and momentum equations. Similar to \eqref{eq:max_entropy_formulation}, this distribution can be obtained by solving the variational optimization problem,

\begin{align}
    \max_{\mathbb{P}} \ -\int \mathbb{P}([\hat{\bm{v}}], [\hat{p}]]) \log \frac{\mathbb{P}([\hat{\bm{v}}], [\hat{p}])}{\mathbb{Q}([\hat{\bm{v}}], [\hat{p}])} \ d\left[\hat{\bm{v}}\right] \ d\left[\hat{p}\right], \quad \text{s.t.} \quad
    \mathbb{E}\left[\mathcal{L}_c\right] \leq \delta_c, \quad \mathbb{E}\left[\mathcal{L}_m\right] \leq \delta_m.
    \label{eq:max_relative_entropy_formulation}
\end{align}

\noindent The solution is given by,

\begin{align}
    \mathbb{P}\left([\hat{\bm{v}}], [\hat{p}]\right) \propto \mathbb{Q}\left([\hat{\bm{v}}], [\hat{p}]\right)  \exp\Biggl\{-\frac{\lambda_c}{2|\Omega^h|}\norm{\bm{M}_p^{-\frac{1}{2}}\bm{B}[\hat{\bm{v}}]}_2^2 - \frac{\lambda_m}{2|\Omega^h|}\norm{\bm{M}_v^{-\frac{1}{2}}\left(\text{Re}^{-1}\tilde{\bm{K}_v}[\hat{\bm{v}}] + \bm{A}\left([\hat{\bm{v}}]\right) -\tilde{\bm{B}}^T[\hat{p}]\right)}_2^2 \Biggr\}.
    \label{eq:full_prior}
\end{align}

\noindent \eqref{eq:full_prior} is guaranteed to be normalizable due to initial smoothness prior $\mathbb{Q}$. The mapping from the pair of moment tolerances $(\delta_c, \delta_m)$ to the pair of multipliers $(\lambda_c, \lambda_m)$ is one-to-one in the region where constraints are active. This is because, no nontrivial affine combination of $\mathcal{L}_c$ and $\mathcal{L}_m$ is almost surely constant which makes the log-partition function of \eqref{eq:full_prior} strictly-convex. Hence, we parameterize the prior directly by the multipliers. The maximum entropy and maximum relative entropy approaches in \eqref{eq:max_entropy_formulation}, \eqref{eq:max_relative_entropy_formulation} are widely used way integrating constraints to Bayesian inference \cite{Skilling1988, Caticha2021}. It is worthwhile to note that, the same prior \eqref{eq:full_prior} can be obtained by introducing virtual observations of the constraint residuals,

\begin{align*}
    \bm{y}_c = \bm{B}[\hat{\bm{v}}] + \bm{\mathcal{E}}_c, \qquad & \bm{\mathcal{E}}_c \sim \mathcal{N}\left(\bm{0}, \frac{|\Omega^h|}{\lambda_c}\bm{M}_{p}\right), \\
    \bm{y}_m = \text{Re}^{-1}\tilde{\bm{K}}_v[\hat{\bm{v}}] + \bm{A}\left([\hat{\bm{v}}]\right) - \tilde{\bm{B}}^T[\hat{p}] + \bm{\mathcal{E}}_m, \qquad & \bm{\mathcal{E}}_m \sim \mathcal{N}\left(\bm{0}, \frac{|\Omega^h|}{\lambda_m}\bm{M}_{v}\right),
\end{align*}

\noindent conditioning on $\bm{y}_c = \bm{y}_m = \bm{0}$, and updating the prior in \eqref{eq:smoothness_prior} by Bayes' theorem \cite{WJ2019}. Here, the multipliers act as inverse noise levels on the virtual constraint observations.

\subsubsection{Likelihood}
\label{sec:likelihood}


We will assume additive Gaussian noise such that,

\begin{align*}
    \left[\tilde{\bm{v}}_i\right]_{i=1}^{\text{NOBS}} = \bm{T}\left[\hat{\bm{v}}_i\right]_{i=1}^{\text{NBV}} + \bm{\mathcal{E}}, \quad \text{with} \quad \bm{\mathcal{E}} \sim \mathcal{N}(\bm{0}_{3\text{NOBS}}, \bm{\Sigma}_o),
\end{align*}

\noindent where $\bm{T} \in \mathbb{R}^{3\text{NOBS}\times 3\text{NN}}$ is a matrix modeling the observation operator and $\bm{\Sigma}_0 \in \mathbb{R}^{3\text{NOBS}\times 3\text{NOBS}}$ is the covariance matrix of the Gaussian noise. Depending on the setting, $\bm{T}$ can either be an interpolation matrix where observations are pointwise evaluations of the velocity field such that,

\begin{align*}
\left[\bm{v}^h\left(\bm{x}_i\right)\right]_{i=1}^{\text{NOBS}} = \bm{T}\left[\hat{\bm{v}}_i\right]_{i=1}^{\text{NBV}},
\end{align*}

\noindent or a voxel averaging matrix where observations are average values of the velocity field over the voxels such that, 

\begin{align*}
\left[\int_ {V_i} \bm{v}^h\left(\bm{x}\right) \ d\bm{x}\right]_{i=1}^{\text{NOBS}} = \bm{T}\left[\hat{\bm{v}}_i\right]_{i=1}^{\text{NBV}},  
\end{align*}

\noindent where $V_i$ is the volume of $i$-th voxel. In either case, $\bm{T}$ will be a sparse matrix whose nonzero entries are basis functions evaluated at observation points for pointwise observations or basis functions evaluated at quadrature points for voxelwise observations. We will also assume that observations are independent implying $\bm{\Sigma}_0 = \sigma_0^2\bm{I}_{\text{NOBS}\times\text{NOBS}}$, for some constant variance parameter $\sigma_0^2$ which will be known to us through machine specifications. If any covariance structure is known about the noise model it can easily be integrated in to non-diagonal entries of $\bm{\Sigma_0}$. Then, the likelihood is given by the Gaussian,

\begin{align*}
     \mathbb{P}\left(\left\{(\bm{x}_i, \bm{v}_i)\right\}_{i=1}^{\text{NOBS}} \Bigm| \left[\hat{\bm{v}}_i\right]_{i=1}^{\text{NVB}}, \ \sigma_0^2\right) \propto \exp\left\{-\frac{1}{2\sigma_0^2}\norm{\left[\tilde{\bm{v}}_i\right]_{i=1}^{\text{NOBS}} - \bm{T}\left[\hat{\bm{v}}_i\right]_{i=1}^{\text{NVB}}}_2^2\right\}.
\end{align*}

\subsubsection{Posterior distribution}
\label{sec:posterior}

The likelihood and the prior are multiplied to yield the posterior distribution,

\begin{align*}
    \mathbb{P}\left(\left[\hat{\bm{v}}\right], \left[\hat{p}\right] \Bigm| \left\{(\bm{x}_i, \bm{v}_i)\right\}_{i=1}^{\text{NOBS}}, \ \sigma_0^2\right) \propto \exp\Biggl\{ -& \frac{1}{2\sigma^2}\norm{\left[\tilde{\bm{v}}\right] -
    \bm{T}\left[\hat{\bm{v}}\right]}_2^2 - \frac{\lambda_v}{2|\Omega^h|}[\hat{\bm{v}}]^T\left(\bm{M}_v + \alpha_v\bm{K}_v\right)[\hat{\bm{v}}] \ + \\
    &\frac{\lambda_p}{2|\Omega^h|}[\hat{p}]^T\left(\bm{M}_p + \alpha_p\bm{K}_p\right)[\hat{p}] - \frac{\lambda_c}{2|\Omega^h|}\norm{\bm{M}_p^{-\frac{1}{2}}\bm{B}[\hat{\bm{v}}]}_2^2 - \\ 
    &\frac{\lambda_m}{2|\Omega^h|}\norm{\bm{M}_v^{-\frac{1}{2}}\left(\text{Re}^{-1}\tilde{\bm{K}_v}[\hat{\bm{v}}] + \bm{A}\left([\hat{\bm{v}}]\right) -\tilde{\bm{B}}^T[\hat{p}]\right)}_2^2 \Biggr\}. 
\end{align*}

\noindent Thus, the most likely velocity and pressure degrees of freedom in light of the observations are given by the solution of the optimization problem,

\begin{equation}
\begin{aligned}
  \left[\hat{\bm{v}}\right]^*, \left[\hat{p}\right]^* &= \argmax_{\left[\hat{\bm{v}}\right] \in \mathbb{R}^{3\text{NVB}}, \ \left[\hat{p}\right] \in \mathbb{R}^{\text{NPB}}} \quad \mathbb{P}\left(\left[\hat{\bm{v}}\right]^*, \left[\hat{p}\right]^* \Bigm| \left\{(\bm{x}_i, \bm{v}_i)\right\}_{i=1}^{\text{NOBS}}, \ \sigma_0^2\right)\\
  &= \argmin_{\left[\hat{\bm{v}}\right] \in \mathbb{R}^{3\text{NVB}}, \ \left[\hat{p}\right] \in \mathbb{R}^{\text{NPB}}} \quad -\log\left\{\mathbb{P}\left(\left[\hat{\bm{v}}\right]^*, \left[\hat{p}\right]^* \Bigm| \left\{(\bm{x}_i, \bm{v}_i)\right\}_{i=1}^{\text{NOBS}}, \ \sigma_0^2\right)\right\} \\
  &= \argmin_{\left[\hat{\bm{v}}\right] \in \mathbb{R}^{3\text{NVB}}, \ \left[\hat{p}\right] \in \mathbb{R}^{\text{NPB}}} \quad \underbrace{\cfrac{1}{2\text{NOBS}}\norm{\left[\tilde{\bm{v}}\right] - \bm{T}\left[\hat{\bm{v}}\right]}_2^2}_{\textbf{Data}} + \underbrace{\cfrac{\mu_s}{2|\Omega^h|} \begin{bmatrix}
    [\hat{\bm{v}}] \\
    [\hat{p}]
\end{bmatrix}^T\begin{bmatrix}
    \bm{M}_v + \alpha_v\bm{K}_v & \bm{0} \\
    \bm{0} & \bm{M}_p + \alpha_p\bm{K}_p
\end{bmatrix} \begin{bmatrix}
    [\hat{\bm{v}}] \\
    [\hat{p}]
\end{bmatrix}}_{\textbf{Smoothness}} \\
    & \quad \quad \quad \quad \quad \quad \quad + \underbrace{\cfrac{\mu_c}{2|\Omega^h|}\norm{\bm{M}_p^{-\frac{1}{2}}\bm{B}\left[\hat{\bm{v}}\right]}_2^2}_{\textbf{Continuity}} + 
    \underbrace{\cfrac{\mu_m}{2|\Omega^h|}\norm{\bm{M}_v^{-\frac{1}{2}}\left(\text{Re}^{-1}\tilde{\bm{K}}\left[\hat{\bm{v}}\right] - \tilde{\bm{B}}^T\left[\hat{p}\right] + \bm{A}(\left[\hat{\bm{v}}\right])\right)}_2^2}_{\textbf{Momentum}} \\[1em]
    &= \argmin_{\left[\hat{\bm{v}}\right] \in \mathbb{R}^{3\text{NVB}}, \ \left[\hat{p}\right] \in \mathbb{R}^{\text{NPB}}} \quad F\left(\left[\hat{\bm{v}}\right], \left[\hat{p}\right]\right).
    \label{eq:MAP}
\end{aligned}
\end{equation}

\noindent Here the velocity values corresponding to the velocity nodes on the vessel walls are set to zero and are not included inside $\left[\hat{\bm{v}}\right]$. Similarly, to remove the constant pressure null-space, we fix the pressure to zero at a single node in our domain and exclude it from $\left[\hat{p}\right]$. We define $\mu_i = \frac{\sigma_0^2 \lambda_i}{\text{NOBS}}$ for $i = s, c, m$ for algebraic convenience. \eqref{eq:MAP} is a large scale, sparse, nonlinear least squares optimization problem. The first term penalizes the data mismatch, the second term penalizes roughness, and the third and fourth terms penalize deviations from the Navier-Stokes equations. The only nonlinearity stems from the discretized advection vector $\bm{A}(\left[\hat{\bm{v}}\right])$ appearing inside the momentum term. Unlike the other terms inside the norm, which are linear functions of degrees of freedom, $\bm{A}(\left[\hat{\bm{v}}\right])$ is quadratic function of $\left[\hat{\bm{v}}\right]$. Its strength is determined by the Reynolds number. The objective function $F\left(\left[\hat{\bm{v}}\right], \left[\hat{p}\right]\right)$ has the same structure with a PINN loss, a data-misfit term together with weighted penalties on the governing-equation residuals. It differs in that the fields are linear in the unknown coefficients, so the objective is a degree-four polynomial in the coefficients rather than a composite nonlinear function of weights. Such structure gives a more favorable optimization landscape compared to PINNs. Expressing the objective function as,

\begin{align*}
    F\left(\left[\hat{\bm{v}}\right], \left[\hat{p}\right]\right) \coloneq \dfrac{1}{2}\norm{\bm{r}\left(\left[\hat{\bm{v}}\right], \left[\hat{p}\right]\right)}_2^2 + \begin{bmatrix}
         \left[\hat{\bm{v}}\right] \\
         \left[\hat{p}\right]
     \end{bmatrix}^T\bm{S}\begin{bmatrix}
         \left[\hat{\bm{v}}\right] \\
         \left[\hat{p}\right]
     \end{bmatrix},
\end{align*}

\noindent where the residual vector $(\bm{r})$ and smoothness matrix $(\bm{S})$ are,

{\begin{align*}
     \bm{r}\left(\left[\hat{\bm{v}}\right], \left[\hat{p}\right]\right) &\coloneqq
     \begin{bmatrix}
         \dfrac{1}{\sqrt{\text{NOBS}}}\left(\bm{T}\left[\hat{\bm{v}}\right] - \left[\tilde{\bm{v}}\right]\right) \\[1.0em]
         \sqrt{\dfrac{\mu_c}{|\Omega^h|}} \bm{M}_p^{-\frac{1}{2}}\bm{B}\left[\hat{\bm{v}}\right] \\[1.0em]
         \sqrt{\dfrac{\mu_m}{|\Omega^h|}} \bm{M}_v^{-\frac{1}{2}}\left(\text{Re}^{-1}\tilde{\bm{K}}\left[\hat{\bm{v}}\right] - \tilde{\bm{B}}^T\left[\hat{p}\right] + \bm{A}(\left[\hat{\bm{v}}\right])\right)
     \end{bmatrix}, \\[1.0em]
     \bm{S} &\coloneqq \sqrt{\frac{\mu_s}{|\Omega^h|}}\begin{bmatrix}
    \bm{M}_v + \alpha_v\bm{K}_v & \bm{0} \\
    \bm{0} & \bm{M}_p + \alpha_p\bm{K}_p
\end{bmatrix},
\end{align*}}

\noindent gradient of the objective function is,

\begin{align*}
   \begin{bmatrix}  \dfrac{\partial F}{\partial \left[\hat{\bm{v}}\right]} \\[1.0em]
     \dfrac{\partial F}{\partial \left[\hat{p}\right]} \end{bmatrix} = \bm{J}\left(\left[\hat{\bm{v}}\right], \left[\hat{p}\right]\right)^T\bm{r}\left(\left[\hat{\bm{v}}\right], \left[\hat{p}\right]\right) + \bm{S}\begin{bmatrix}
         \left[\hat{\bm{v}}\right] \\
         \left[\hat{p}\right]
     \end{bmatrix}.
\end{align*}

\noindent Here $\bm{J}$ is the Jacobian of the residual vector taking the form,

{\begin{align*}
    \bm{J}\left(\left[\hat{\bm{v}}\right], \left[\hat{p}\right]\right)
     &\coloneqq \begin{bmatrix}
         \dfrac{1}{\sqrt{\text{NOBS}}}\bm{T} & \bm{0} \\[1.0em]
         \sqrt{\dfrac{\mu_c}{|\Omega^h|}}\bm{M}_p^{-\frac{1}{2}}\bm{B} & \bm{0} \\[1.0em]
          \sqrt{\dfrac{\mu_m}{|\Omega^h|}} \bm{M}_v^{-\frac{1}{2}}\left(\text{Re}^{-1}\tilde{\bm{K}} + \dfrac{\partial \bm{A}(\left[\hat{\bm{v}}\right])}{\partial \left[\hat{\bm{v}}\right]}\right) & - \sqrt{\dfrac{\mu_m}{|\Omega^h|}}\tilde{\bm{B}}^T
     \end{bmatrix}.
\end{align*}}

\noindent The optimality condition for pressure degrees of freedom ($\frac{\partial F}{\partial \left[\hat{p}\right]} = \bm{0}$) require,

\begin{align}
      \left[\hat{p}\right]^* = \bm{\Pi}\left([\hat{\bm{v}}]^*\right) \coloneqq \left(\mu_s(\bm{M}_p + \alpha_p\bm{K}_p) + \mu_m \tilde{\bm{B}} \bm{M}_v^{-1} \tilde{\bm{B}}^T\right)^{-1}\mu_m \tilde{\bm{B}} \bm{M}_v^{-1}\left(\text{Re}^{-1}\tilde{\bm{K}}[\hat{\bm{v}}]^* + \bm{A}([\hat{\bm{v}}]^*)\right).
      \label{eq:pressure_elimination}
\end{align}

\noindent This constraint can be employed to reduce the dimensionality of the optimization problem \eqref{eq:MAP} by eliminating the pressure degrees of freedom. Then, the reduced problem becomes,  

\begin{align*}
    \left[\hat{\bm{v}}\right]^* \coloneqq \arg\min_{\left[\hat{\bm{v}}\right]} \quad F\left(\left[\hat{\bm{v}}\right], \bm{\Pi}\left([\hat{\bm{v}}]\right) \right) = \arg\min_{\left[\hat{\bm{v}}\right]} \quad \tilde{F}\left(\left[\hat{\bm{v}}\right]\right).
\end{align*}

\noindent Note that, evaluating the gradient of the reduced objective function $\tilde{F}\left(\left[\hat{\bm{v}}\right]\right)$ at $[\hat{\bm{v}}]$ is equivalent to evaluating the gradient of the original objective function at $\left([\hat{\bm{v}}], \bm{\Pi}\left([\hat{\bm{v}}]\right)\right)$ because,

\begin{align}
    \frac{\partial \tilde{F}}{\partial [\hat{\bm{v}}]} = \frac{\partial F}{\partial [\hat{\bm{v}}]}\Bigg\vert_{\left([\hat{\bm{v}}], \ \bm{\Pi}\left([\hat{\bm{v}}]\right)\right)} + \left(\frac{\partial \bm{\Pi}}{\partial [\hat{\bm{v}}]}\right)^T\cancelto{0}{\frac{\partial F}{\partial [\hat{p}]}\Bigg\vert_{\left([\hat{\bm{v}}], \ \bm{\Pi}\left([\hat{\bm{v}}]\right)\right)}} = \frac{\partial F}{\partial [\hat{\bm{v}}]}.
\end{align}

\noindent We solve \eqref{eq:MAP} in reduced space as it yields a better conditioned problem and accelerates the convergence of the numerical solution. The solution to \eqref{eq:MAP} corresponds the mode of the posterior and is called the MAP estimate. We use the MAP estimate degrees of freedom along with basis functions to form our reconstructions. To quantify the uncertainty in the estimated degrees of freedom we need the posterior covariance. In principle, this can be obtained via Markov chain Monte Carlo (MCMC) methods. However, in high-dimensional settings, where the unknowns correspond to finite-element degrees of freedom, generating sufficiently many posterior samples becomes intractable. Variational inference methods, which minimize the divergence between assumed distributions and the posterior, offer an alternative. Yet, they often require the solution of an optimization problem more difficult than \eqref{eq:MAP}. We employ the Laplace approximation to obtain the variance of the posterior distribution. Laplace approximation estimates the density function of the posterior with a Gaussian using the second order Taylor expansion of log posterior at the MAP such that,

\begin{align}
    \mathbb{P}\left(\left[\hat{\bm{v}}\right], \left[\hat{p}\right] \Bigm| \left\{(\bm{x}_i, \bm{v}_i)\right\}_{i=1}^{\text{NOBS}}, \ \sigma_0^2\right) \propto \exp\left\{-\frac{1}{2}\left(\dfrac{\text{NOBS}}{\sigma_0^2}\right)\begin{bmatrix}
        \left[\hat{\bm{v}}\right] - \left[\hat{\bm{v}}\right]^*  \\
        \left[\hat{p}\right] - \left[\hat{p}\right]^* \end{bmatrix}^T \nabla^2 F\left(\left[\hat{\bm{v}}\right]^*, \left[\hat{p}\right]^*\right) \begin{bmatrix}
        \left[\hat{\bm{v}}\right] - \left[\hat{\bm{v}}\right]^* \\
        \left[\hat{p}\right] - \left[\hat{p}\right]^* 
    \end{bmatrix} \right\}.
    \label{eq:Laplace_approximation}
\end{align}

\noindent $\nabla^2 F\left(\left[\hat{\bm{v}}\right]^*, \left[\hat{p}\right]^*\right)$ is the Hessian of the objective function in \eqref{eq:MAP} evaluated at the MAP and is given by,

\begin{align}
    \nabla^2 F\left(\left[\hat{\bm{v}}\right]^*, \left[\hat{p}\right]^*\right) = \bm{J}\left(\left[\hat{\bm{v}}\right]^*, \left[\hat{p}\right]^*\right)^T\bm{J}\left(\left[\hat{\bm{v}}\right]^*, \left[\hat{p}\right]^*\right) + \bm{S} + \begin{bmatrix}
       \dfrac{\partial^2 \bm{\xi}^T\left(\bm{A}\left(\left[\hat{\bm{v}}\right]\right)\right)}{\partial \left[\hat{\bm{v}}\right]}\Bigg\vert_{\left[\hat{\bm{v}}\right] = \left[\hat{\bm{v}}\right]^*} & \bm{0} \\[1.0em]
       \bm{0} & \bm{0}
    \end{bmatrix},
    \label{eq:obj_Hessian}
\end{align}

\noindent with $\bm{\xi} = \cfrac{\mu_m}{|\Omega^h|}\bm{M}_v^{-1}\left(\text{Re}^{-1}\tilde{\bm{K}}\left[\hat{\bm{v}}\right] - \tilde{\bm{B}}^T\left[\hat{p}\right] + \bm{A}(\left[\hat{\bm{v}}\right])\right)$. This approach is particularly attractive for large-scale problems, as it leverages quantities already available from the optimization procedure like the Jacobian matrix and the residual vector. The accuracy of the Laplace approximation depends on how well the posterior can be represented locally by a Gaussian distribution. The true posterior deviates from a Gaussian due to cubic and quartic $\left[\hat{\bm{v}}\right]$ terms coming out of the momentum term. We expect the large number of observations to concentrate the posterior around the MAP estimate, assigning small probability mass to regions where the Gaussian approximation is poor. Consequently, the Laplace approximation can provide accurate estimates of posterior moments even when it does not accurately represent the global posterior shape. Moreover, for moderate Reynolds numbers the advection effects will be weak compared to the viscous contribution, causing these higher-order terms to remain small in the vicinity of the MAP estimate. Since the momentum term is penalized it is common to expect that $\norm{\bm{\xi}}_2$ is small and neglect the last term in \eqref{eq:obj_Hessian} as well giving the \textit{Gauss-Newton} approximation of the Hessian. This also guarantees that the Hessian is positive semi-definite, which is required for the covariance matrix of the Laplace approximation of the posterior to be well-defined. Then, the covariance matrix of the Laplace approximation ($\bm{\Sigma}$) is given by,

\begin{align*}
    \bm{\Sigma} = \frac{\sigma_0^2}{\text{NOBS}}\left(\nabla^2 F\left(\left[\hat{\bm{v}}\right]^*, \left[\hat{p}\right]^*\right)\right)^{-1}.
\end{align*}

\noindent Under Laplace approximation, the posterior distribution on nodal degrees of freedom define a Gaussian process on the reconstructed velocity and pressure fields which is expressed by,

\begin{align*}
\begin{bmatrix}
    \bm{v}^h(\bm{x})\\
    p^h(\bm{x})
\end{bmatrix} \sim
\mathcal{GP}\left(\bm{N}\left(\bm{x}\right) \begin{bmatrix}
    \left[\hat{\bm{v}}\right]^{*} \\
    \left[\hat{p}\right]^{*}
\end{bmatrix}, \
\bm{N}\left(\bm{x}\right)
\bm{\Sigma}\bm{N}\left(\bm{x}'\right)^T\right).
\end{align*}.

\noindent The mean function of this Gaussian process provides the reconstructed velocity and pressure fields, while the covariance function can be used to evaluate the marginal variance fields associated with the reconstructions,

\begin{align*}
    \begin{bmatrix}
        \mathbb{V}\left[\bm{v}^h(\bm{x})\right]\\
        \mathbb{V}\left[p^h(\bm{x})\right]
    \end{bmatrix} = 
\bm{N}\left(\bm{x}\right)\bm{\Sigma}\bm{N}\left(\bm{x}\right)^T,
\end{align*}

\noindent which quantify the uncertainty on the reconstructed fields at any point $\bm{x}\in\Omega^h$. This uncertainty can be propagated to various hemodynamic quantities of interest (QoI). In this work, the QoIs we will consider are: flow rate and average pressure over specified planes, pressure drop over vessel centerline, and WSS magnitude profile.
Under Laplace approximation, QoIs that are linear functionals of velocity or pressure will also follow a Gaussian distribution. If $\bm{Q}$ is the matrix representation of the linear mapping from degrees of freedom to QoIs, then,

\begin{align}
    \textbf{QoI} \sim \mathcal{N}\left(\bm{Q}\begin{bmatrix}
    \left[\hat{\bm{v}}\right]^* \\
    \left[\hat{p}\right]^*
    \end{bmatrix}, \bm{Q}\bm{\Sigma}\bm{Q}^T\right),
    \label{eq:Laplace_qoi}
\end{align}
    
\noindent implying that the mean and covariance of the posterior QoI are analytically available. The diagonal entries of $\bm{Q}\bm{\Sigma}\bm{Q}^T$ are the marginal variances of the QoIs and quantify the uncertainty in their predictions. For the $i$-th QoI, the variance is obtained by solving,

\begin{align*}
\left(\frac{\text{NOBS}}{\sigma_0^2}\nabla^2 F\right)\bm{u} = \bm{q}_i,
\end{align*}

\noindent and computing $\bm{q}_i^T\bm{u}$ where $\bm{q}_i$ is the $i$-th row of $\bm{Q}$. When sufficient memory is available, a single Cholesky factorization of $\nabla^2F$ can be reused for all QoIs. Otherwise, the systems may be solved iteratively using Krylov subspace methods with a reusable preconditioner. For average pressure $(\langle P \rangle)$ and flow-rate $(Q)$ over a specified planar surface $S$, we have,

\begin{align*}
    \begin{bmatrix}
        Q \\
        \langle P \rangle
    \end{bmatrix} =
    \begin{bmatrix}
        \displaystyle\int\limits_{S} \bm{v}^h \cdot \bm{n} \ dS\\[1.5em]
        \displaystyle\dfrac{1}{|S|} \int\limits_{S} p^h \ dS
    \end{bmatrix} =
    \underbrace{\begin{bmatrix}
    \displaystyle\bm{n}^T\left(\left[\left(\int\limits_{S}\phi_i(\bm{x}) \ dS\right)\bm{I}_{3\times 3}\right]_{i = 1}^{\text{NVB}}\right)^T & \bm{0}\\[1.0em]
     \bm{0} & \displaystyle \frac{1}{|S|}\left(\left[\int_{S}\psi_i(\bm{x}) \ dS\right]_{i=1}^{\text{NPB}}\right)^T 
    \end{bmatrix}}_{\displaystyle\bm{Q}}
    \begin{bmatrix}
    \left[\hat{\bm{v}}\right] \\
    \left[\hat{p}\right]
    \end{bmatrix}. 
\end{align*}

\noindent We obtain the WSS field by $L^2$-projecting the tangential traction derived from the continuous, piecewise-quadratic velocity field onto a continuous, piecewise-linear finite element space on the vessel wall such that,

\begin{align*}
    \bm{\tau}_{\text{wall}}^h(\bm{x}) \coloneqq \arg\min_{\bm{\tau}(\bm{x}) \in \mathcal{T}^h} \quad \norm{\bm{\tau}(\bm{x}) - \frac{2}{\text{Re}}\left(\bm{I} - \bm{n} \otimes \bm{n}\right)\bm{\varepsilon}\left(\bm{v}^h\right)\bm{n}}_{L^2(\Gamma_{\text{wall}})}^2,
\end{align*}

\noindent with

\begin{align*}
\mathcal{T}^h
&\coloneq
\left\{p^h \in L^2(\Gamma_{\text{wall}}^h) \
 \Big\vert \ \bm{\tau}^h|_{K_i} \in \mathbb{P}_1(K_i) \ \ \forall K_i \in \Gamma_{\text{wall}}^h \right\}.
 \label{eq:spaces}
\end{align*}

\noindent Therefore,

\begin{align*}
    \bm{\tau}_{\text{wall}}^h(\bm{x}) = \left(\left[\psi_i(\bm{x})\bm{I}_{3\times 3}\right]_{i = 1}^{\text{NPB}}\right)^T\left[\hat{\bm{\tau}}_i\right]_{i = 1}^{\text{NPB}} = 
    \underbrace{\left(\left[\psi_i(\bm{x})\bm{I}_{3\times 3}\right]_{i = 1}^{\text{NPB}}\right)^T \begin{bmatrix}
       \bm{M}_{\tau}^{-1}\bm{D} & \bm{0}
    \end{bmatrix}}_{\displaystyle \bm{Q}}
    \begin{bmatrix}
    \left[\hat{\bm{v}}\right] \\
    \left[\hat{p}\right]
    \end{bmatrix}, 
\end{align*}

\noindent where, $\bm{M}_{\tau}$ and $\bm{D}$ are matrices defined by the bilinear forms,

\begin{align*}
\left(\left[\hat{\bm{t}}_i\right]_{i=1}^{\text{NPB}}\right)^T\bm{M}_{\tau}\left[\hat{\bm{\tau}}_i\right]_{i=1}^{\text{NPB}} &\coloneqq \int_{\Gamma_{\text{wall}}^h} \bm{t}^h \cdot \bm{\tau}^h \ d\Gamma_{\text{wall}}^h, \\
\left(\left[\hat{\bm{t}}_i\right]_{i=1}^{\text{NPB}}\right)^T\bm{D}\left[\hat{\bm{v}}_i\right]_{i=1}^{\text{NVB}} &\coloneqq \int_{\Gamma_{\text{wall}}^h} \bm{t}^h \cdot  \frac{2}{\text{Re}}\left(\bm{I} - \bm{n} \otimes \bm{n}\right)\bm{\varepsilon}\left(\bm{v}^h\right)\bm{n} \ d\Gamma_{\text{wall}}^h.
\end{align*}

\noindent Here, $\bm{t}^h$ is just an arbitrary test function in $\mathcal{T}^h$ and $\left[\hat{\bm{t}}_i\right]_{i=1}^{\text{NPB}}$ is the vector of its degrees of freedom. When the QoIs are nonlinear functions of the degrees of freedom, we estimate their statistics by Monte Carlo sampling $\left(\left[\hat{\bm{v}}\right], \left[\hat{p}\right]\right)$ from the Laplace approximation \eqref{eq:Laplace_approximation} and propagating the samples through the QoI map. For instance, to quantify uncertainty in the mean WSS magnitude, we draw samples of $\left[\hat{\bm{\tau}}\right]$ from \eqref{eq:Laplace_qoi}. For each sample, we evaluate the mean WSS magnitude by integrating the WSS magnitude over a selected region of vessel wall. The posterior mean and variance of the QoI are then estimated from the resulting Monte Carlo samples. To generate samples from \eqref{eq:Laplace_approximation} we can either use the Cholesky decomposition or low-rank eigen-decomposition of the Hessian. When QoIs are pointwise quantities evaluated on a high-resolution grid, storing all Monte Carlo samples can be memory intensive. Then, the sample mean and variance can be accumulated online using \textit{Welford's algorithm}, avoiding the storage of the ensemble. It is worth noting that, unlike the proposed full-space UQ framework, adjoint-based UQ approaches parameterized by a small number of control variables require repeated solutions of the steady Navier--Stokes equations to generate posterior samples. Consequently, the dominant computational cost in adjoint-based methods stems from repeated PDE solves, whereas in the proposed approach, it stems from one-time factorization of the large Hessian matrix.

\subsection{Divergence-constraining}

As opposed to the momentum equation, which contains an assumed rheological model and its associated parameters, incompressibility represents an exact kinematic constraint of blood flow, given accurate vessel segmentation. Moreover, incompressibility imposes a linear constraint on the velocity field, making it easier to incorporate into a probabilistic framework. Specifically, the discrete divergence matrix can be treated as a linear observation matrix, and mass conservation can be enforced by conditioning the velocity prior in \eqref{eq:smoothness_prior} on noiseless zero-divergence observations. This yields a reconstruction framework that gives weakly divergence-free velocity fields up to machine precision. The posterior assigns probability mass only on the divergence-free subspace defined by $\bm{B}\left[\hat{\bm{v}}\right] = \bm{0}$. Then, the MAP is given by the constrained optimization problem, 

\begin{equation}
\begin{aligned}
    \left[\hat{\bm{v}}\right]^*, \left[\hat{p}\right]^* = &\argmin_{\left[\hat{\bm{v}}\right] \in \mathbb{R}^{3\text{NVB}}, \ \left[\hat{p}\right] \in \mathbb{R}^{\text{NPB}}} \quad \cfrac{1}{2\text{NOBS}}\norm{\left[\tilde{\bm{v}}\right] - \bm{T}\left[\hat{\bm{v}}\right]}_2^2 + \cfrac{\mu_s}{2|\Omega^h|} \begin{bmatrix}
    [\hat{\bm{v}}] \\
    [\hat{p}]
\end{bmatrix}^T\begin{bmatrix}
    \bm{M}_v + \alpha_v\bm{K}_v & \bm{0} \\
    \bm{0} & \bm{M}_p + \alpha_p\bm{K}_p
\end{bmatrix} \begin{bmatrix}
    [\hat{\bm{v}}] \\
    [\hat{p}]
\end{bmatrix} \\
    & \quad \quad \quad \quad \quad \quad \quad \quad + 
    \cfrac{\mu_m}{2|\Omega^h|}\norm{\bm{M}_v^{-\frac{1}{2}}\left(\text{Re}^{-1}\tilde{\bm{K}}\left[\hat{\bm{v}}\right] - \tilde{\bm{B}}^T\left[\hat{p}\right] + \bm{A}(\left[\hat{\bm{v}}\right])\right)}_2^2 \\[1em]
    & \quad \ \ \ \text{such that} \quad \bm{B}\left[\hat{\bm{v}}\right] = \bm{0}.
    \label{eq:div_free_MAP}
\end{aligned}
\end{equation}

\noindent \eqref{eq:div_free_MAP} can be solved as an unconstrained problem with respect to $\left[\hat{\bm{u}}\right]$ by introducing the transformation,

\begin{align*}
    \left[\hat{\bm{v}}\right] = \left(\bm{I} - \bm{B}^T\left(\bm{B}\bm{B}^T\right)^{-1}\bm{B}\right)\left[\hat{\bm{u}}\right],
\end{align*}

\noindent along with the pressure elimination \eqref{eq:pressure_elimination}. As before, we can use the Laplace approximation of the posterior about the MAP to quantify uncertainties in our reconstructions. We can obtain the covariance matrix of the Laplace approximation via Schur complement as,

\begin{align*}
    \bm{\Sigma} \;=\; \bm{H}^{-1} - \bm{H}^{-1}\hat{\bm{B}}^{\mathsf{T}}\left(\hat{\bm{B}}\,\bm{H}^{-1}\hat{\bm{B}}^{\mathsf{T}}\right)^{-1}\hat{\bm{B}}\,\bm{H}^{-1},
\end{align*}

\noindent where $\hat{\bm{B}} = \left[\bm{B} \ \bm{0} \right]$, $\bm{H} = \frac{\text{NOBS}}{\sigma_0^2}{}\nabla^2G\left(\left[\hat{\bm{v}}\right]^*, \left[\hat{p}\right]^*\right)$, and $G$ is the objective function in \eqref{eq:div_free_MAP}. We can propagate the divergence-free posterior to QoIs in the same way as in section (\ref{sec:posterior}). In this framework, if the velocity and pressure finite-element spaces satisfy $\nabla \cdot \mathcal{V}^h = \mathcal{P}^h$, then the reconstructed velocity field is pointwise divergence-free. Examples of such exactly divergence-free finite-element pairs include: Scott–Vogelius, Raviart–Thomas, and Brezzi–Douglas–Marini elements. The latter two are more suitable when the velocity is subject to normal-flux boundary conditions.

\subsection{Baselines}

We consider two baseline methods for comparison. The first is the Lekien--Marsden tricubic interpolation \cite{Lekien2005} after Gaussian smoothing, and the second is a fully connected multilayer perceptron PINN trained using only observational data and strong-form governing-equation residuals. The former does not require the solution of a large-scale optimization problem and represents a computationally inexpensive reconstruction approach that is widely used in practice. The latter involves a large-scale optimization problem similar to that of the proposed method, incorporates the governing equations through penalty terms, and serves as a widely adopted physics-informed reconstruction framework.

\subsubsection{Lekien-Marsden tricubic interpolation}

We first denoised the velocity observations via a multidimensional Gaussian filter implemented as a sequence of 1-D convolution filters. We performed oracle tuning where we selected the standard deviation parameters of the Gaussian filter by minimizing the root mean squared error between the true and the denoised observations over all observation points. Although, this relies on the knowledge of the ground-truth solution which is only available in a synthetic study, it ensures near-optimal configuration of the denoising procedure. Then, we apply Lekien-Marsden tricubic interpolation on the optimally denoised observations to obtain a $C^1$ velocity interpolant ($\bm{v}_{\mathcal{I}}$). We project the interpolant to a weakly divergence-free finite element space by solving,

\begin{equation}
    \underset{v_{\mathcal{I}}^h \in \mathcal{V}^h}{\argmin} \quad \frac{1}{2} \int_{\Omega} \norm{\bm{v}_{\mathcal{I}} - \bm{v}_{\mathcal{I}}^h}_2^2 \ d\Omega \qquad \text{s.t.} \qquad \int_{\Omega} \nabla\cdot\bm{v}_{\mathcal{I}}^h \ d\Omega = 0.
    \label{eq:divergence_projection}
\end{equation}

\noindent This projection enhances the accuracy of both velocity and pressure predictions. Finally, we reconstruct the pressure field by solving the variational problem,
    
\begin{equation}
    \underset{p^h \in \mathcal{P}^h}{\argmin} \quad \frac{1}{2}\int_{\Omega} \norm{\nabla p^h - (\text{Re} \nabla \cdot \nabla \bm{v}_{\mathcal{I}}^h - (\nabla \bm{v}_{\mathcal{I}}^h)\bm{v}_{\mathcal{I}}^h)}_2^2 \ d\Omega.
    \label{eq:pressure reconstruction}
\end{equation}

\noindent Solving  \eqref{eq:pressure reconstruction} is equivalent to solving a pressure Poisson problem, as the optimality condition of \eqref{eq:pressure reconstruction} yields the weak-form of the Poisson problem with the boundary condition $\left(\nabla p^h - (\text{Re} \nabla \cdot \nabla \bm{v}_{\mathcal{I}}^h - (\nabla \bm{v}_{\mathcal{I}}^h)\bm{v}_{\mathcal{I}}^h)\right)\cdot \bm{n} = 0$. We pin the pressure to be zero at an arbitrary node of the mesh to eliminate the pressure null-space.

\subsubsection{Physics-informed neural networks}

We configure a fully-connected MLP that jointly predicts velocity and pressure such that,

\begin{equation*}
   [\bm{v}_{\text{MLP}}(\bm{x}), \ p_{\text{MLP}}(\bm{x})] = \text{MLP}(\bm{x}; \{w_i\}_{i=1}^{\text{NW}})
\end{equation*}

\noindent where $\{w_i\}_{i=1}^{\text{NW}}$ are the weights of the  neural network. We train it on the loss,

\begin{align*}
    \mathcal{L} &= \mathcal{L}_{\text{data}} + \lambda_{\text{continuity}}\mathcal{L}_{\text{continuity}} + \lambda_{\text{momentum}}\mathcal{L}_{\text{momentum}} + \lambda_{\text{boundary}}\mathcal{L}_{\text{boundary}} \\
    \mathcal{L}_{\text{data}} &= \frac{1}{\text{ND}}\sum_{i = 1}^{\text{ND}} \norm{\tilde{\bm{v}} - \bm{v}_{\text{MLP}}(\bm{x}_i)}_2^2, \\
    \mathcal{L}_{\text{continuity}} &= \frac{1}{\text{NC}}\sum_{i = 1}^{\text{NC}} \left(\nabla \cdot \bm{v}_{\text{MLP}}(\bm{x}_i)\right)^2 \quad \text{s.t.} \quad \bm{x}_i \in \Omega, \\
    \mathcal{L}_{\text{momentum}} &= \frac{1}{\text{NM}}\sum_{i = 1}^{\text{NM}} \norm{\left(\nabla\bm{v}_{\text{MLP}}(\bm{x}_i)\right)\bm{v}_{\text{MLP}}(\bm{x}_i) + \nabla p_{\text{MLP}}(\bm{x}_i) - \text{Re}^{-1}\nabla \cdot \nabla\bm{v}_{\text{MLP}}(\bm{x}_i)}_2^2 \quad \text{s.t.} \quad \bm{x}_i \in \Omega, \\
    \mathcal{L}_{\text{boundary}} &= \frac{1}{\text{NB}}\sum_{i = 1}^{\text{NB}} \norm{\bm{v}_{\text{MLP}}(\bm{x}_i)}_2^2 \quad \text{s.t.} \quad \bm{x}_i \in \Gamma_{\text{wall}},
\end{align*}

\noindent with an initial optimization using Adam followed by L-BFGS refinement. At each optimization step, we used all the data points and collocation points to evaluate the losses.

\subsection{Implementation}

We assemble all finite element vectors and matrices in DOLFINx environment \cite{Baratta2023} using Basix \cite{Scroggs2022v1, Scroggs2022v2} and UFL \cite{Alnaes2014} forms. Optimization algorithms and all linear algebra subroutines are implemented using PETSc TAO \cite{Dalcin2011, Balay2025} via petsc4py. The constant matrices $(\bm{T}, \bm{M}_p, \bm{M}_v, \bm{K}_p, \bm{K}_v, \tilde{\bm{K}}_v, \bm{B}, \tilde{\bm{B}})$  are assembled only once outside the optimization loop. Inverses and square roots of $\bm{M}_p$ and $\bm{M}_v$ are approximated through mass lumping. For pressure mass matrix we use row-sum mass lumping and for velocity mass matrix we use HRZ (Hinton–Rock–Zienkiewicz) mass lumping. We perform the inversion in pressure elimination in \eqref{eq:pressure_elimination} through a one-time Cholesky-decomposition and backward substitutions in objective and gradient evaluation routines. We use Quasi-Newton method with More-Thuente line search to solve the optimization problem with respect to $[\hat{\bm{v}}]$. The Jacobian of the discretized advection vector is never assembled during optimization. The action of its transpose on an arbitrary vector $\left[\hat{\bm{u}}\right]$ is assembled using the identity,

\begin{align*}
    \left[\hat{\bm{w}}\right]^T\left(\frac{\partial \bm{A}(\left[\hat{\bm{v}}\right])}{\partial \left[\hat{\bm{v}}\right]}\right)^T\left[\hat{\bm{u}}\right] = \int_{\Omega^h} \bm{u} \cdot \left(\left(\nabla \bm{v}\right)\bm{w} + \left(\nabla \bm{w}\right)\bm{v}\right),
\end{align*}

\noindent in which $\bm{u} = \left(\left[\phi_i(\bm{x})\bm{I}_{3\times 3}\right]_{i = 1}^{\text{NVB}}\right)^T \left[\hat{\bm{u}}_i\right]_{i = 1}^{\text{NVB}}$ and  $\bm{w} = \left(\left[\phi_i(\bm{x})\bm{I}_{3\times 3}\right]_{i = 1}^{\text{NVB}}\right)^T \left[\hat{\bm{w}}_i\right]_{i = 1}^{\text{NVB}}$. We start the optimization from an initial guess of $[\hat{\bm{v}}] = \bm{0}$, which is the minima of the objective function without the data term. For all experiments, we let $\mu_s = 10^{-6}, \ \mu_c = 200, \ \mu_m = 200, \ \alpha_v = 1$ and $\alpha_p = 1$. We validate this choice via Morozov's discrepancy principle so that,

\begin{align*}
    \left|\frac{\norm{\left[\tilde{\bm{v}}\right] - \bm{T}\left[\hat{\bm{v}}\right]^*}_2^2}{\text{NOBS}\ \sigma_0^2} - 1 \right| \leq 0.05.
\end{align*}

\noindent We allow optimizer to take a maximum of 10000 iterations and terminate it when the norm of the gradient falls below $5(10)^{-5}$. For UQ, we assemble $\nabla^2 F\left(\left[\hat{\bm{v}}\right]^*, \left[\hat{p}\right]^*\right)$, including the discretized advection vector Jacobian, and compute its Cholesky factorization. In the divergence constrained case, we assemble the augmented Hessian of \eqref{eq:div_free_MAP} at the MAP,

\begin{align*}
    \hat{\bm{H}} = \begin{bmatrix}
        \nabla \bm{H} & \hat{\bm{B}}^T\\
        \hat{\bm{B}} & \bm{0} 
    \end{bmatrix},
\end{align*}

\noindent compute its Cholesky factorization, and use only the velocity and pressure blocks.
During tricubic interpolation, we use SciPy multidimensional image processing (scipy.ndimage) package for the initial Gaussian smoothing and ARBInterp package \cite{Walker2019} for the Lekien-Marsden interpolation. We solve both \eqref{eq:divergence_projection} and \eqref{eq:pressure reconstruction} using DOLFINx on the same mesh that we use for our reconstruction approach. For the PINN baseline, we use 10 hidden layers each having 128 neurons giving a total of 116612 trainable weights. When computing the loss, we set $\lambda_{\text{continuity}} = \lambda_{\text{momentum}} = \lambda_{\text{boundary}} = 1$ by using Morozov's discrepancy principle so that the data loss term will be $O(1)$. We use the nodes of the reconstruction finite element mesh as collocation points for evaluating the momentum and divergence losses. We let Adam and LBFGS both take 20000 steps, where LFBGS uses 1000 outer and 20 inner iterations. We use an exponential learning rate scheduler per iteration with a decay factor of 0.9995. The code for this work will be made available in \url{https://www.github.com/cgormezano/finite_element_regression} upon publication. The proposed approach and the tricubic interpolation baseline were executed on a single Intel Xeon Gold 6330 CPU while the PINN was trained on an NVIDIA A5000 GPU.

\subsection{Data generation}

We produced synthetic, noisy, under-resolved measurements by adding independent and identically distributed Gaussian noise to downsampled CFD velocity fields, consistent with the likelihood model described in Section~\ref{sec:likelihood}. The
CFD solution, denoted by $(\bm{v}_{\text{truth}})$, is treated as the true unobserved velocity field. We parameterize the noise standard deviation $\sigma_0$ through signal-to-noise ratio,
\begin{align*}
    \text{SNR}
    = \frac{\norm{\bm{v}_{\text{truth}}}_{L^2(\Omega)}}
           {\sigma_0 \, \sqrt{\abs{\Omega}}}.
\end{align*}

\begin{figure}[h!]
\center
\begin{subfigure}[t]{.32\textwidth}
  \centering
  \includegraphics[width=0.99\linewidth]{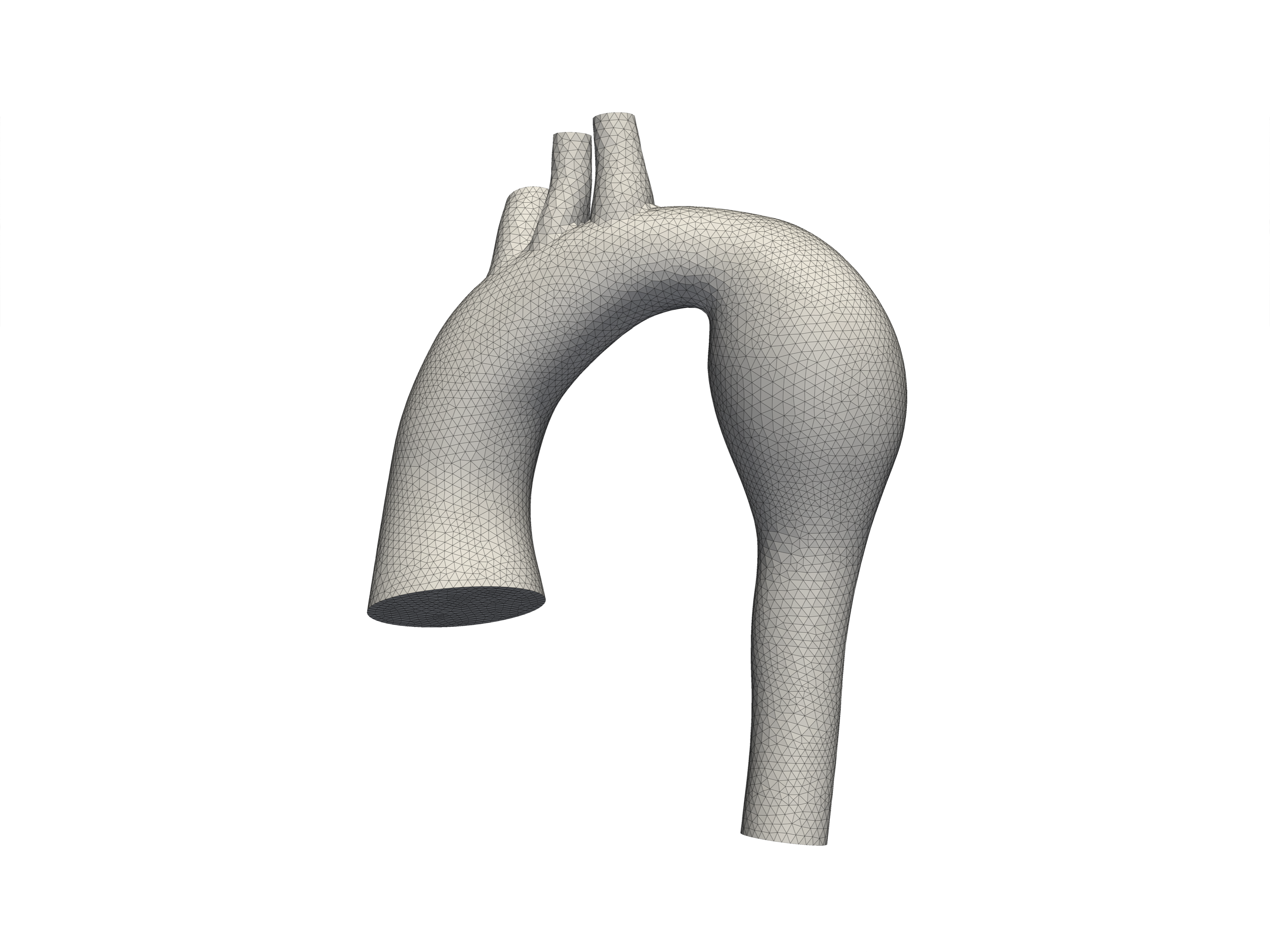}
\caption{Aortic aneurysm.}
\end{subfigure}
\hfill
\begin{subfigure}[t]{.32\textwidth}
  \centering
  \includegraphics[width=0.99\linewidth]{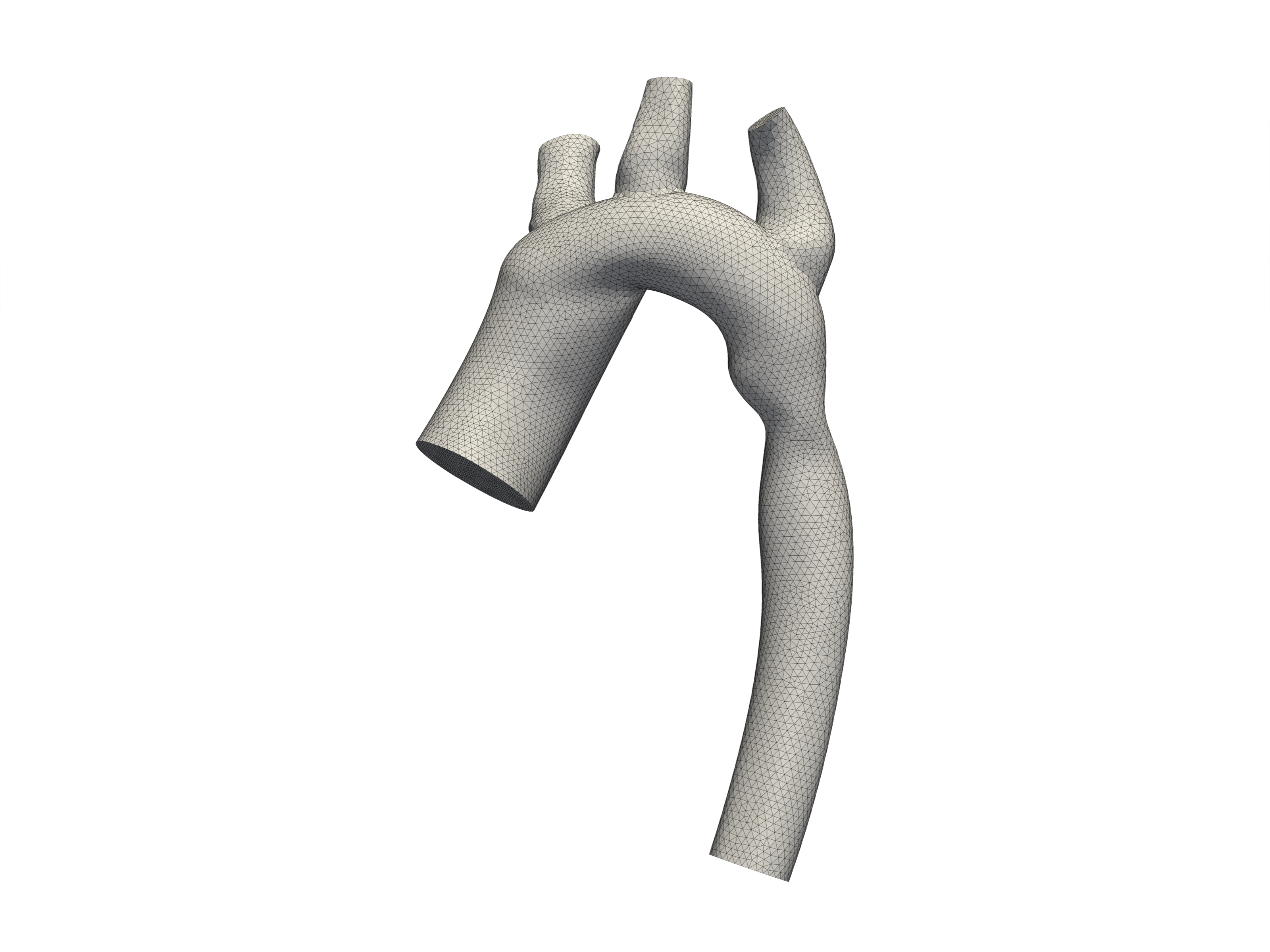}
\caption{Aortic coarctation.}
\end{subfigure}
\hfill
\begin{subfigure}[t]{.32\textwidth}
  \centering
  \includegraphics[width=0.99\linewidth]{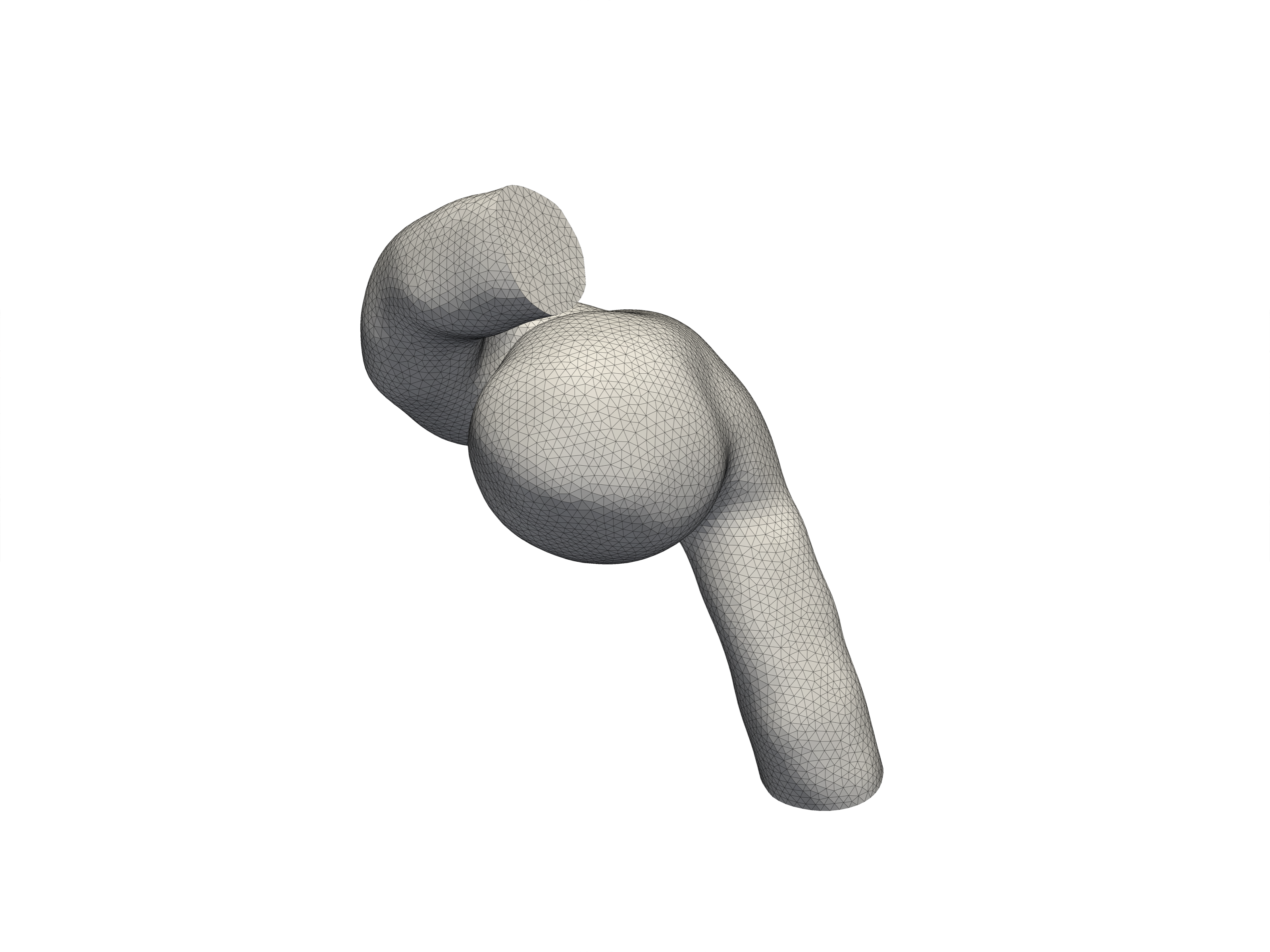}
\caption{Cerebral aneurysm.}
\end{subfigure}
\caption{Geometries used in the numerical experiments.}
\label{fig:models}
\end{figure}

We generated flows over three different patient-specific vascular geometries whose configurations are summarized in Table \ref{tab:test_case_configurations}. These flows were obtained by solving the steady incompressible Navier--Stokes equations using a finite element solver implemented in DOLFINx with Taylor-Hood ($P_2$/$P_1$) elements. For the inlet boundary conditions, we prescribed an inlet flow profile orthogonal to the inlet cap. The profile was constructed by solving the Poisson equation $\nabla^2 \omega = 1$ on the inlet cap subject to homogeneous Dirichlet boundary conditions on the rim. The solution ($\omega$) was scaled to achieve a target flow rate ($Q_{t}$), giving $\bm{v}\vert_{\Gamma_{\text{in}}^h} = \left(Q_{t} \big/ \int_{\Gamma_{\text{in}}^h} \omega \ d\Gamma^h\right)\omega\hat{\bm{n}}$. For aortic models, we prescribed resistance boundary conditions at the outlets to meet physiological flow splits. For cerebral aneurysm mode, we prescribed zero traction at the outlet. To reduce the mesh size and computational cost, we truncated the aortic geometries downstream of the region of interest at the descending aortic outlet. All three geometries are illustrated in Figure (\ref{fig:models}).

\begin{table}[htbp]
  \centering
  \begin{tabular}{|l||c|c|c|}
    \hline
                       & \textbf{Aortic aneurysm} & \textbf{Aortic coarctation} & \textbf{Cerebral aneurysm} \\
    \hhline{|=||=|=|=|}
    \textbf{VMR model name} & 0021\_H\_AO\_MFS  & 0235\_H\_AO\_COA & 0203\_H\_CERE\_CA \\
    \hline
    \textbf{Inlet flow rate} [$\text{cm}^3/\text{s}$] & $70.0$ & $100.0$ & $4.0$ \\
    \hline
    \textbf{Inlet Re}  & $1044$  &  $849$  &  $308$  \\
    \hline
    \textbf{BC type} & Inflow-Resistance &  Inflow-Resistance & Inflow-Traction \\
    \hline
    \textbf{Number of elements} & $181609$ & $180741$ &  $180434$ \\
    \hline
  \end{tabular}
  \caption{Test case configurations. Reynolds numbers are based on the inlet hydraulic diameter. In general, local Re values at the coarctation or aneurysm will be different than the Re values at the inlet.}
  \label{tab:test_case_configurations}
\end{table}

\section{Results}

In this section, we present reconstructions using our proposed approach and its weakly divergence-free variant. We report uncertainty estimates for the reconstructed fields and propagate these uncertainties to several quantities of interest. Since pressure is only recoverable up to an additive constant, all pressure fields presented in this section have been shifted to have zero spatial mean. We compare the accuracy of our approach with tricubic interpolation and PINN baselines across a range of SNRs and spatial resolutions ($\Delta x$).

\subsection{Reconstructions}

\begin{figure}[htbp]
\centering
\begin{subfigure}[t]{.24\textwidth}
  \centering
  \includegraphics[width=.99\linewidth]{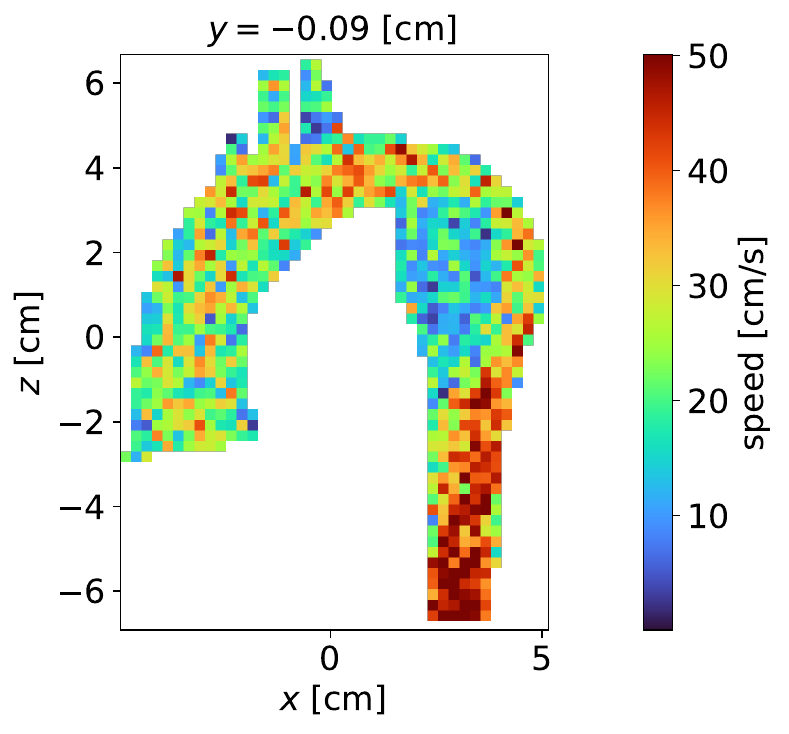}
  \caption{Measurement.}
\end{subfigure}
\hfill
\begin{subfigure}[t]{.24\textwidth}
  \centering
  \includegraphics[width=.99\linewidth]{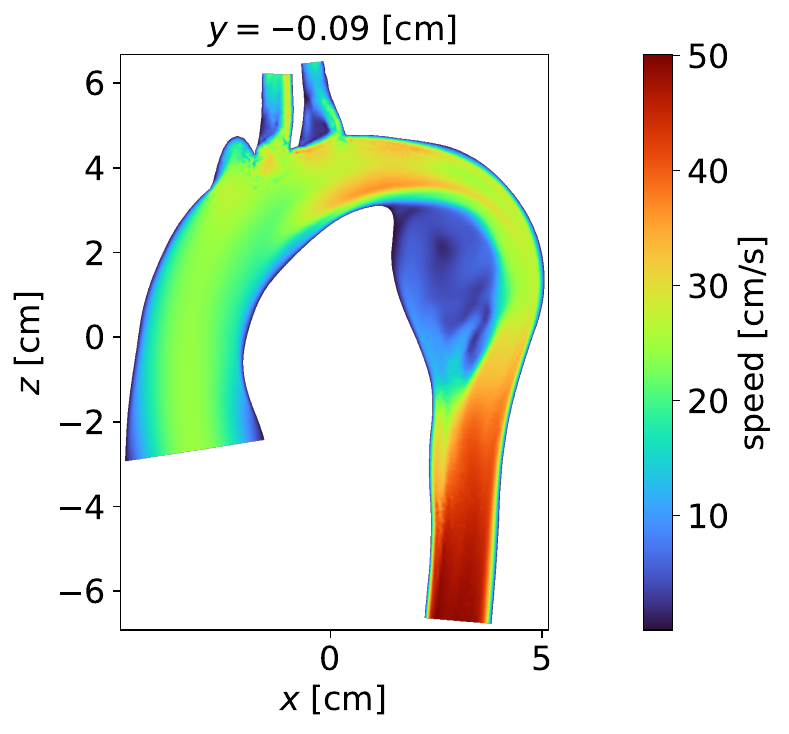}
  \caption{Truth.}
\end{subfigure}
\hfill
\begin{subfigure}[t]{.24\textwidth}
  \centering
  \includegraphics[width=.99\linewidth]{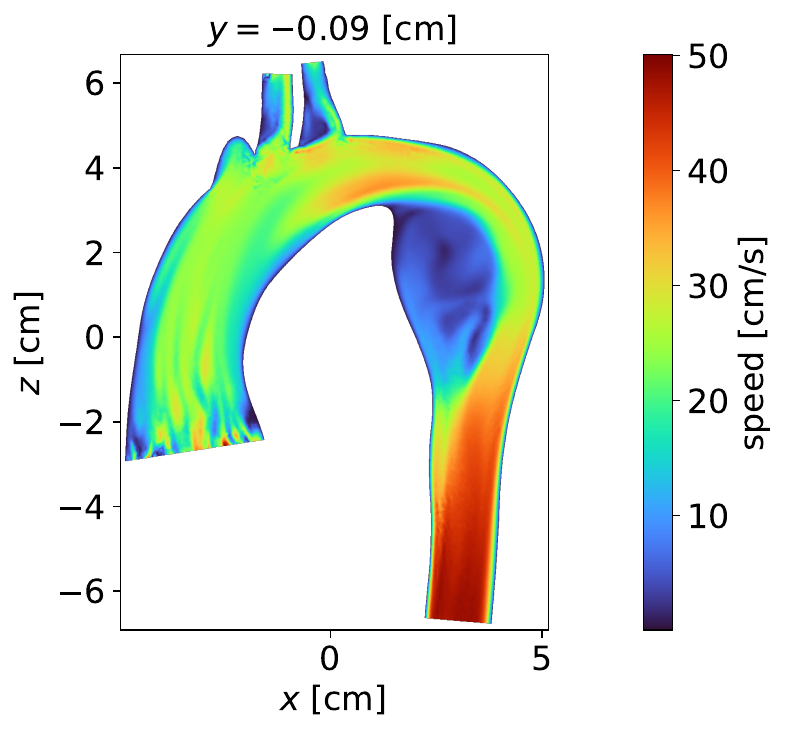}
  \caption{Reconstruction.}
\end{subfigure}
\hfill
\begin{subfigure}[t]{.24\textwidth}
  \centering
  \includegraphics[width=.99\linewidth]{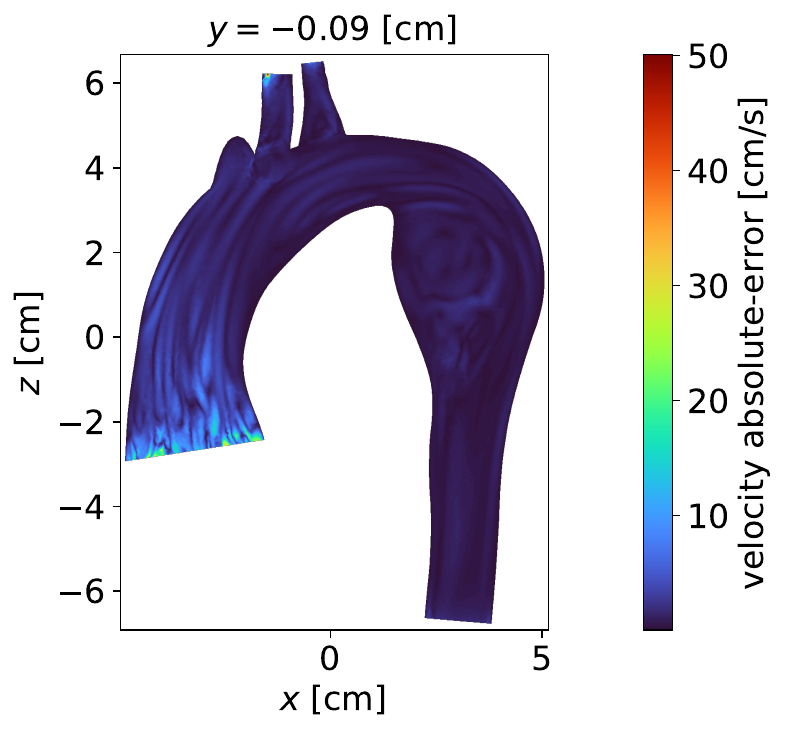}
  \caption{Error. Global RMS error = 11\%. Local RMS error = 4\%.}
\end{subfigure}

\vspace{0.5em}

\hspace*{.25\textwidth}
\begin{subfigure}[t]{.24\textwidth}
  \centering
  \includegraphics[width=.99\linewidth]{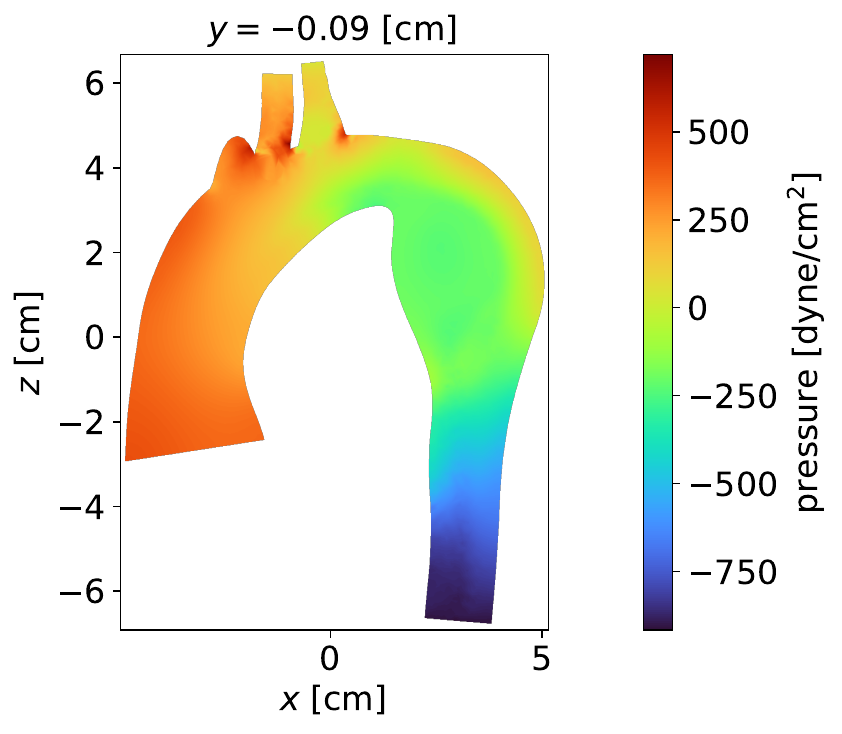}
  \caption{Truth.}
\end{subfigure}
\hfill
\begin{subfigure}[t]{.24\textwidth}
  \centering
  \includegraphics[width=.99\linewidth]{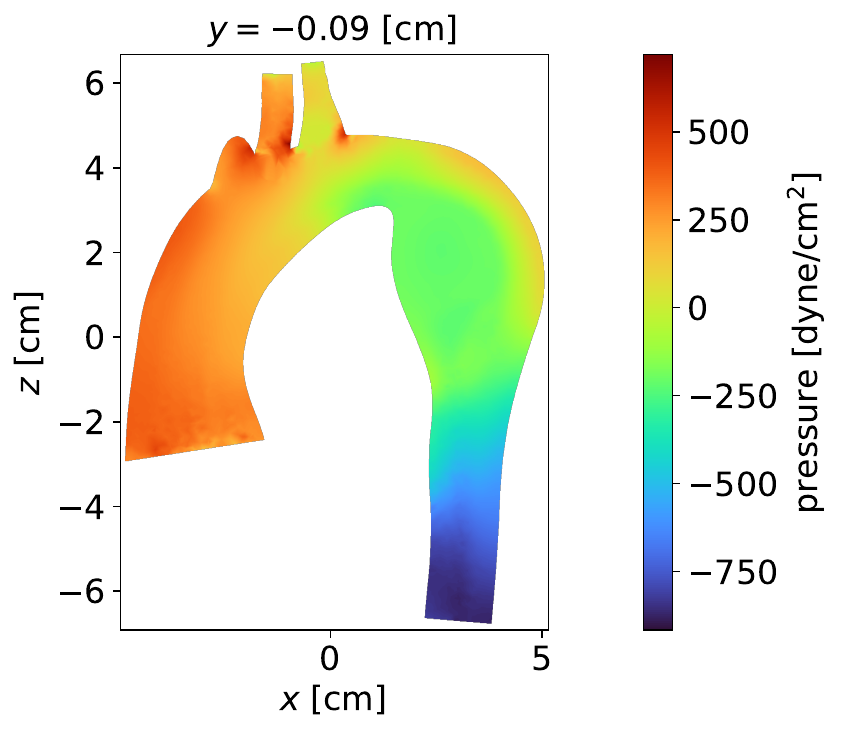}
  \caption{Reconstruction.}
\end{subfigure}
\hfill
\begin{subfigure}[t]{.24\textwidth}
  \centering
  \includegraphics[width=.99\linewidth]{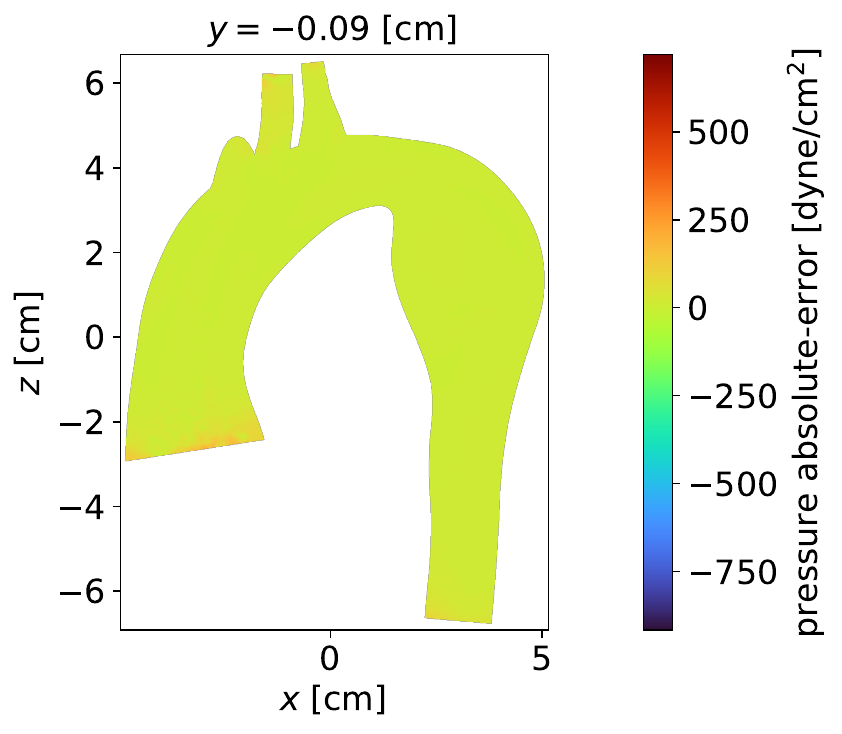}
  \caption{Error. Global RMS error = 5\%. Local RMS error = 4\%.}
\end{subfigure}
\caption{Observations, reconstructions, and corresponding error fields on a planar slice through the aortic aneurysm geometry for $\Delta x = 0.250$ cm and $\mathrm{SNR}=2.5$.}
\label{fig:aa_reconstructions}
\end{figure}

\begin{figure}[htbp]
\centering
\begin{subfigure}[t]{.24\textwidth}
  \centering
  \includegraphics[width=.99\linewidth]{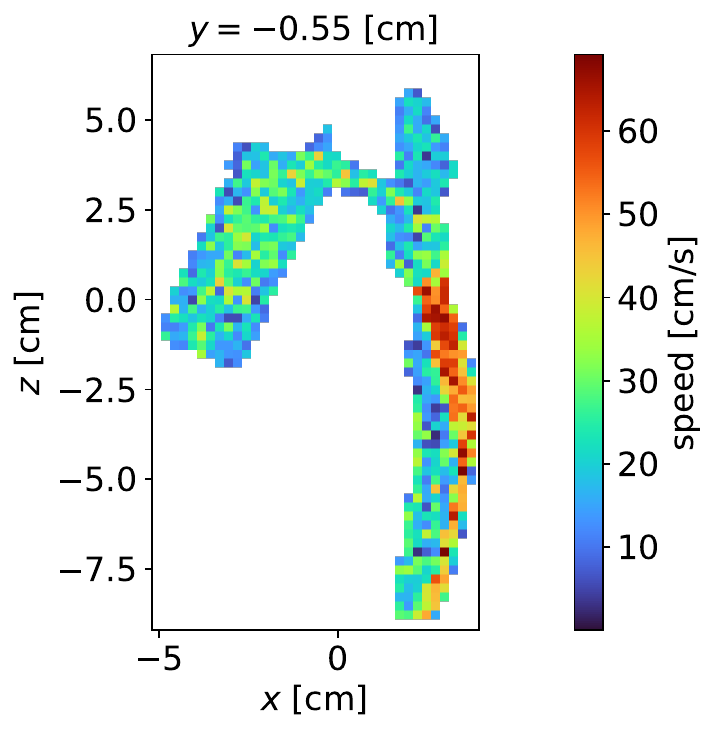}
  \caption{Measurement.}
\end{subfigure}
\hfill
\begin{subfigure}[t]{.24\textwidth}
  \centering
  \includegraphics[width=.99\linewidth]{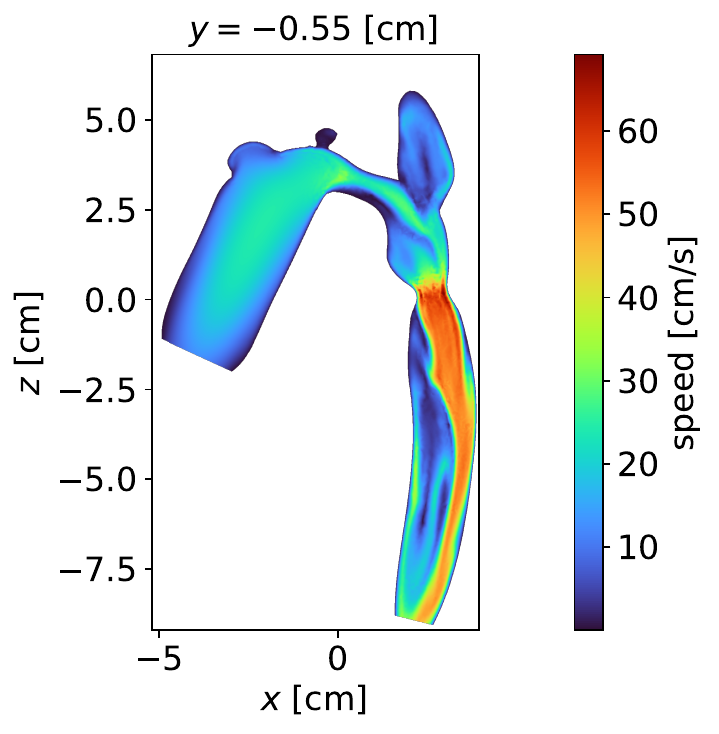}
  \caption{Truth.}
\end{subfigure}
\hfill
\begin{subfigure}[t]{.24\textwidth}
  \centering
  \includegraphics[width=.99\linewidth]{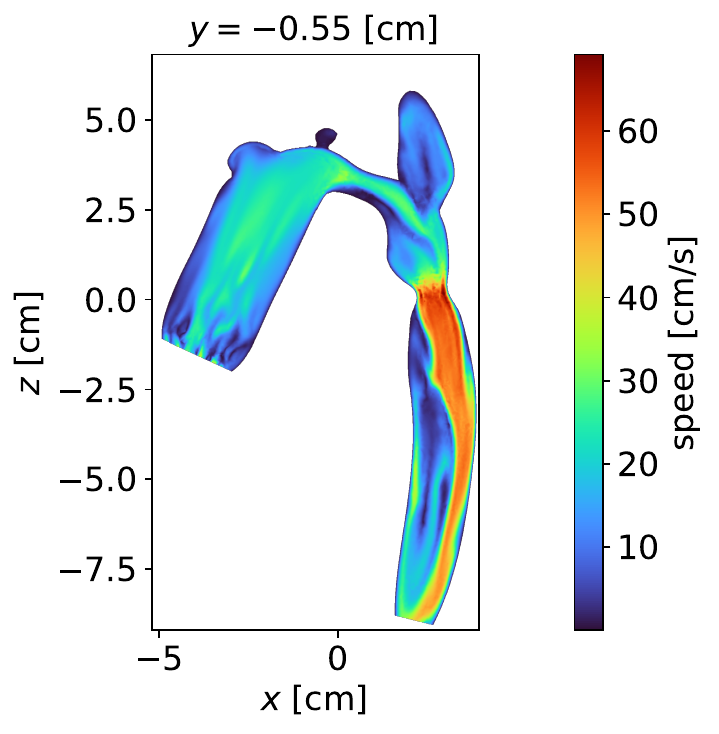}
  \caption{Reconstruction.}
\end{subfigure}
\hfill
\begin{subfigure}[t]{.24\textwidth}
  \centering
  \includegraphics[width=.99\linewidth]{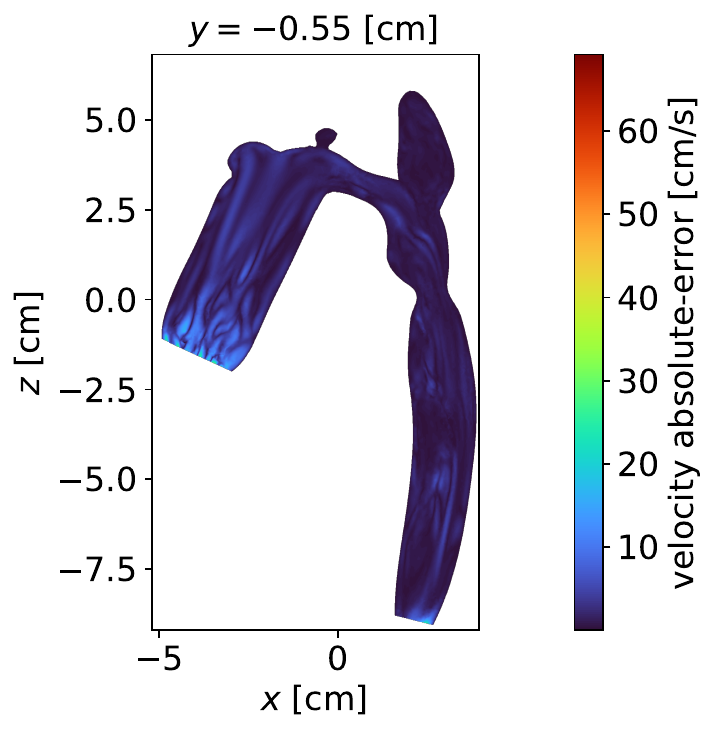}
  \caption{Error. Global RMS error = 14\%. Local RMS error = 4\%.}
\end{subfigure}

\vspace{0.5em}

\hspace*{.25\textwidth}
\begin{subfigure}[t]{.24\textwidth}
  \centering
  \includegraphics[width=.99\linewidth]{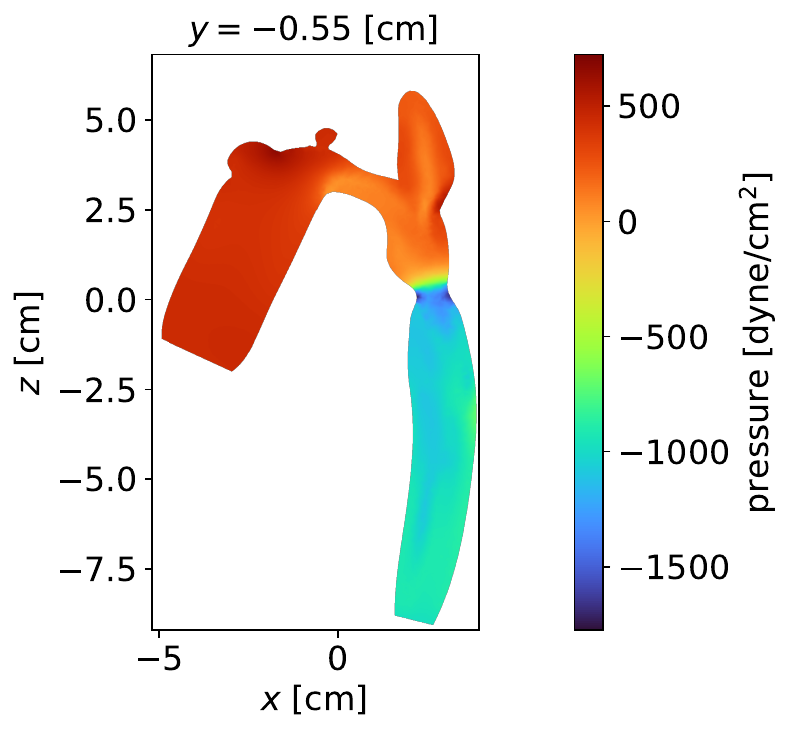}
  \caption{Truth.}
\end{subfigure}
\hfill
\begin{subfigure}[t]{.24\textwidth}
  \centering
  \includegraphics[width=.99\linewidth]{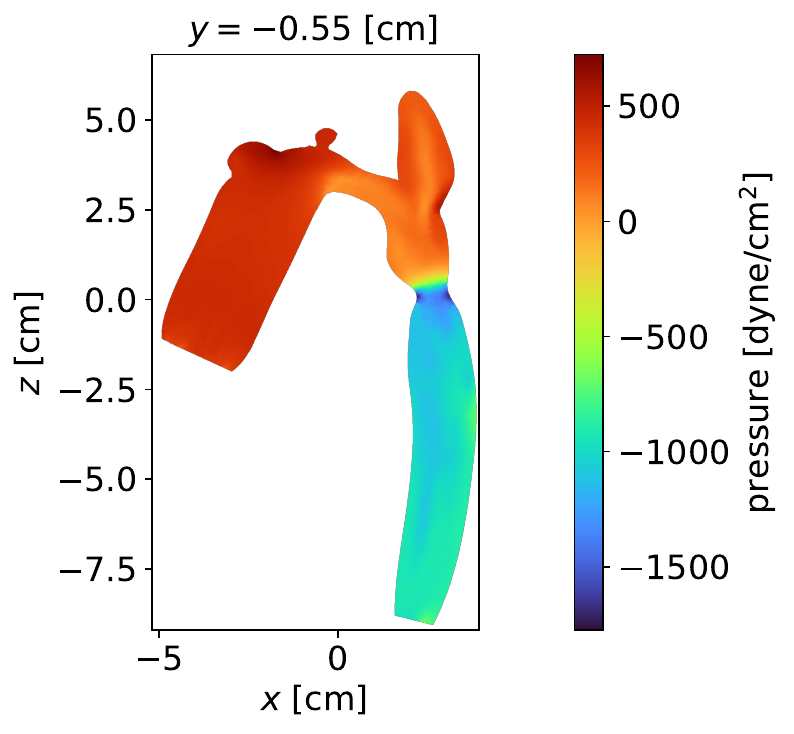}
  \caption{Reconstruction.}
\end{subfigure}
\hfill
\begin{subfigure}[t]{.24\textwidth}
  \centering
  \includegraphics[width=.99\linewidth]{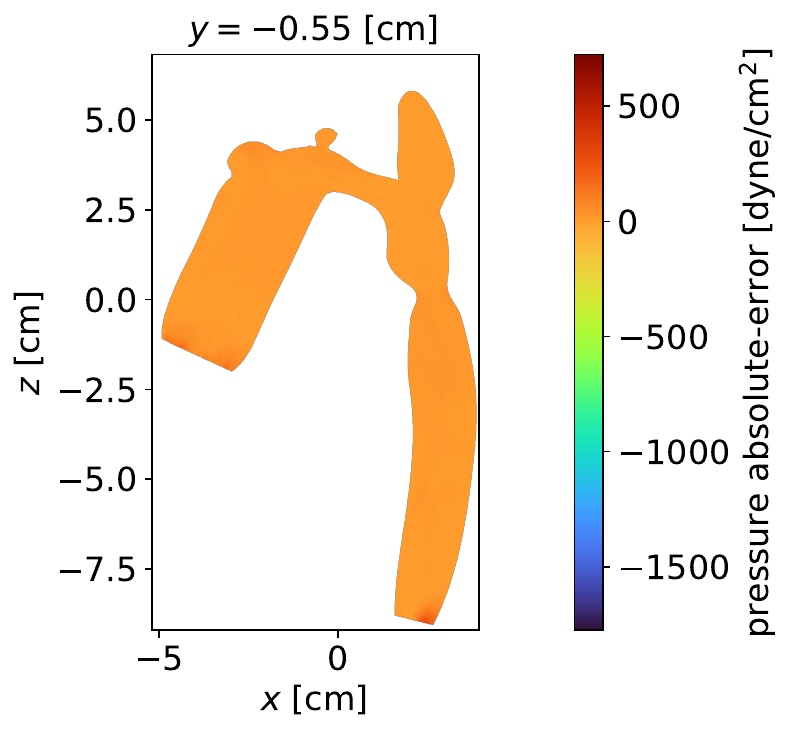}
  \caption{Error. Global RMS error = 3\%. Local RMS error = 2\%.}
\end{subfigure}
\caption{Observations, reconstructions, and corresponding error fields on a planar slice through the aortic aneurysm coarctation for $\Delta x = 0.250$ cm and $\mathrm{SNR}=2.5$.}
\label{fig:ac_reconstructions}
\end{figure}

\begin{figure}[htbp]
\centering
\begin{subfigure}[t]{.24\textwidth}
  \centering
  \includegraphics[width=.99\linewidth]{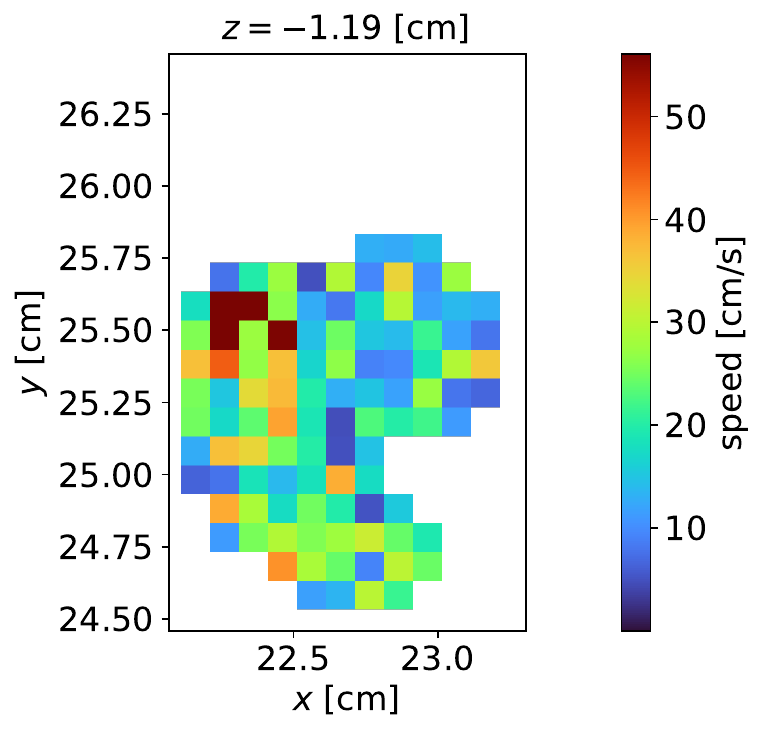}
  \caption{Measurement.}
\end{subfigure}
\hfill
\begin{subfigure}[t]{.24\textwidth}
  \centering
  \includegraphics[width=.99\linewidth]{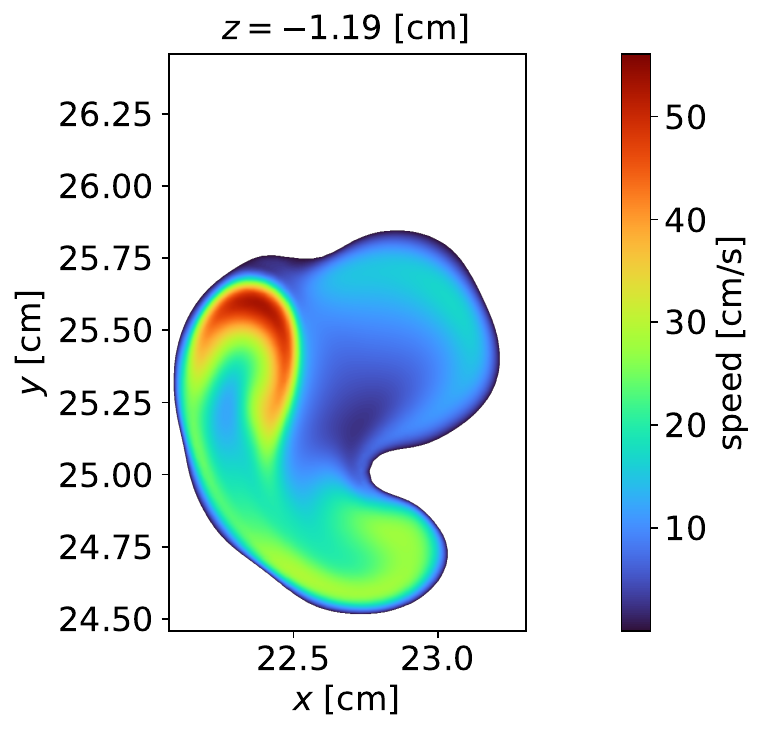}
  \caption{Truth.}
\end{subfigure}
\hfill
\begin{subfigure}[t]{.24\textwidth}
  \centering
  \includegraphics[width=.99\linewidth]{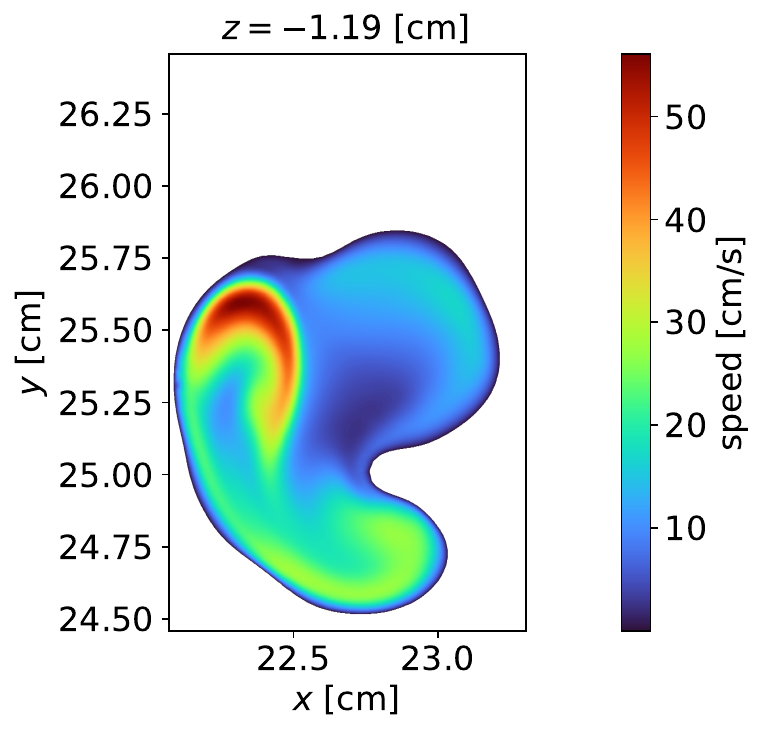}
  \caption{Reconstruction.}
\end{subfigure}
\hfill
\begin{subfigure}[t]{.24\textwidth}
  \centering
  \includegraphics[width=.99\linewidth]{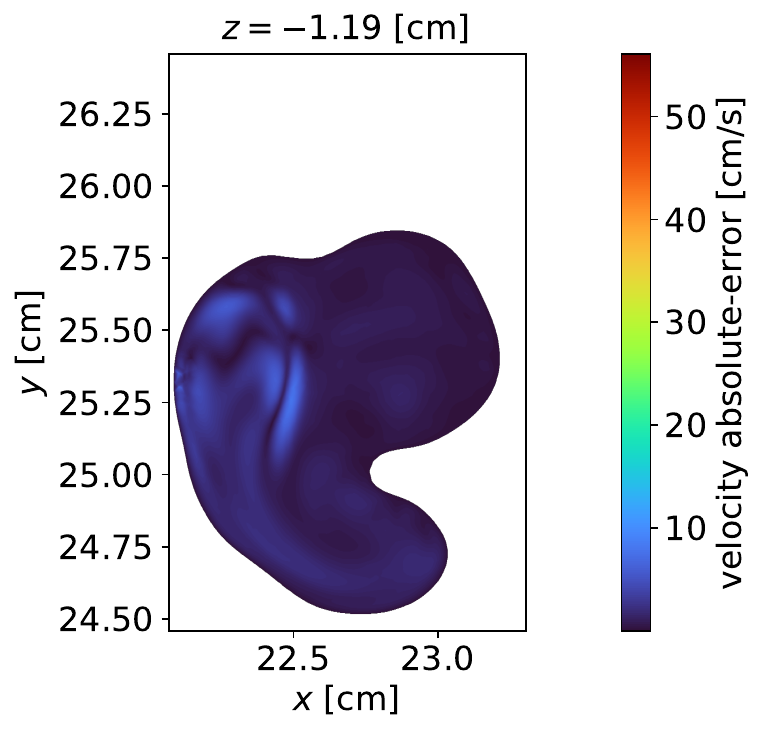}
  \caption{Error. Global RMS error = 17\%. Local RMS error = 7\%.}
\end{subfigure}

\vspace{0.5em}

\hspace*{.25\textwidth}
\begin{subfigure}[t]{.24\textwidth}
  \centering
  \includegraphics[width=.99\linewidth]{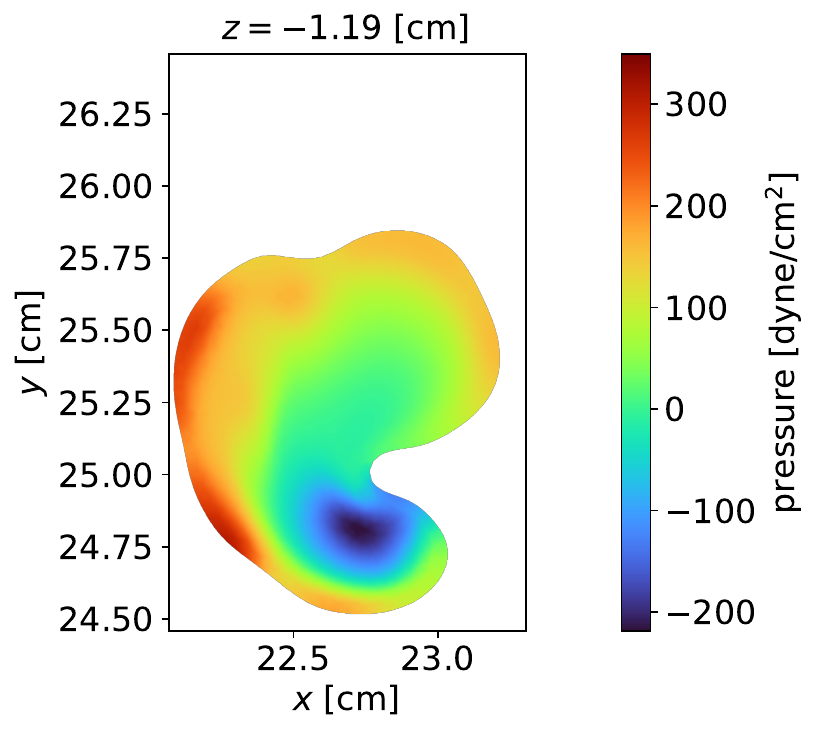}
  \caption{Truth.}
\end{subfigure}
\hfill
\begin{subfigure}[t]{.24\textwidth}
  \centering
  \includegraphics[width=.99\linewidth]{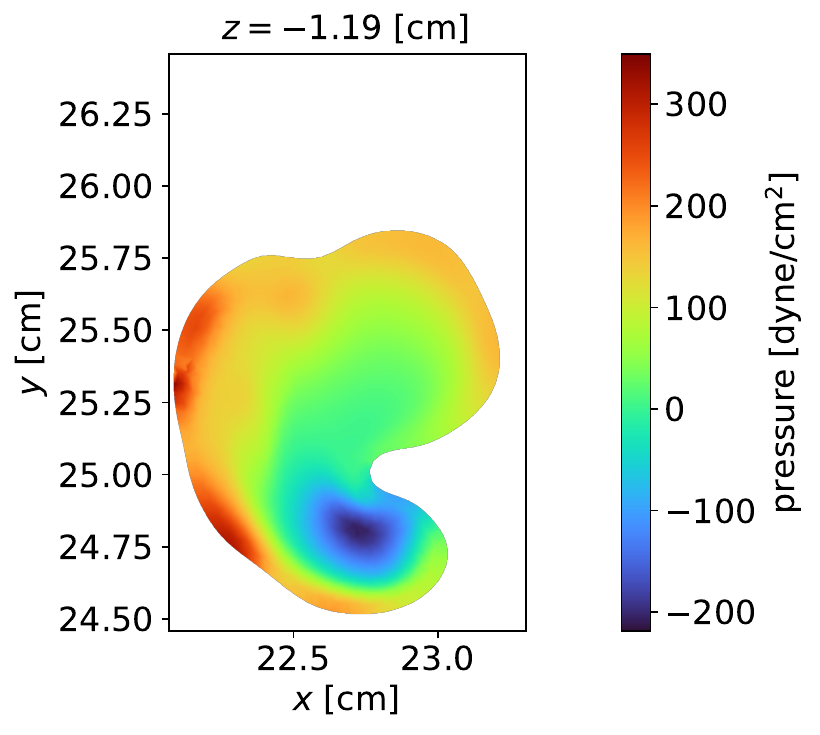}
  \caption{Reconstruction.}
\end{subfigure}
\hfill
\begin{subfigure}[t]{.24\textwidth}
  \centering
  \includegraphics[width=.99\linewidth]{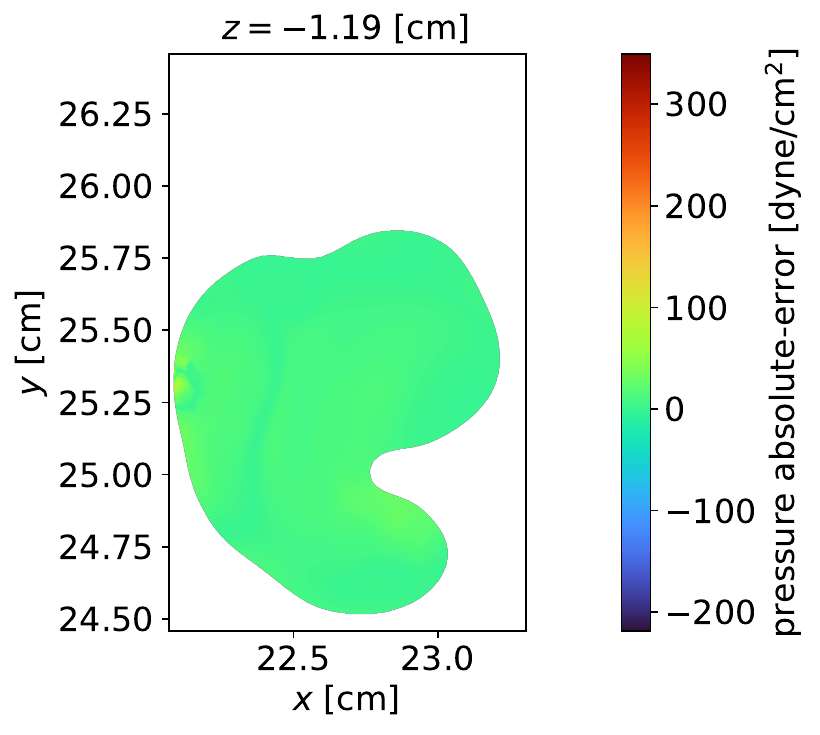}
  \caption{Error. Global RMS error = 16\%. Local RMS error = 7\%.}
\end{subfigure}
\caption{Observations, reconstructions, and corresponding error fields on a planar slice through the cerebral aneurysm geometry for $\Delta x = 0.100$ cm and $\mathrm{SNR}=2.5$.}
\label{fig:ca_reconstructions}
\end{figure}

Figures (\ref{fig:aa_reconstructions}-\ref{fig:ca_reconstructions}) illustrate the noisy under-resolved velocity magnitude measurements, true velocity magnitude and pressure fields, reconstructed velocity magnitude and pressure fields, as well as the absolute error fields associated with velocity and pressure, over a specified plane for different geometries. The spatial resolution and noise levels were chosen as $(\Delta x = 0.250 \ \mathrm{cm}, \ \mathrm{SNR}=2.5)$ for the aortic aneurysm, $(\Delta x = 0.250 \ \mathrm{cm}, \ \mathrm{SNR}=2.5)$ for the aortic coarctation, and $(\Delta x = 0.200 \ \mathrm{cm}, \ \mathrm{SNR}=2.5)$ for the cerebral aneurysm. These settings represent challenging low-resolution and low-SNR imaging conditions.  We observe that the reconstructed fields match with true fields well everywhere except at the vicinity of the inlet. This mismatch is more significant for velocity than pressure. To quantify the discrepancy between truth and reconstructed fields, we define the relative RMS errors in velocity, pressure, and WSS over a given volume $\Omega$ or surface $\Gamma$ as,

\begin{align}
\varepsilon_{\bm{v}}(\Omega) = \sqrt{\cfrac{\norm{\bm{v}^h-\bm{v}_{\text{truth}}}_{L^2(\Omega)}^2}
{\norm{\bm{v}_{\text{truth}}}_{L^2(\Omega)}^2}}, \quad 
\varepsilon_{p}(\Omega) = \sqrt{\cfrac{\left|p^h-p_{\text{truth}}\right|_{L^2(\Omega)}^2}
{\left|p_{\text{truth}}\right|_{L^2(\Omega)}^2}}, \quad \text{and} \quad 
\varepsilon_{\text{wss}}(\Gamma) = \sqrt{\cfrac{\norm{\bm{\tau}^h-\bm{\tau}_{\text{truth}}}_{L^2(\Gamma)}^2}
{\norm{\bm{\tau}_{\text{truth}}}_{L^2(\Gamma)}^2}}.
\label{eq:error_definitions}
\end{align}

\begin{figure}[htbp]
\center
\begin{subfigure}[t]{.32\textwidth}
  \centering
  \includegraphics[width=0.99\linewidth]{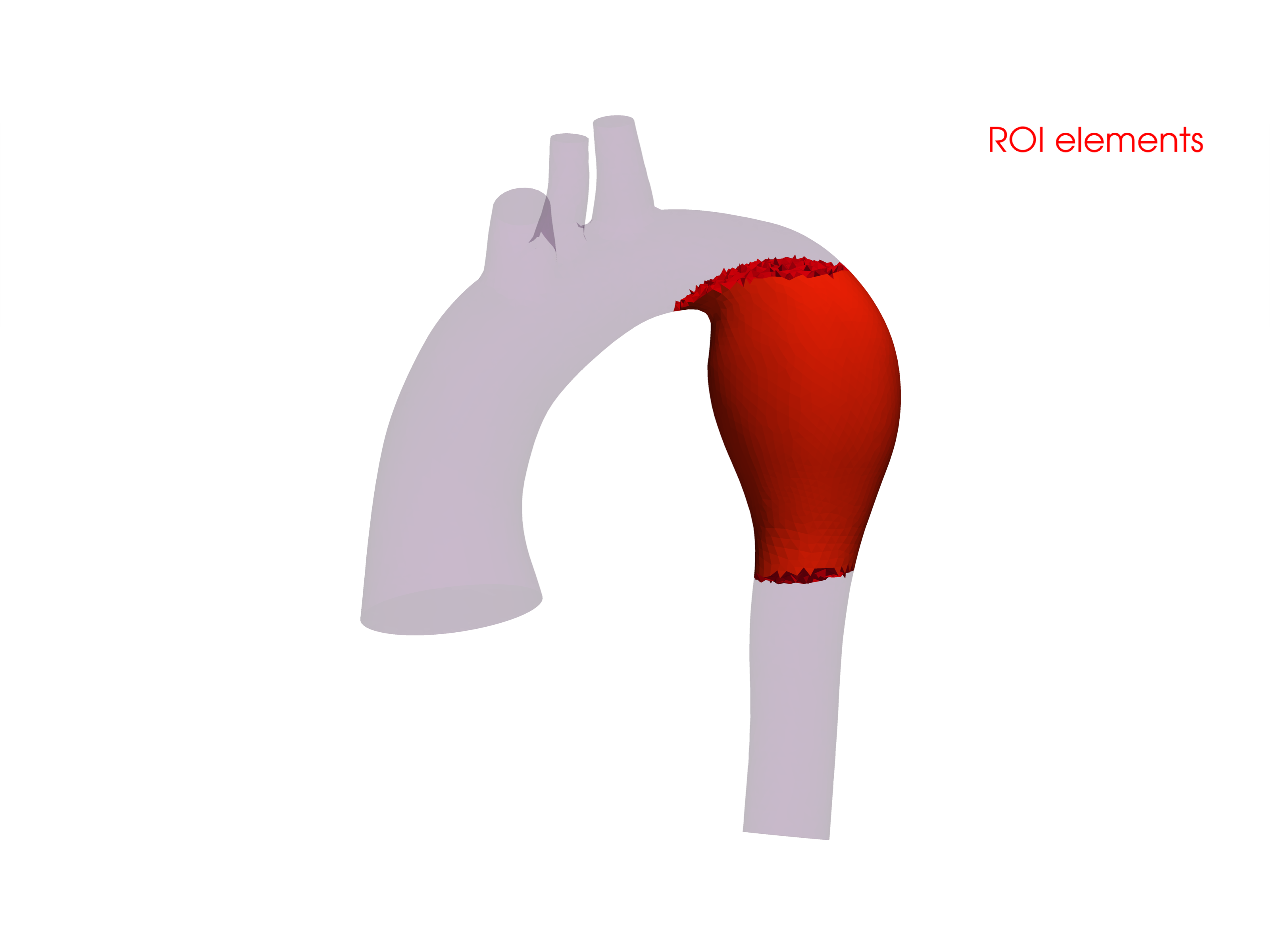}
\caption{Aortic aneurysm.}
\end{subfigure}
\hfill
\begin{subfigure}[t]{.32\textwidth}
  \centering
  \includegraphics[width=0.99\linewidth]{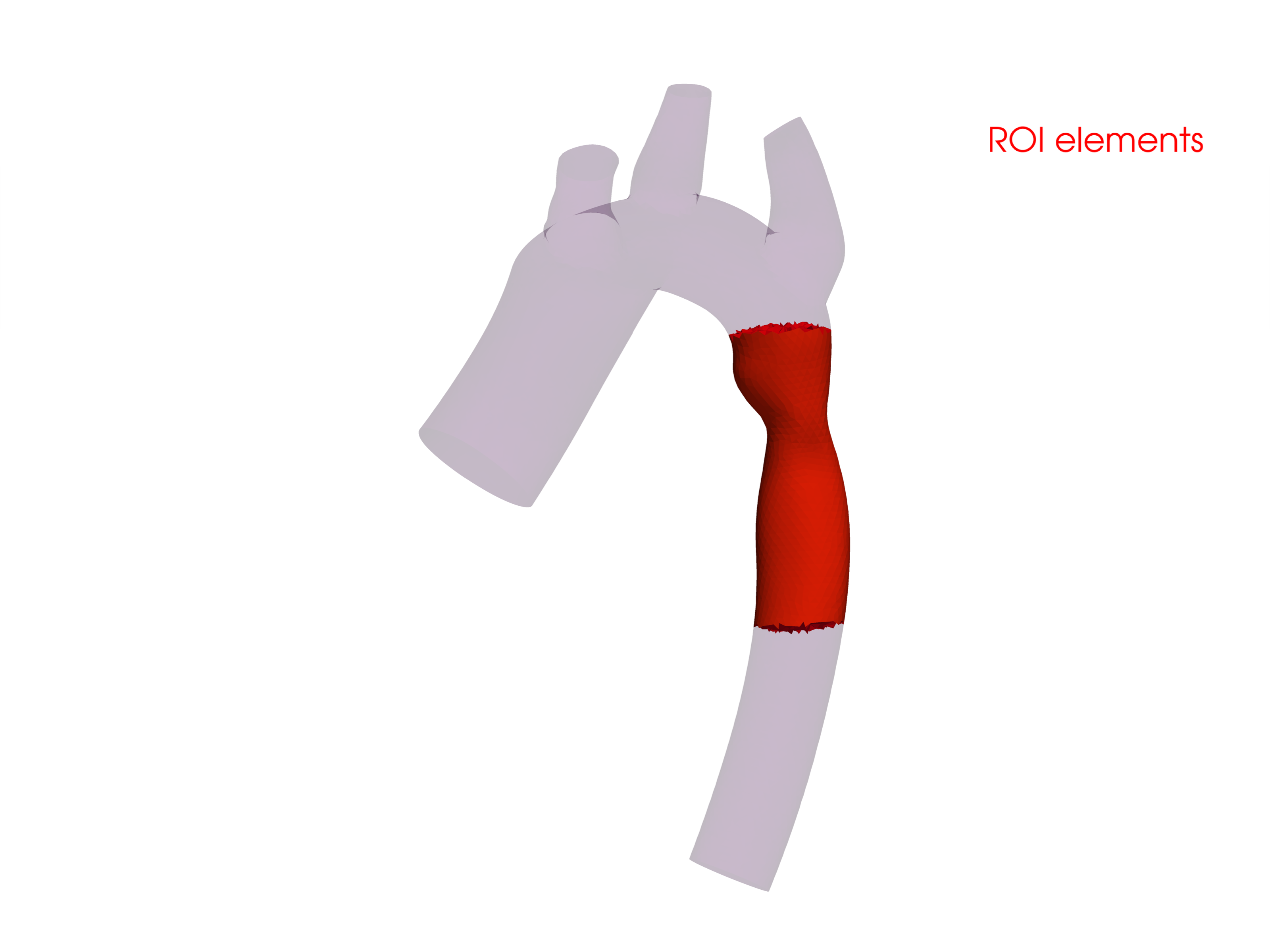}
\caption{Aortic coarctation.}
\end{subfigure}
\hfill
\begin{subfigure}[t]{.32\textwidth}
  \centering
  \includegraphics[width=0.99\linewidth]{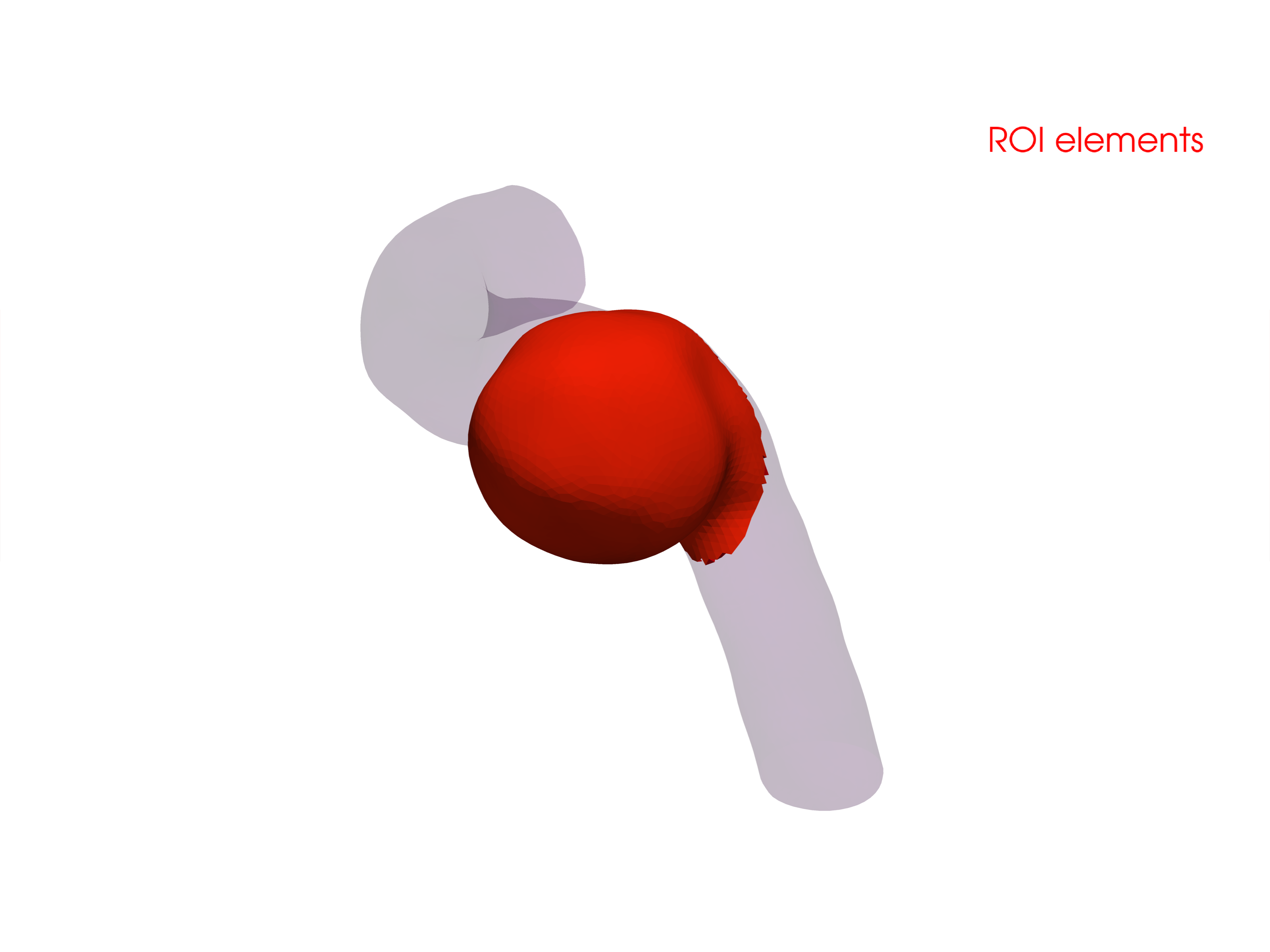}
\caption{Cerebral aneurysm.}
\end{subfigure}
\caption{Regions of interest for different geometries.}
\label{fig:regions_of_interest}
\end{figure}

\noindent We evaluate \eqref{eq:error_definitions} over both the entire geometry (global) and a region of interest (ROI), where accurate reconstruction is more important for downstream diagnostic tasks. These regions correspond to the aneurysm sacs in the aneurysm geometries and the stenotic neck in the coarctation geometry, as illustrated in Figure (\ref{fig:regions_of_interest}).

\noindent We see that velocity error is much smaller in the region of interest than the entire volume indicating that most of the global error originates from poor reconstruction at the inlet. A similar trend is observed for pressure, yet the differences between local and global errors are less pronounced. These results demonstrate that the proposed approach accurately reconstructs both the low-velocity recirculation zones within aneurysm sacs and the high-velocity, high-pressure-gradient jet regions in the coarctation geometry.

\FloatBarrier

\subsection{Uncertainty quantification}

\begin{figure}[h!]
\centering
\begin{subfigure}[t]{.24\textwidth}
  \centering
  \includegraphics[width=0.99\linewidth]{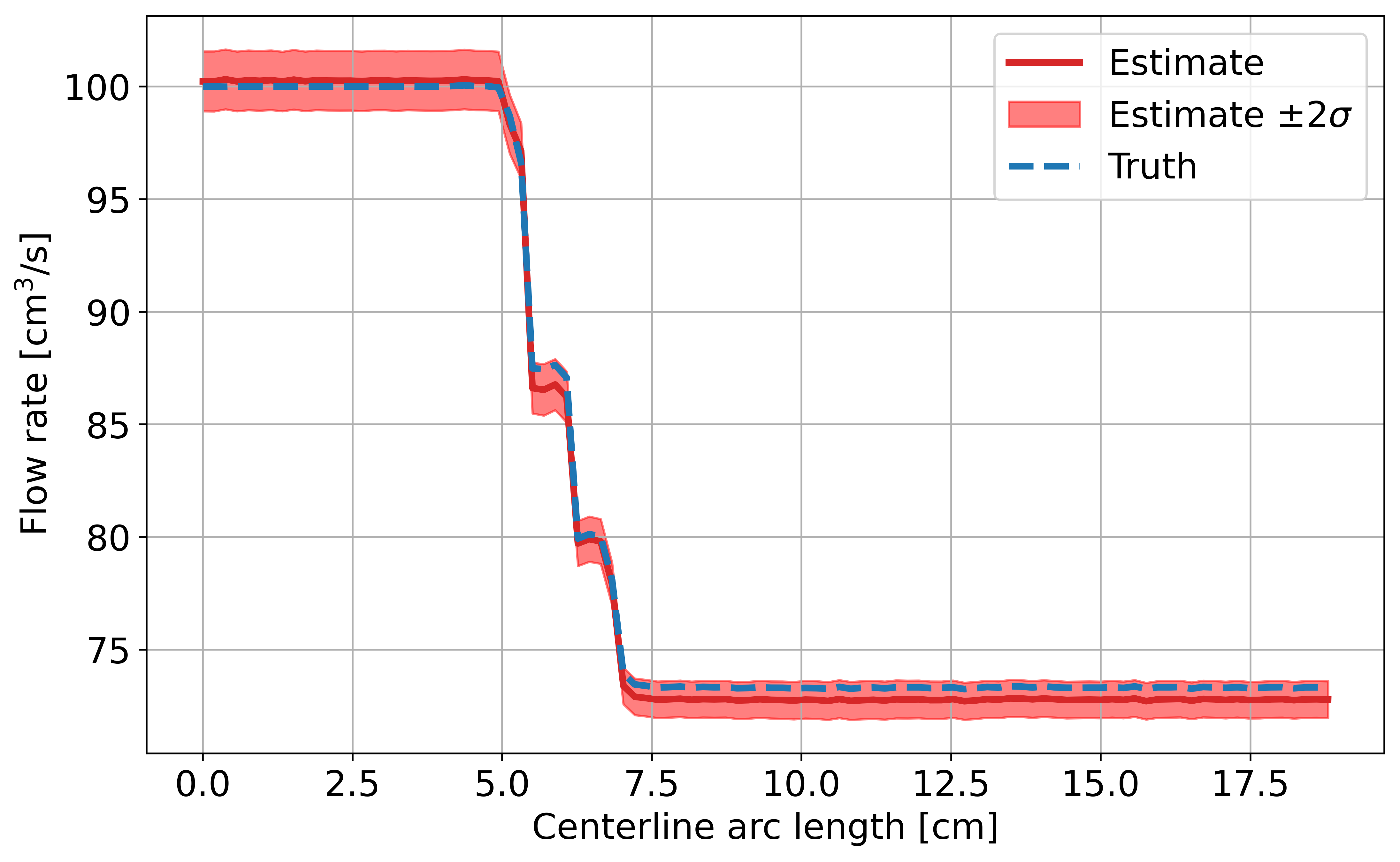}
  \caption{Aortic aneurysm.}
\end{subfigure}
\hfill
\begin{subfigure}[t]{.24\textwidth}
  \centering
  \includegraphics[width=0.99\linewidth]{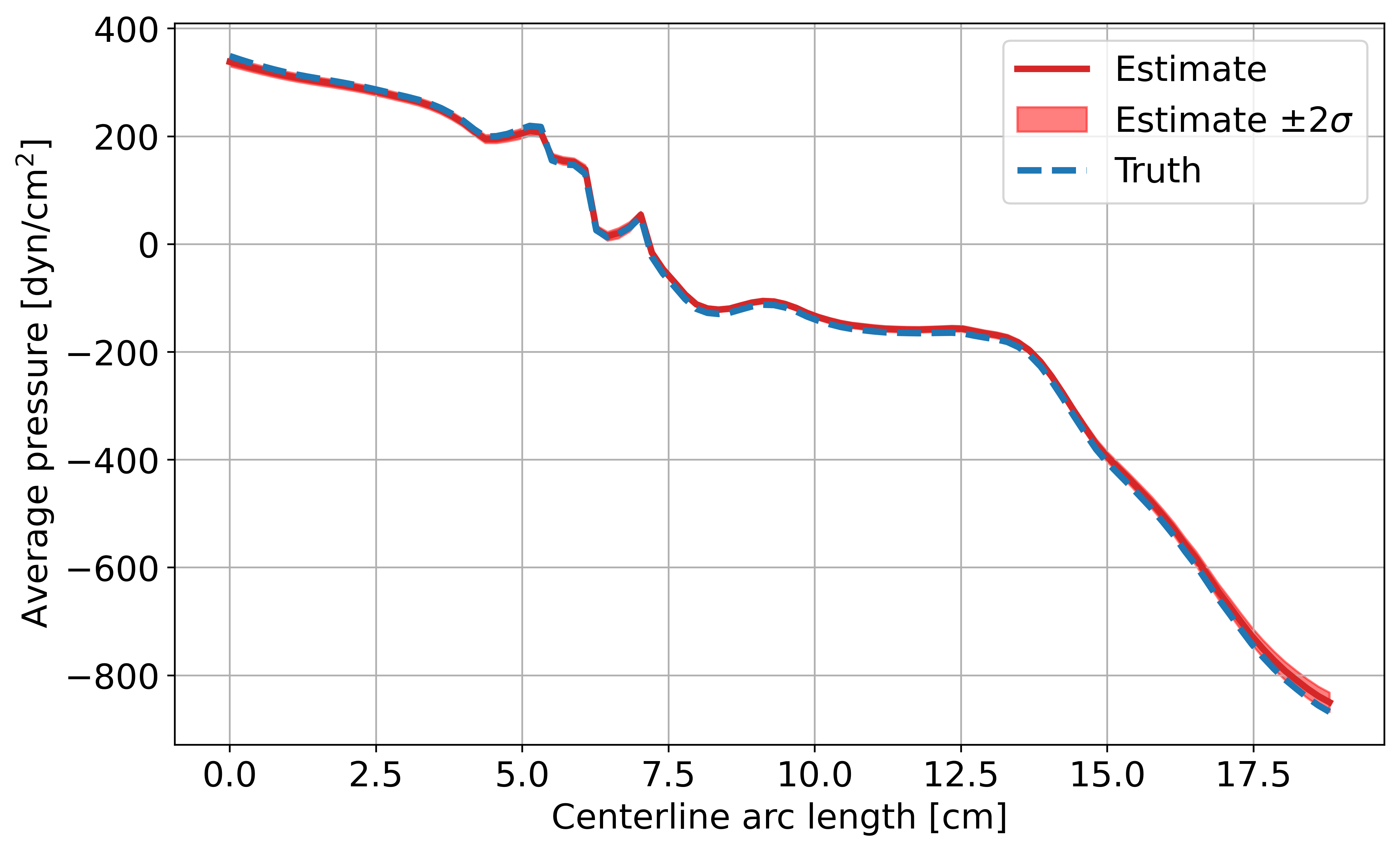}
  \caption{Aortic aneurysm.}
\end{subfigure}
\hfill
\begin{subfigure}[t]{.24\textwidth}
  \centering
  \includegraphics[width=0.99\linewidth]{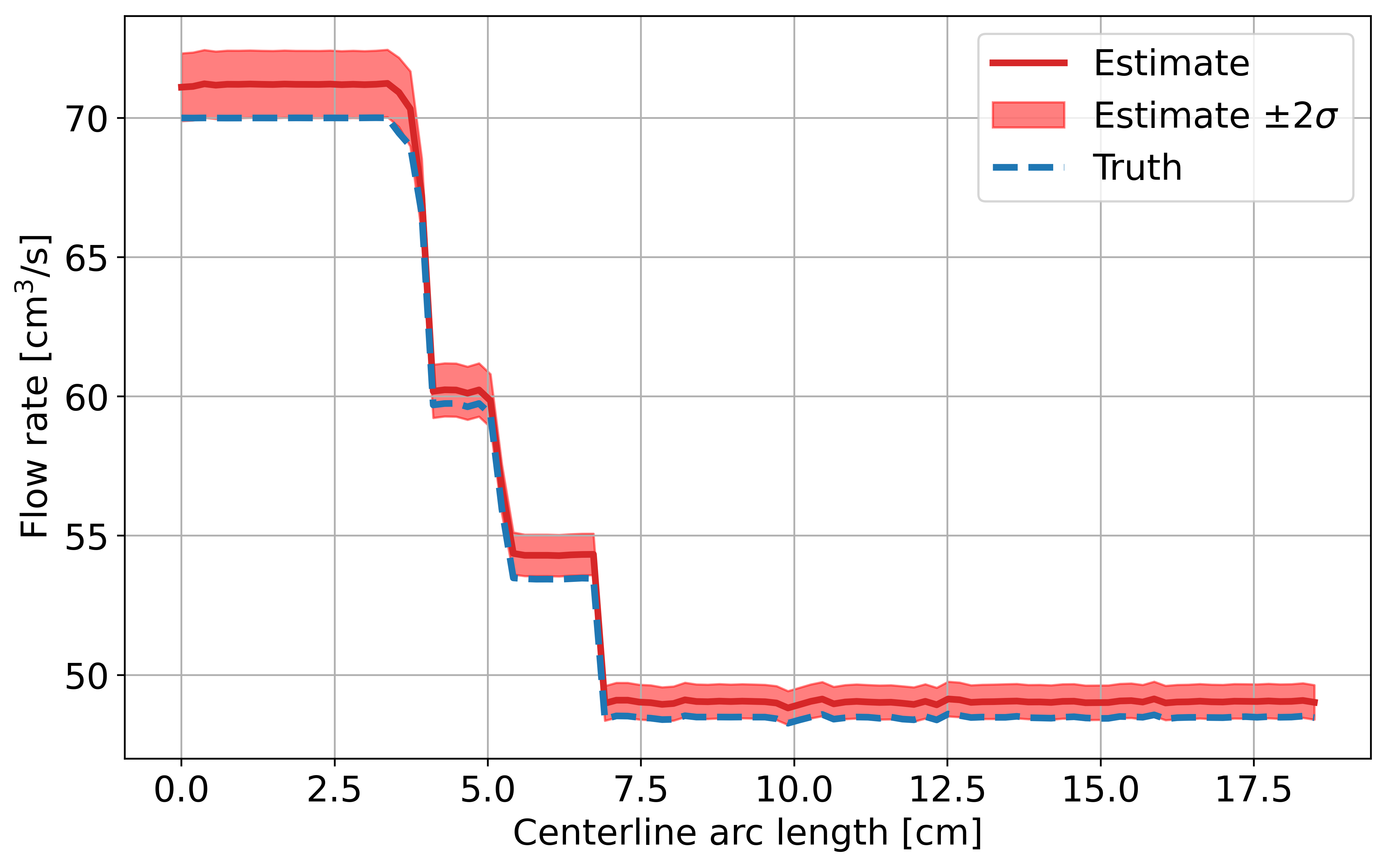}
  \caption{Aortic coarctation.}
\end{subfigure}
\hfill
\begin{subfigure}[t]{.24\textwidth}
  \centering
  \includegraphics[width=0.99\linewidth]{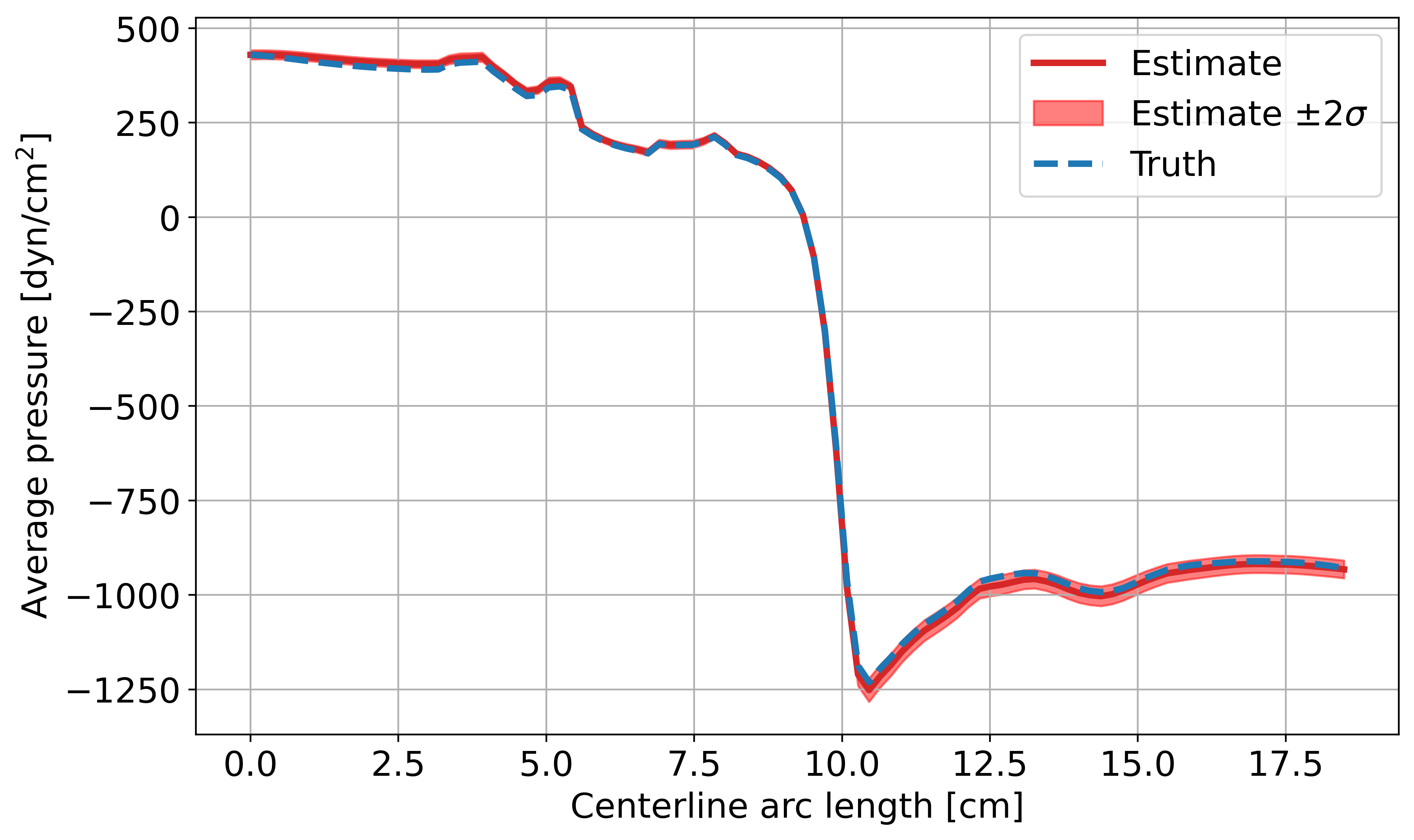}
  \caption{Aortic coarctation.}
\end{subfigure}
\caption{Flow rates and average pressures, with corresponding uncertainty estimates, evaluated on planes orthogonal to the vessel centerline in the aortic aneurysm and aortic coarctation geometries for $\Delta x = 0.200$ cm and $\mathrm{SNR}=5.0$.}
\label{fig:centerline_uq}
\end{figure}

Figure (\ref{fig:centerline_uq}) shows the true and estimated flow rates average pressures, together with their associated uncertainties, on planes uniformly distributed in arc length and oriented orthogonally to the centerline connecting the inlet and descending aorta outlet. The spatial resolution and noise level were set to ($\Delta x=0.250~\mathrm{cm},,\mathrm{SNR}=2.5$) for both the aortic aneurysm and aortic coarctation geometries. As expected, the flow rate decreases along the centerline as blood branches through the BCA, LCCA, and LSA. As dictated by physics, average pressure decreases along the centerline due to viscous dissipation in aneurysm and due to flow acceleration in coarctation. Under the Laplace approximation, the flow rates and average pressures are normally distributed. and $\pm2\sigma$ bands in Figure (\ref{fig:centerline_uq}) represent 95\% credible intervals. It can be observed that, both the true flow-rate and average-pressure profiles, lie within their corresponding uncertainty bands across the centerline. Moreover, the uncertainty in the average pressure is consistently smaller than that of the flow rate.

\begin{figure}[h!]
\centering
\begin{subfigure}[t]{.32\textwidth}
  \centering
  \includegraphics[width=0.99\linewidth]{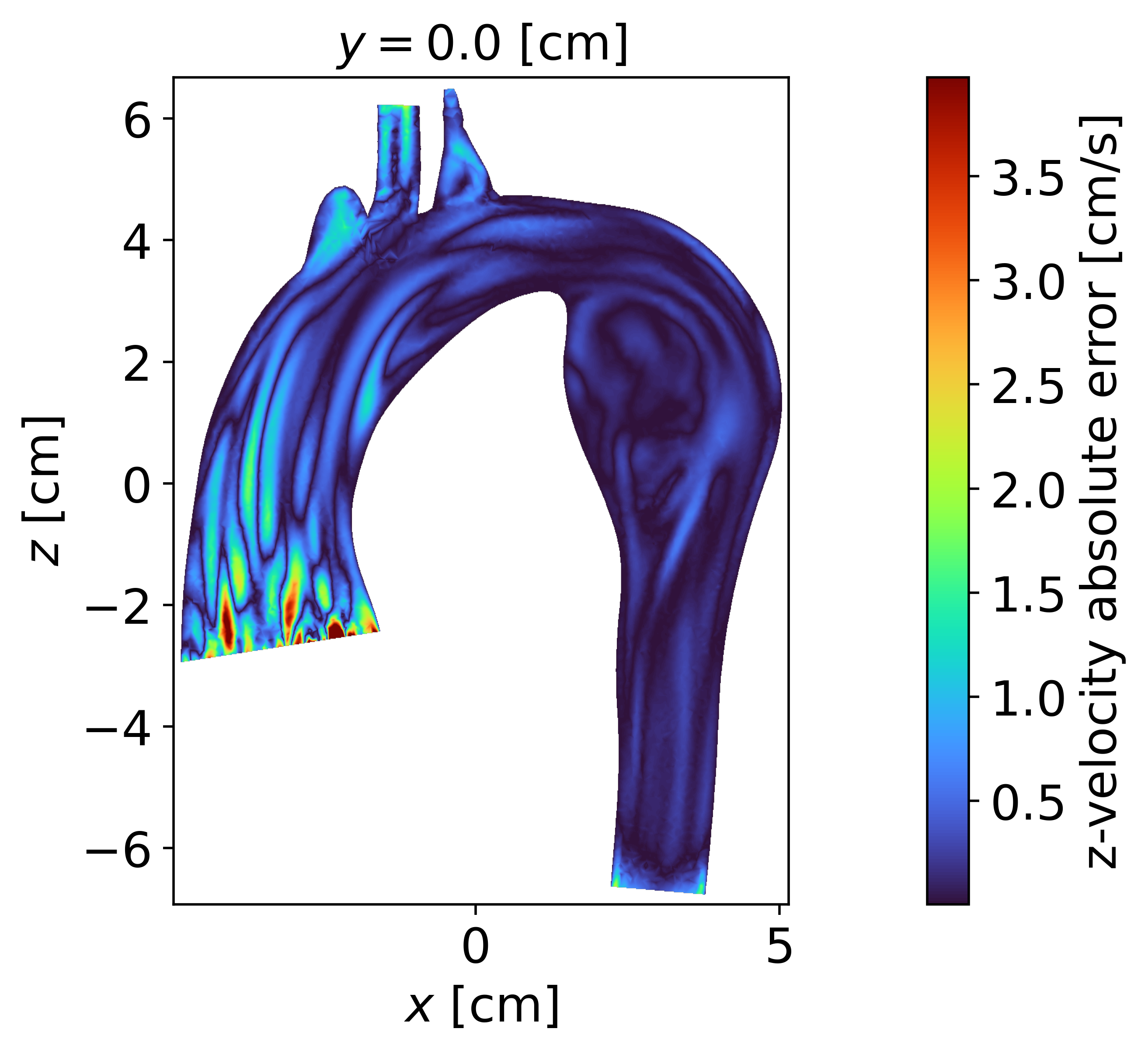}
  \caption{Error}
\end{subfigure}
\hfill
\begin{subfigure}[t]{.32\textwidth}
  \centering
  \includegraphics[width=0.99\linewidth]{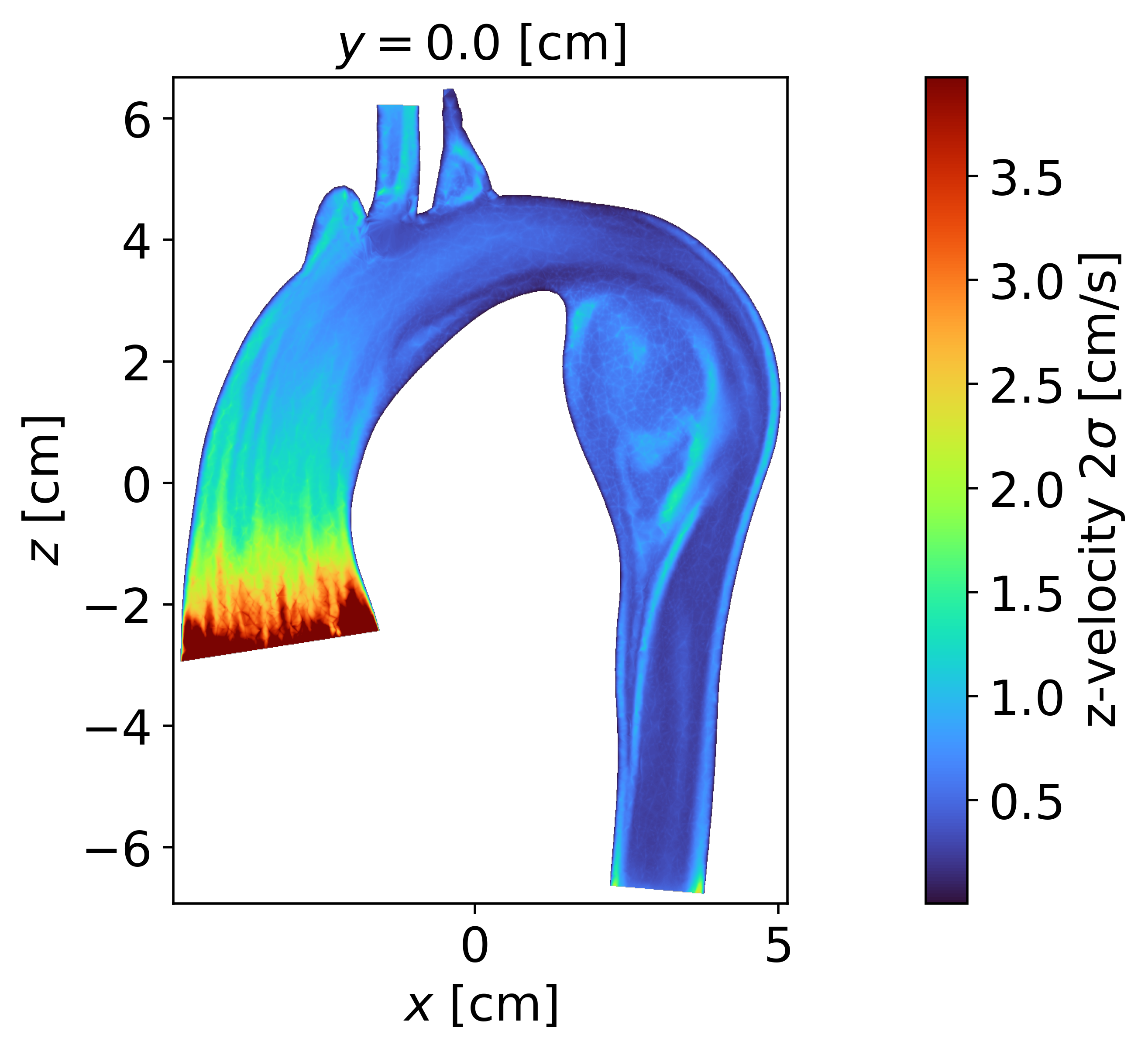}
  \caption{Predicted uncertainty.}
\end{subfigure}
\hfill
\begin{subfigure}[t]{.32\textwidth}
  \centering
  \includegraphics[width=0.99\linewidth]{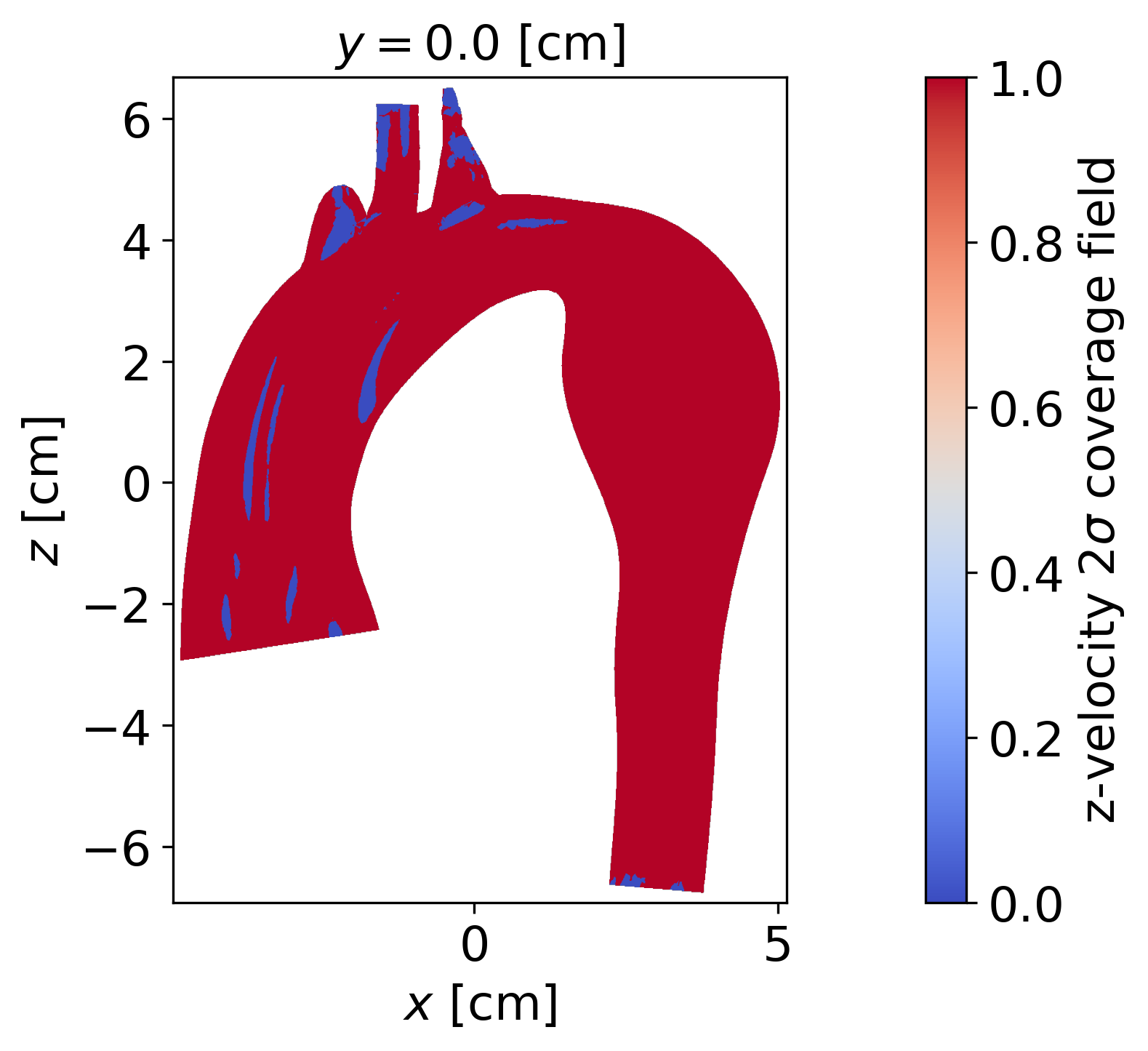}
  \caption{95\% coverage.}
\end{subfigure}

\vspace{0.5em}

\begin{subfigure}[t]{.32\textwidth}
  \centering
  \includegraphics[width=0.99\linewidth]{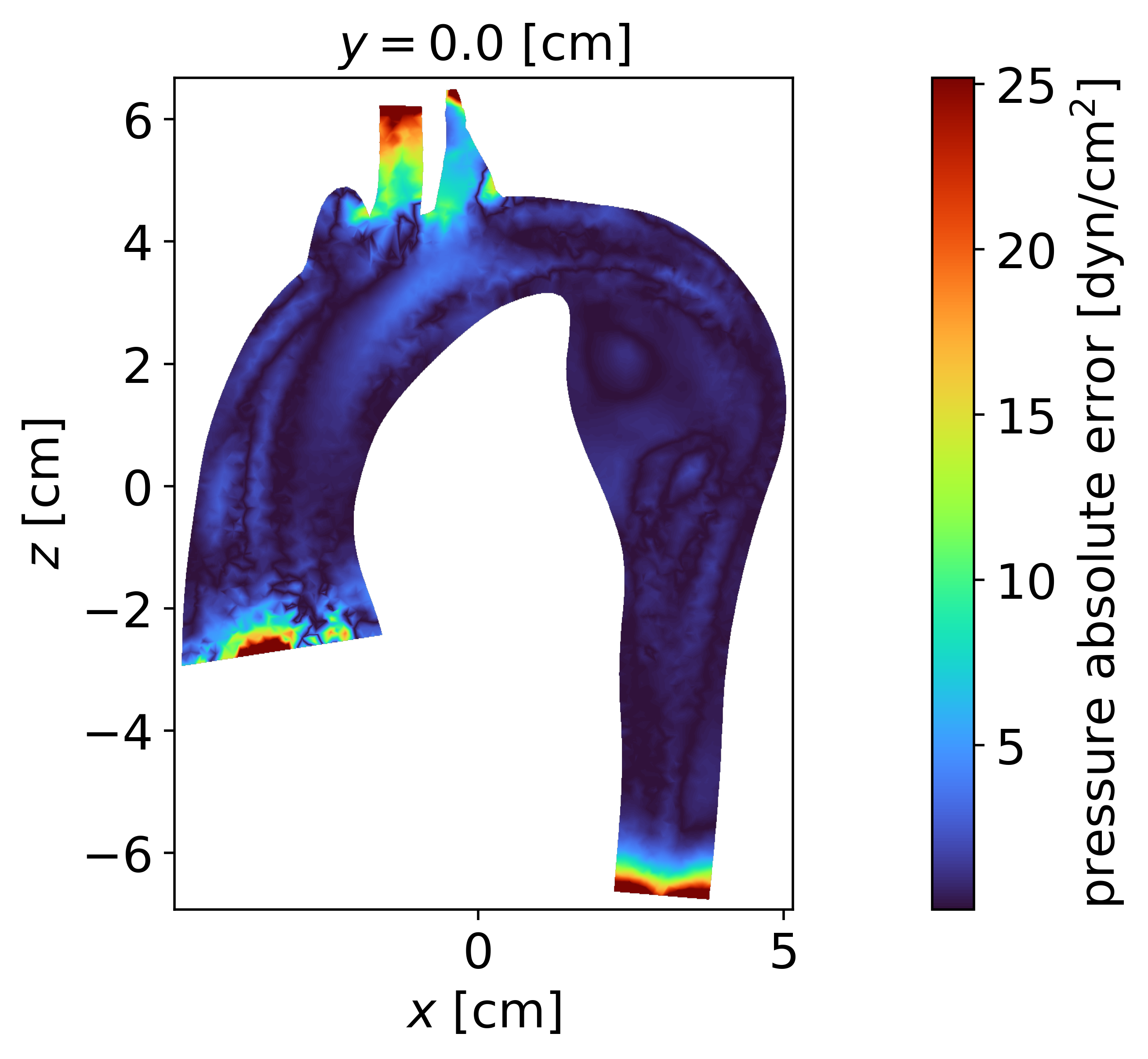}
  \caption{Error.}
\end{subfigure}
\hfill
\begin{subfigure}[t]{.32\textwidth}
  \centering
  \includegraphics[width=0.99\linewidth]{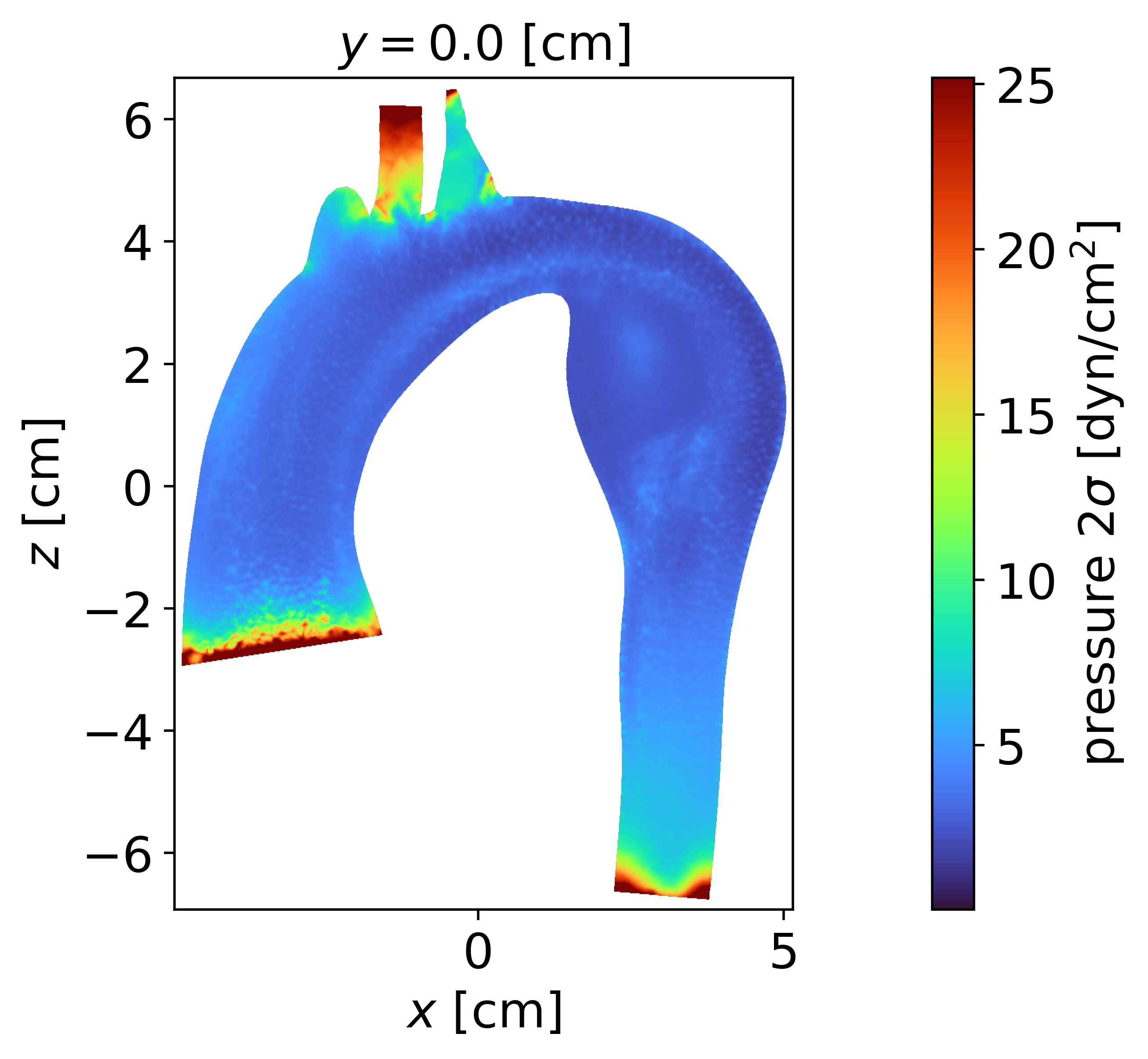}
  \caption{Predicted uncertainty.}
\end{subfigure}
\hfill
\begin{subfigure}[t]{.32\textwidth}
  \centering
  \includegraphics[width=0.99\linewidth]{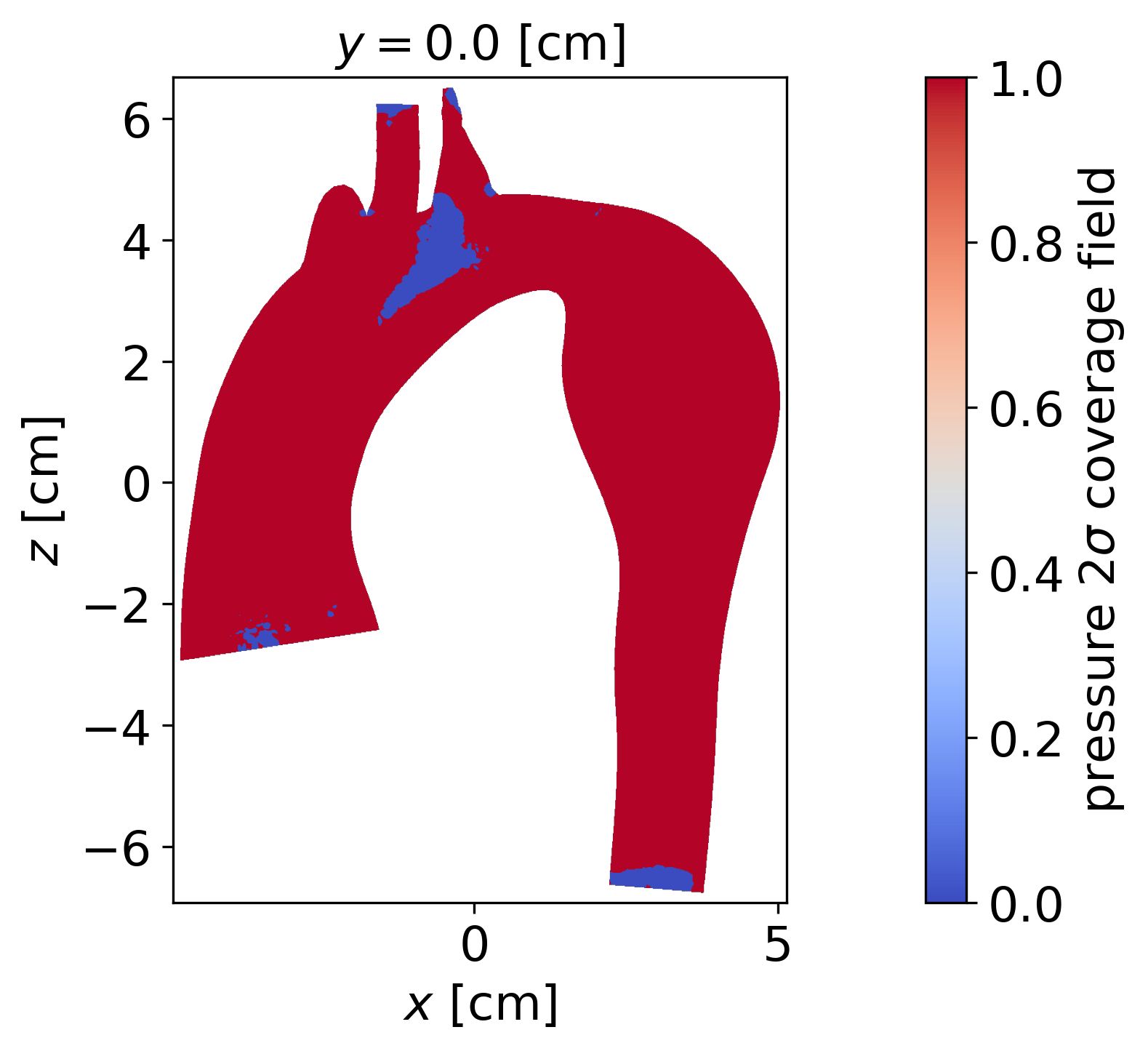}
  \caption{95\% Coverage.}
\end{subfigure}

\vspace{0.5em}

\begin{subfigure}[t]{.32\textwidth}
  \centering
  \includegraphics[width=0.99\linewidth]{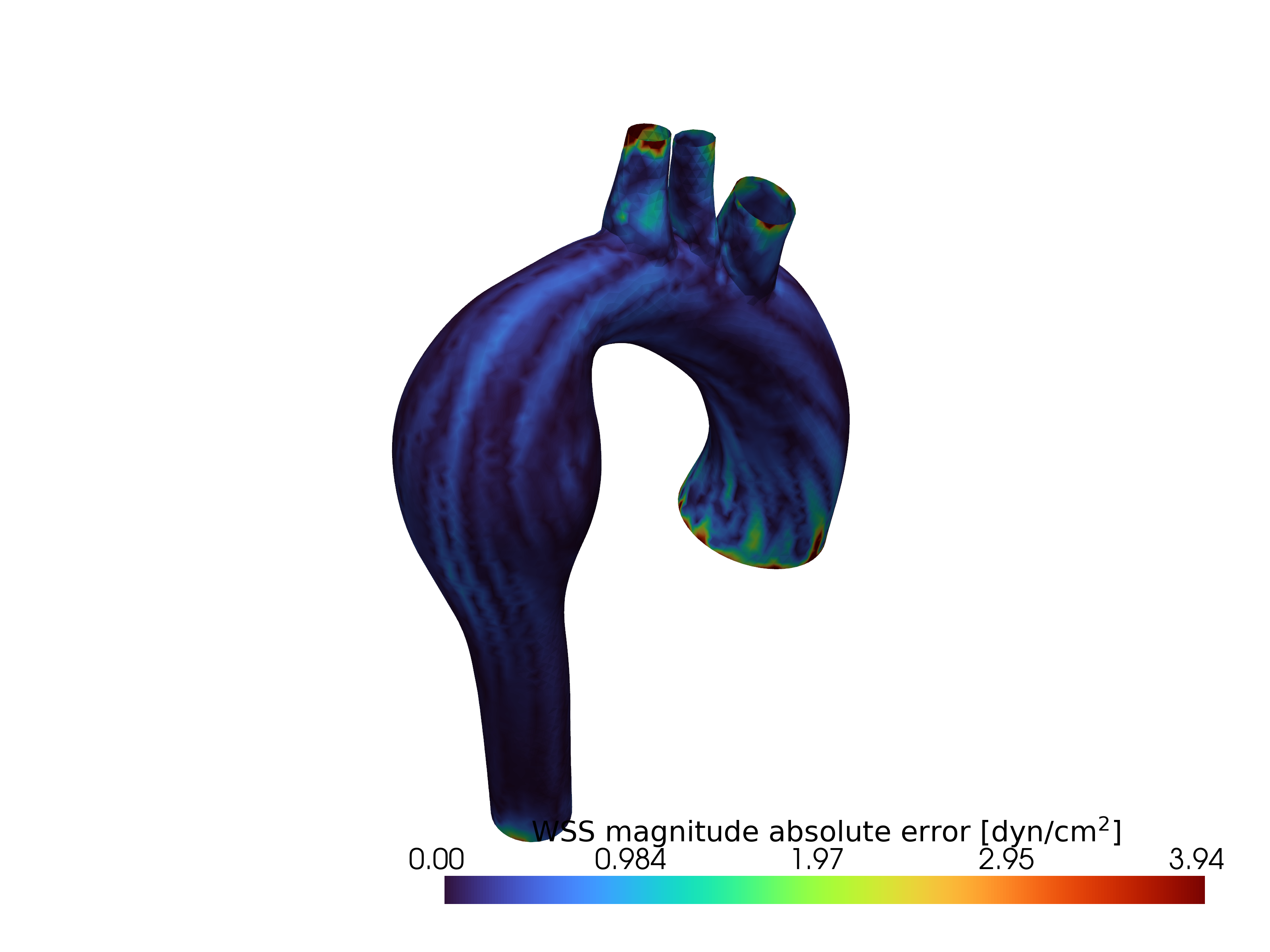}
  \caption{Error.}
\end{subfigure}
\hfill
\begin{subfigure}[t]{.32\textwidth}
  \centering
  \includegraphics[width=0.99\linewidth]{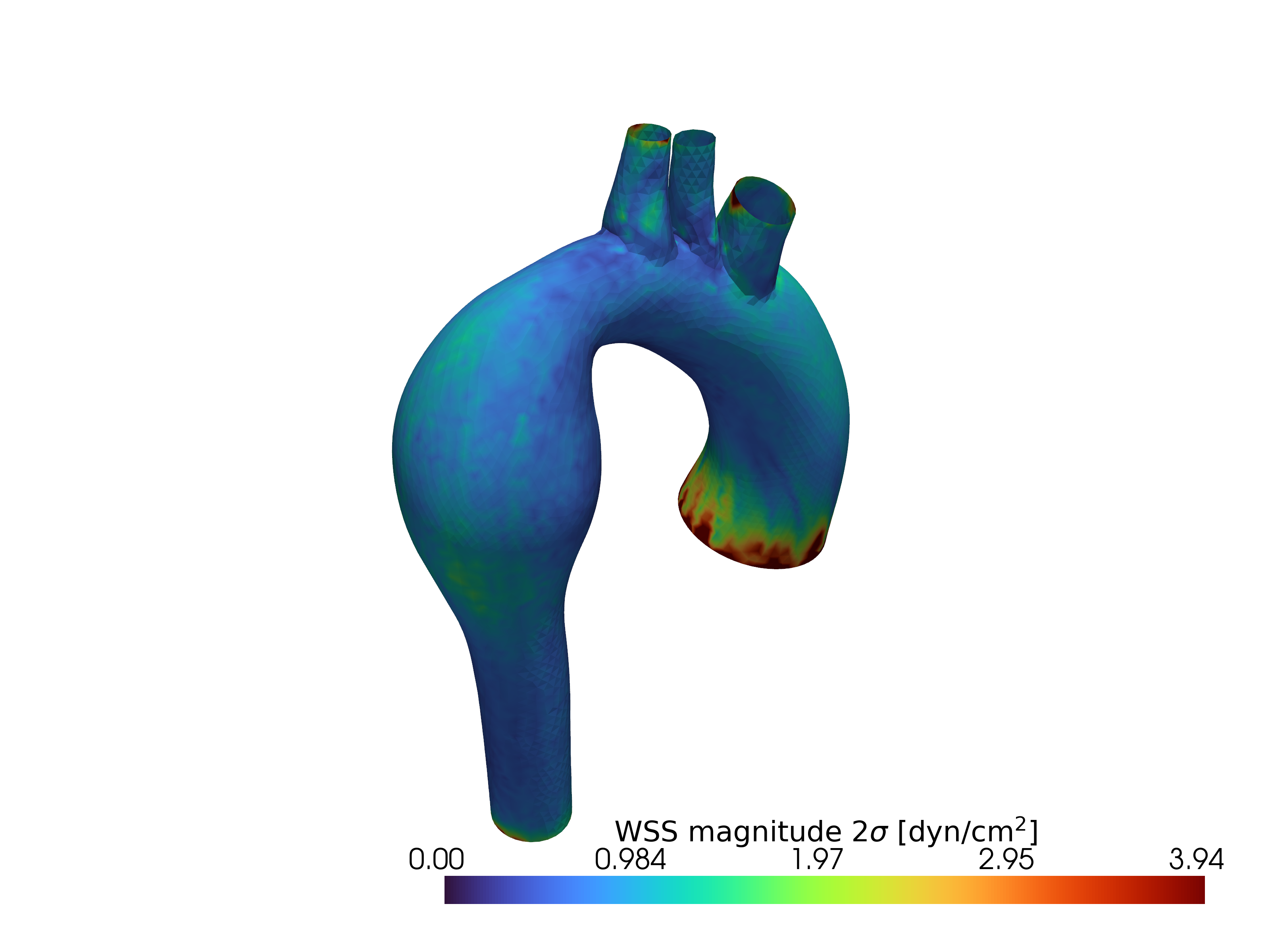}
  \caption{Predicted uncertainty}
\end{subfigure}
\hfill
\begin{subfigure}[t]{.32\textwidth}
  \centering
  \includegraphics[width=0.99\linewidth]{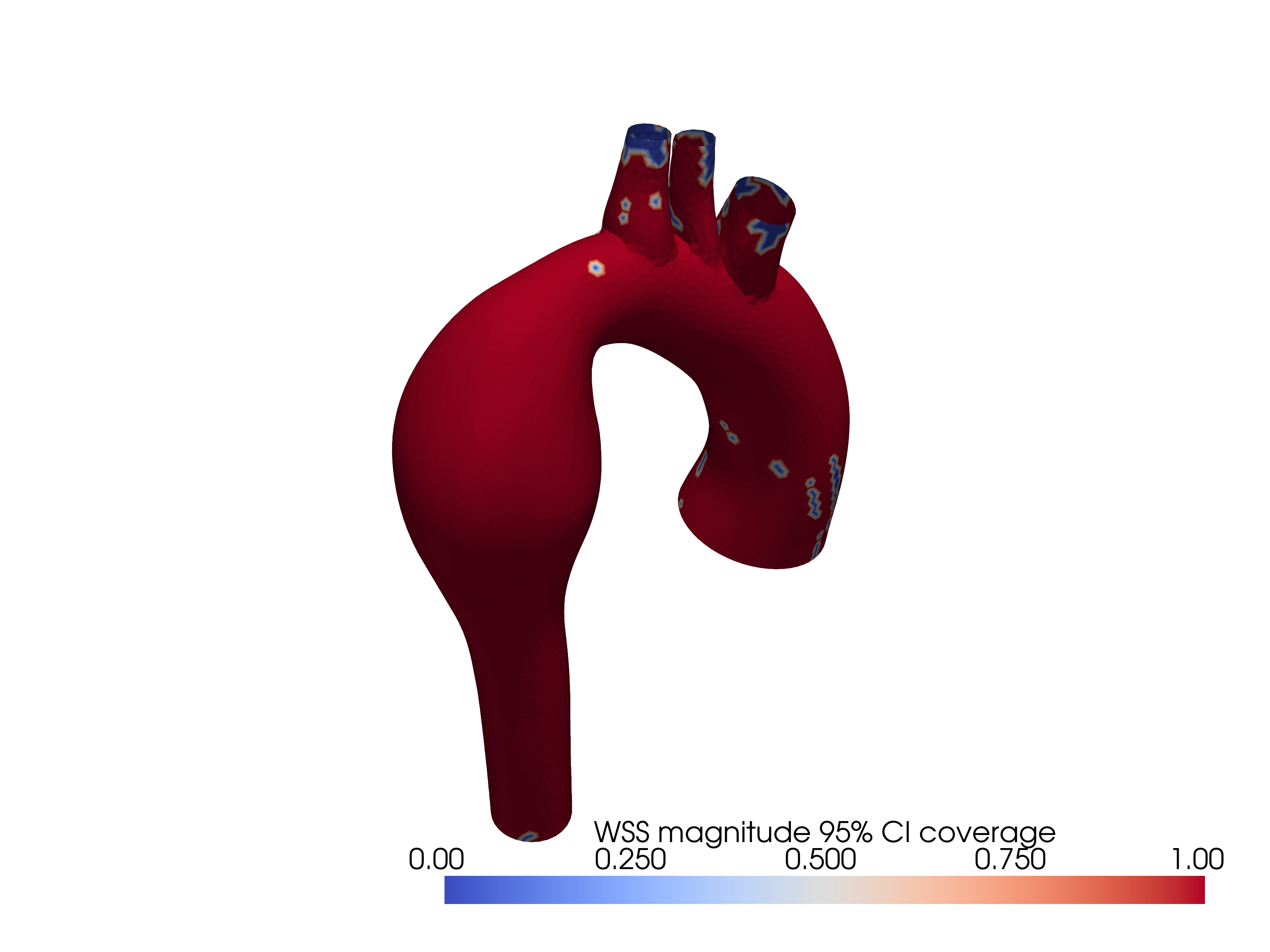}
  \caption{95\% Coverage.}
\end{subfigure}
\caption{Reconstruction errors, posterior uncertainties, and coverage fields for the z-velocity, pressure, and WSS magnitude in the aortic aneurysm geometry, for $\Delta x = 0.200$ cm and $\mathrm{SNR}=5.0$.}
\label{fig:aa_uq}
\end{figure}

\begin{figure}
\centering
\begin{subfigure}[t]{.32\textwidth}
  \centering
  \includegraphics[width=0.99\linewidth]{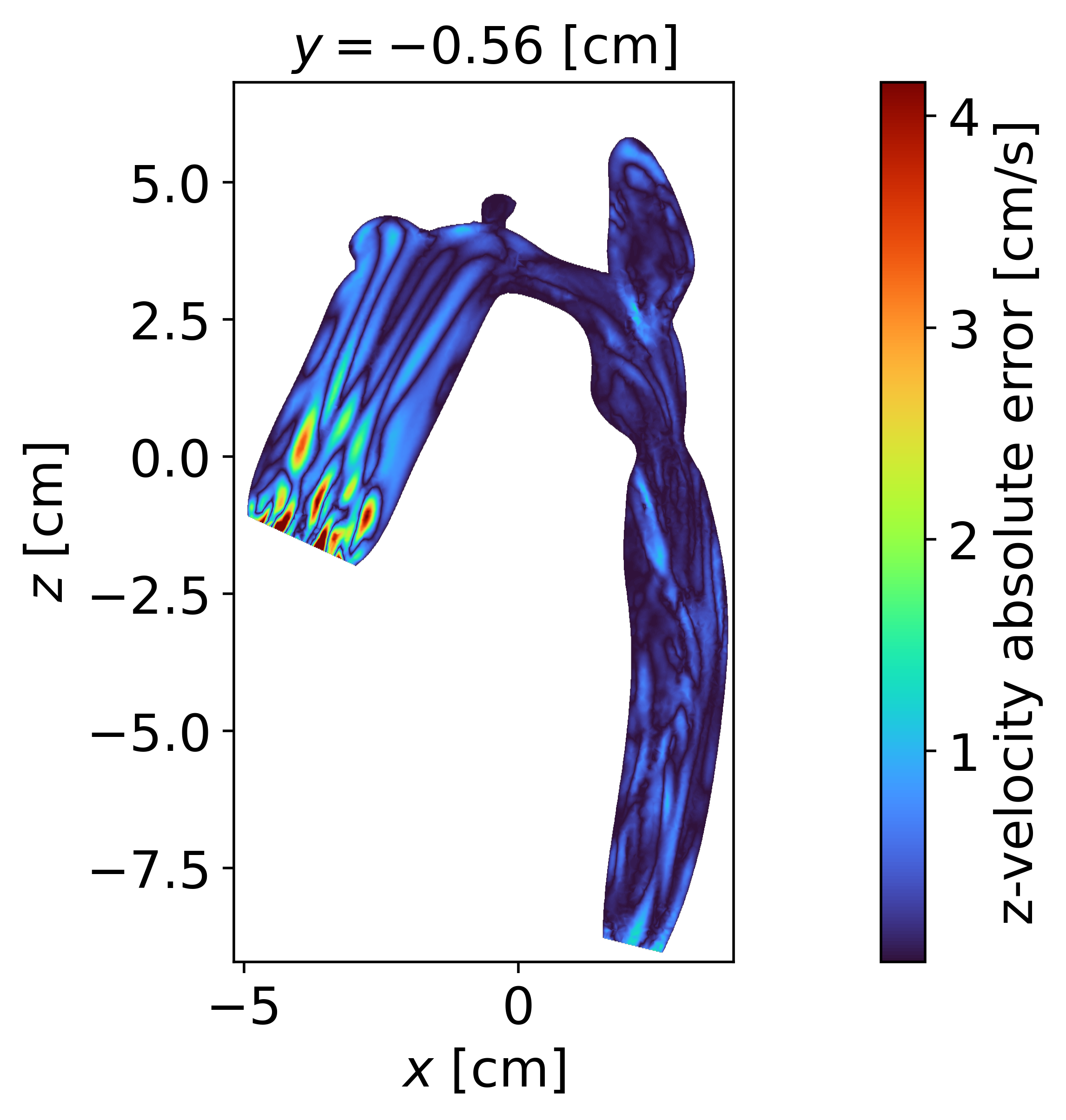}
  \caption{Error}
\end{subfigure}
\hfill
\begin{subfigure}[t]{.32\textwidth}
  \centering
  \includegraphics[width=0.99\linewidth]{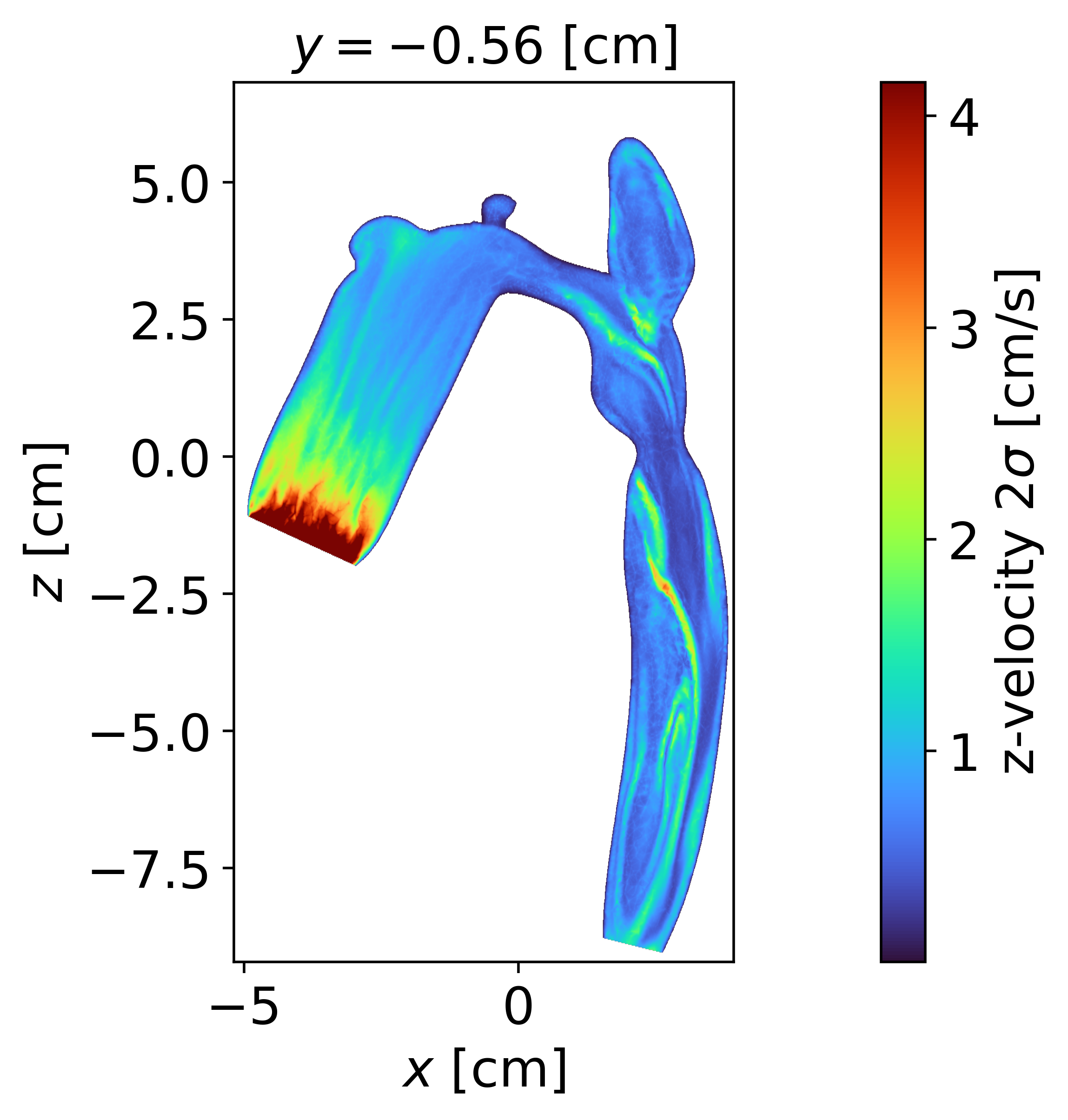}
  \caption{Predicted uncertainty.}
\end{subfigure}
\hfill
\begin{subfigure}[t]{.32\textwidth}
  \centering
  \includegraphics[width=0.99\linewidth]{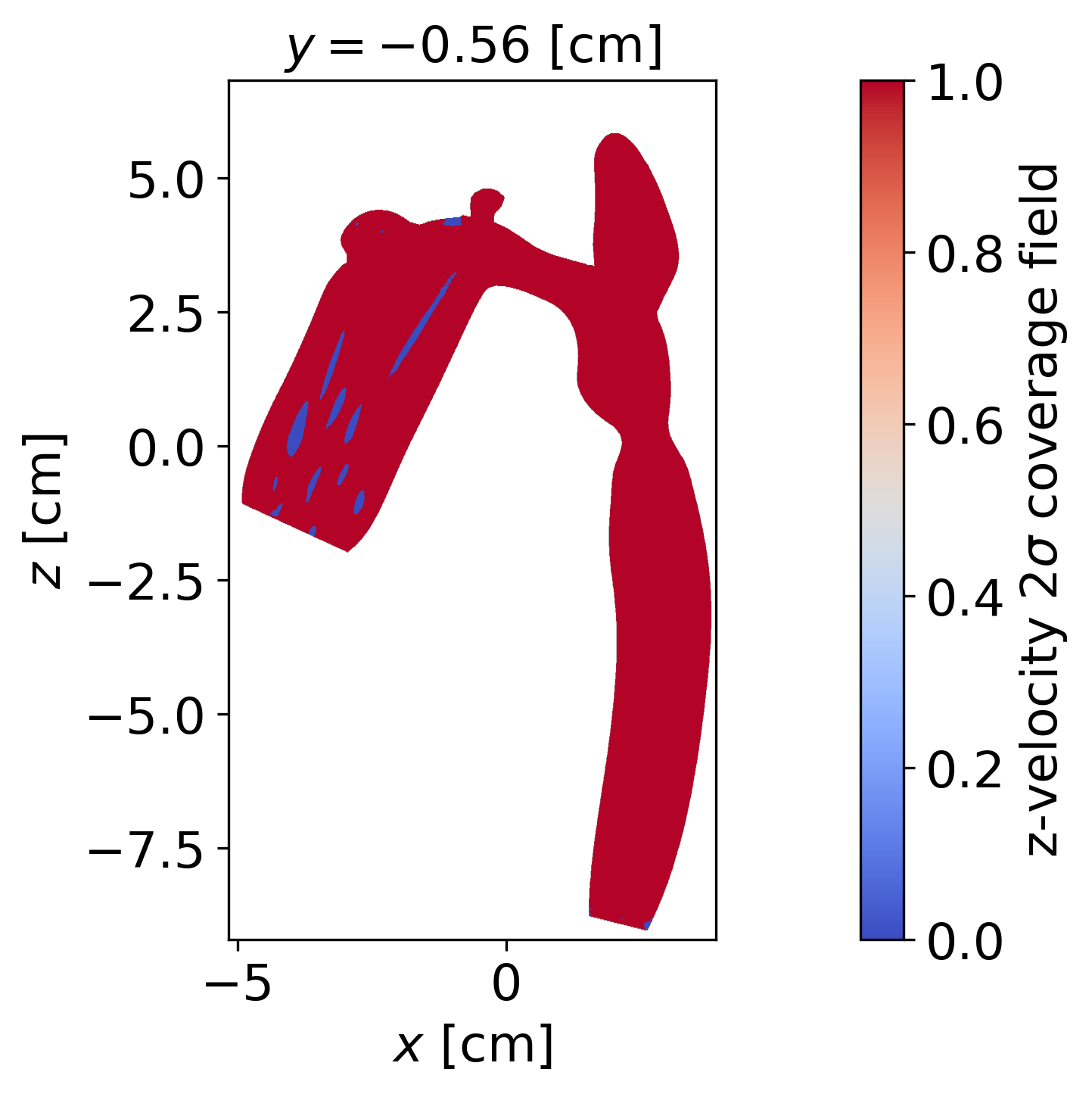}
  \caption{95\% coverage.}
\end{subfigure}

\vspace{0.5em}

\begin{subfigure}[t]{.32\textwidth}
  \centering
  \includegraphics[width=0.99\linewidth]{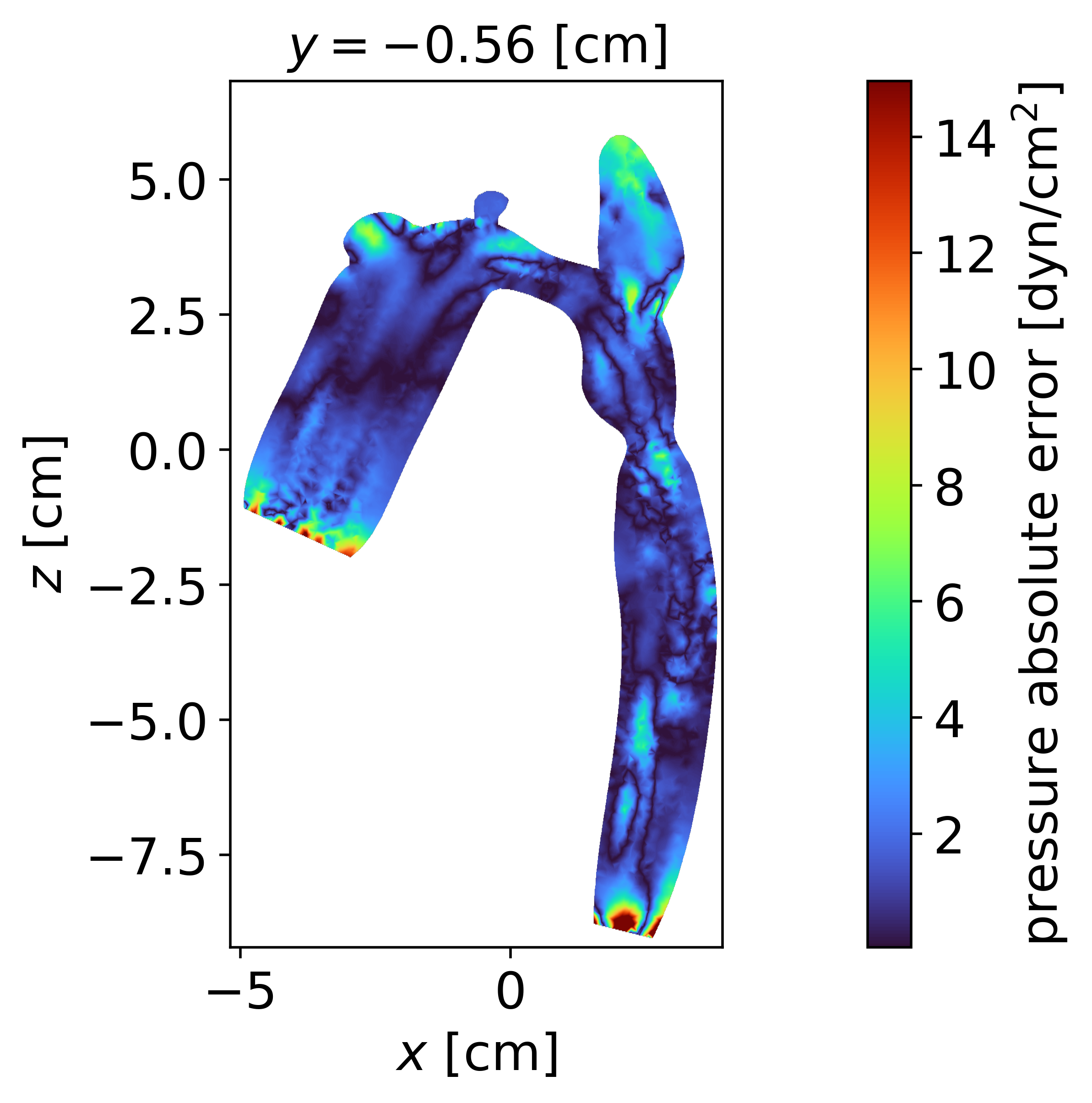}
  \caption{Error.}
\end{subfigure}
\hfill
\begin{subfigure}[t]{.32\textwidth}
  \centering
  \includegraphics[width=0.99\linewidth]{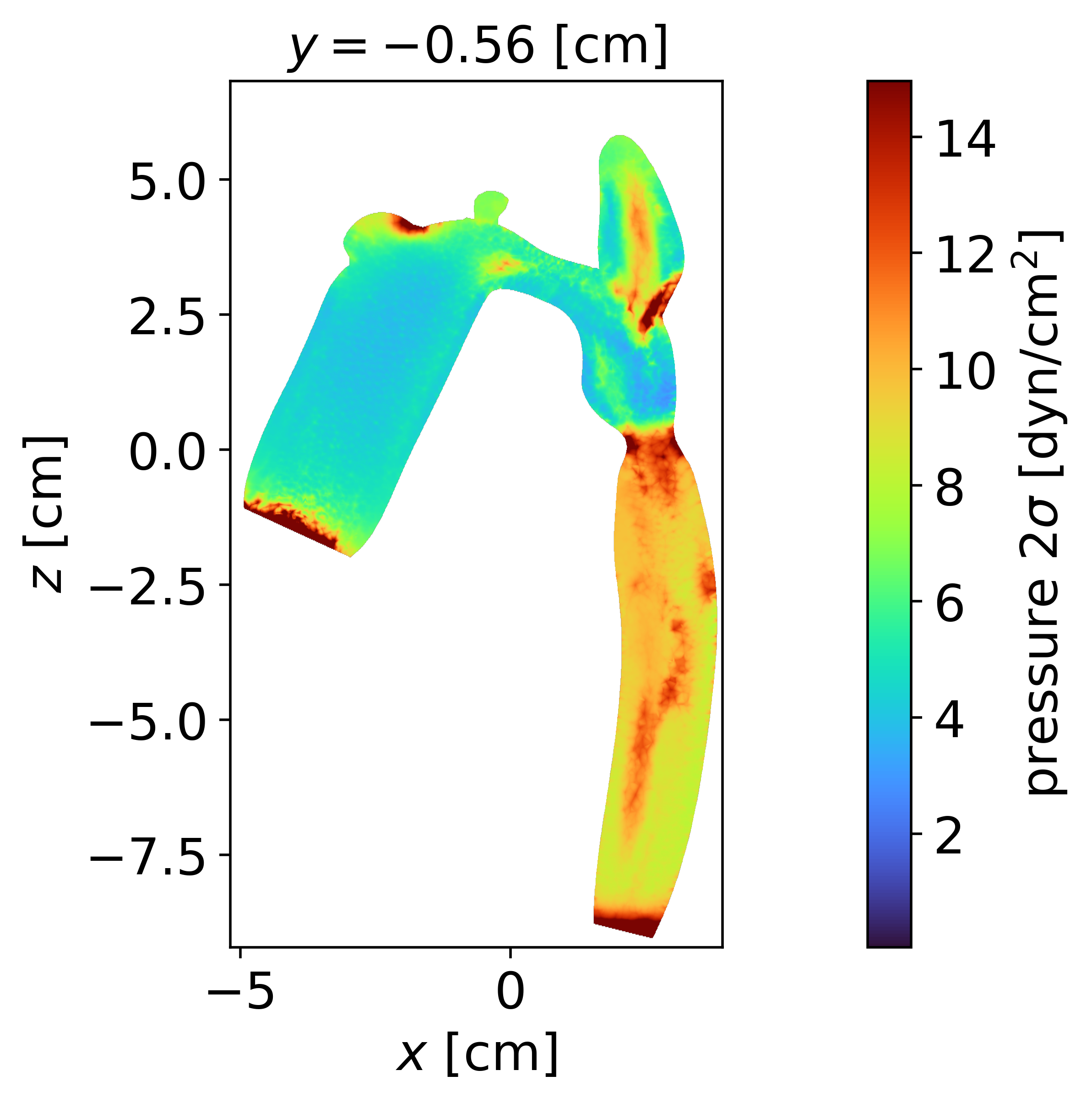}
  \caption{Predicted uncertainty.}
\end{subfigure}
\hfill
\begin{subfigure}[t]{.32\textwidth}
  \centering
  \includegraphics[width=0.99\linewidth]{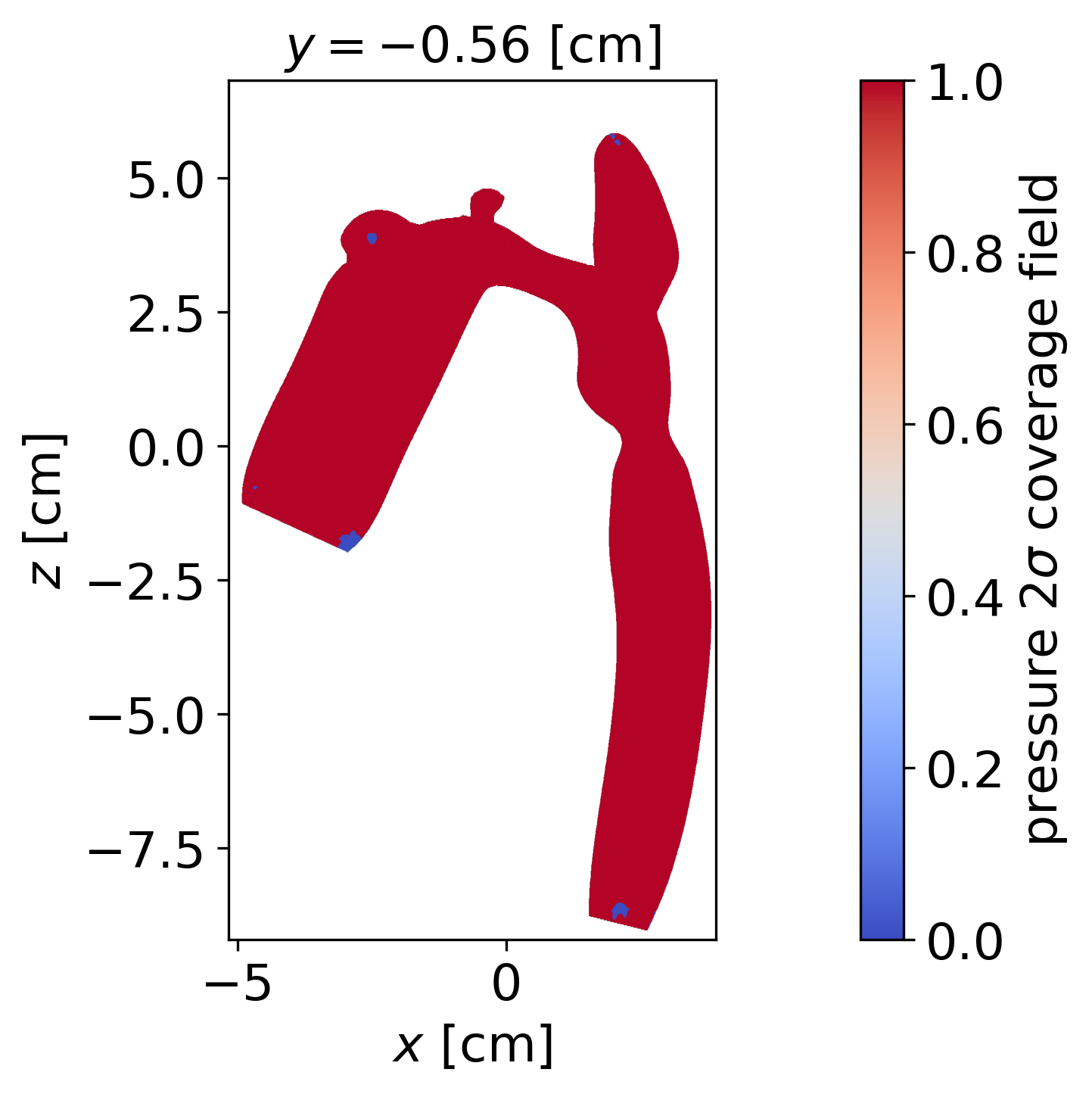}
  \caption{95\% Coverage.}
\end{subfigure}

\vspace{0.5em}

\begin{subfigure}[t]{.32\textwidth}
  \centering
  \includegraphics[width=0.99\linewidth]{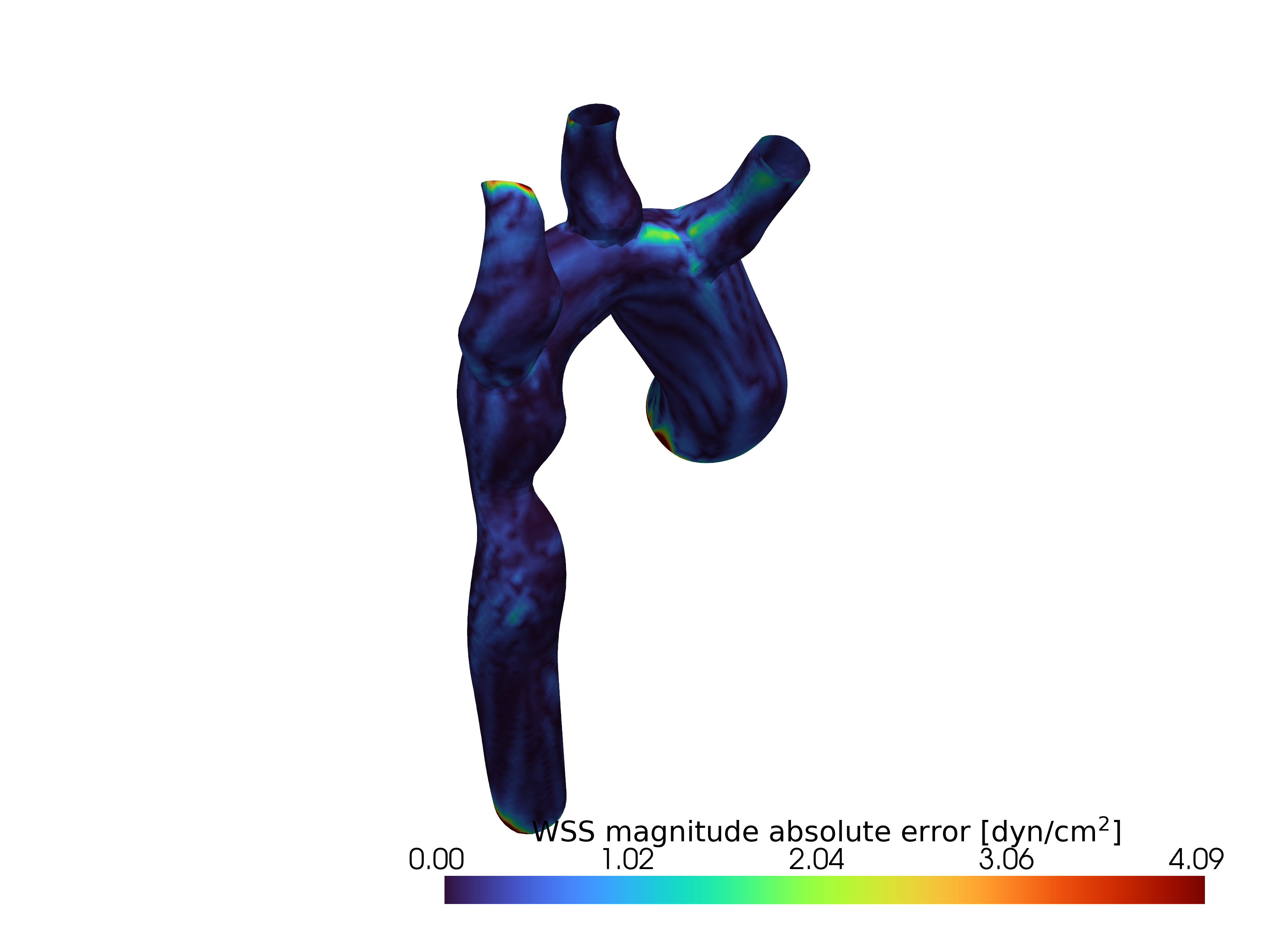}
  \caption{Error.}
\end{subfigure}
\hfill
\begin{subfigure}[t]{.32\textwidth}
  \centering
  \includegraphics[width=0.99\linewidth]{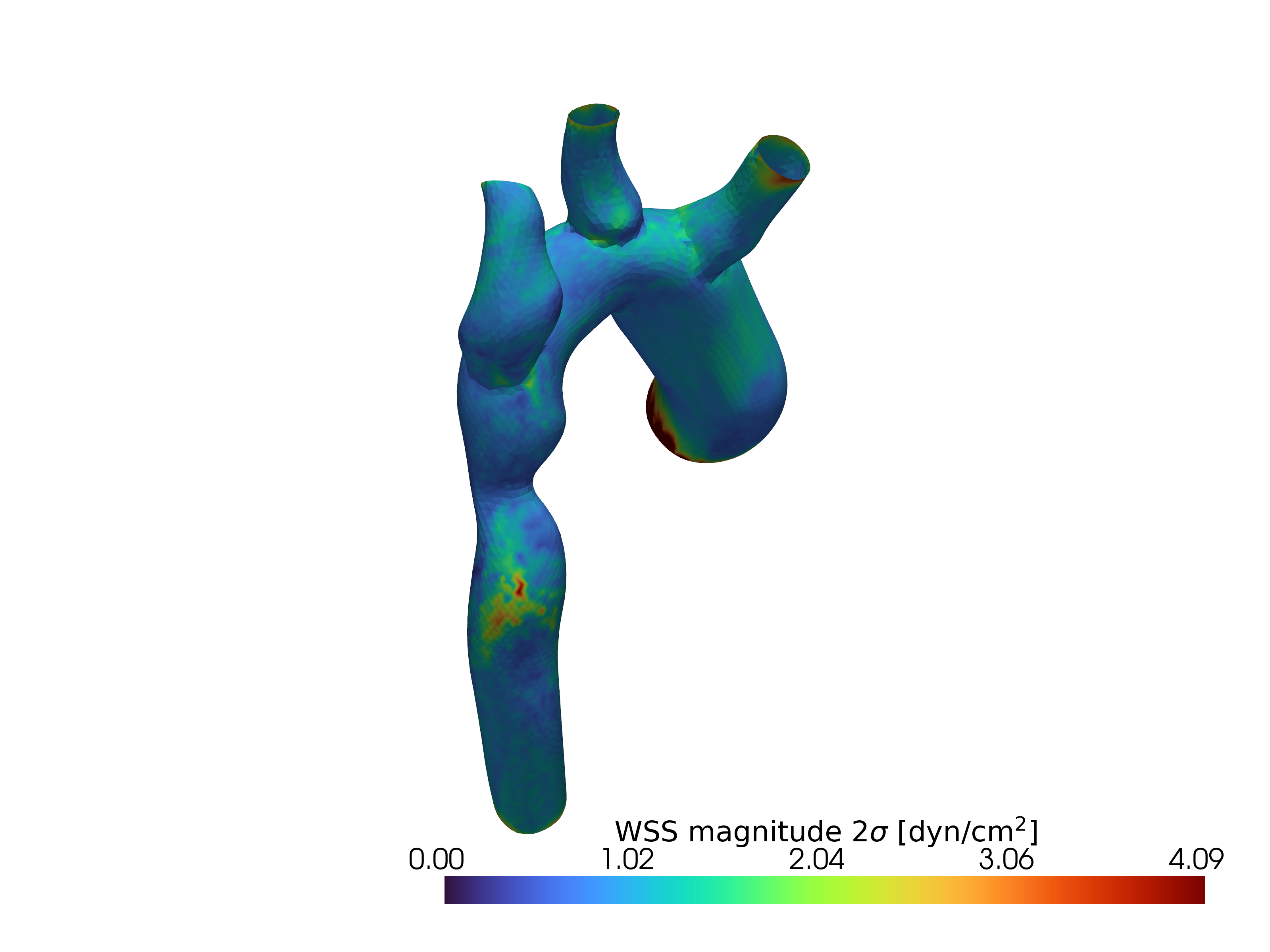}
  \caption{Predicted uncertainty}
\end{subfigure}
\hfill
\begin{subfigure}[t]{.32\textwidth}
  \centering
  \includegraphics[width=0.99\linewidth]{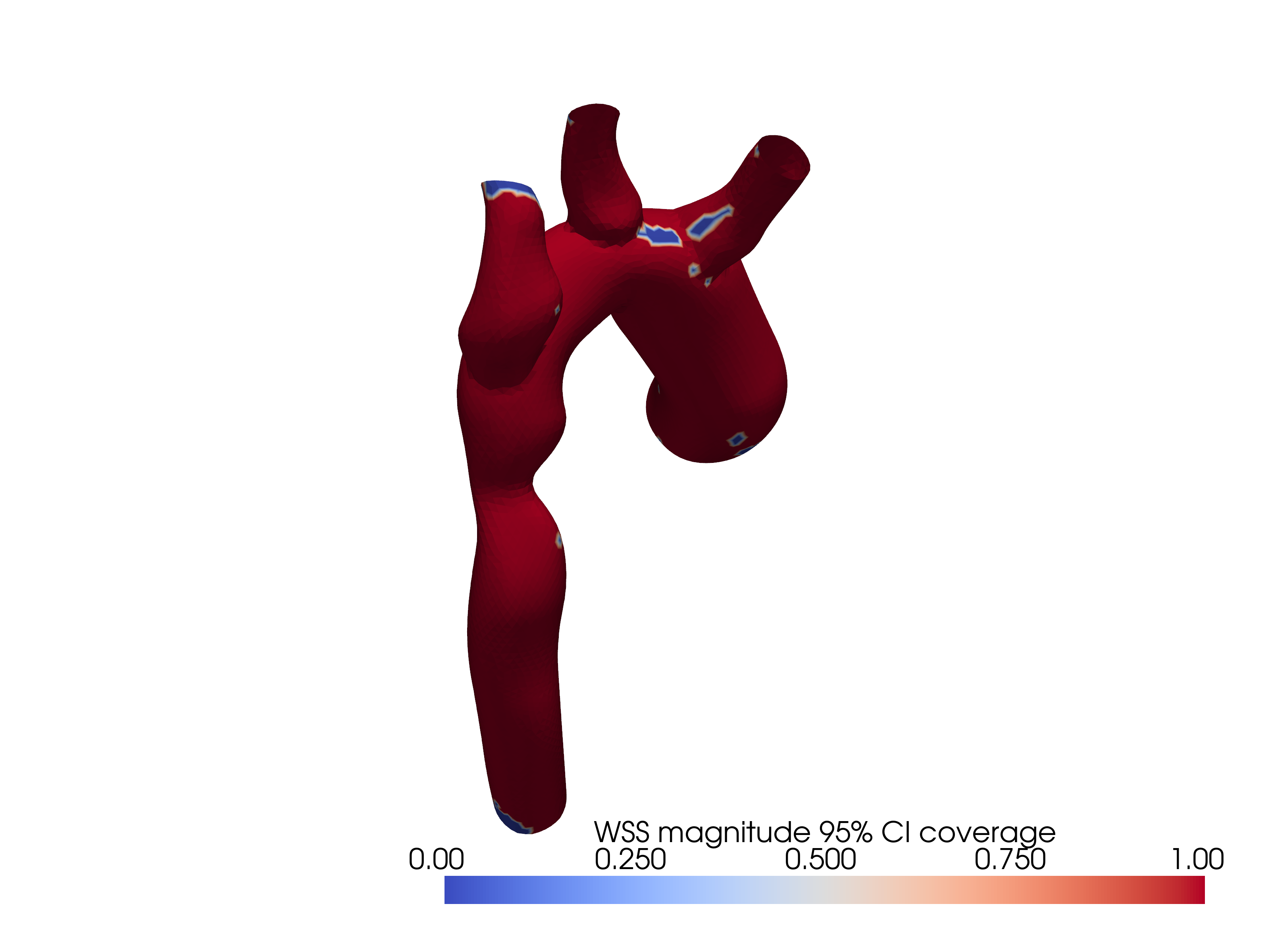}
  \caption{95\% Coverage.}
\end{subfigure}
\caption{Reconstruction errors, posterior uncertainties, and coverage fields for the z-velocity, pressure, and WSS magnitude in the aortic coarctation geometry, for $\Delta x = 0.200$ cm and $\mathrm{SNR}=5.0$.}
\label{fig:ac_uq}
\end{figure}

Figures (\ref{fig:aa_uq}--\ref{fig:ac_uq}) depict the reconstruction error, uncertainty (marginal standard deviation, $\sigma$), and empirical coverage for the $z$-velocity, pressure, and WSS magnitude in the aortic aneurysm, aortic coarctation, and cerebral aneurysm geometries. In all cases, regions of large standard deviation generally coincide with regions of large reconstruction error, indicating that the proposed uncertainty quantification framework successfully identifies areas where the reconstruction is less reliable. For all geometries, the posterior standard deviation predicts poor accuracy in the $z$-velocity field near the inlet, which is consistent with the reported reconstruction error fields. Similarly, the pressure uncertainty is largest near the inlet and outlet boundaries, where larger pressure errors are reported. For the WSS magnitude, the posterior standard deviation is highest near the inlet region, again corresponding to regions of increased reconstruction error. In the aortic coarctation and aortic aneurysm geometries, slightly elevated uncertainty is also observed near the stenosis and aneurysm, respectively, where the reconstruction errors in $z$-velocity, pressure, and WSS magnitude are accordingly larger. Under the Laplace approximation, the posterior distributions of velocity and pressure are Gaussian. Consequently, approximately 95\% of realizations are expected to lie within ($\pm 2\sigma$) of the posterior mean. In contrast, WSS magnitude is a nonlinear functional of the velocity degrees of freedom and therefore does not follow a Gaussian distribution. While the posterior standard deviation still provides a useful measure of local uncertainty, we computed the credible intervals for WSS magnitude by applying the probabilistic integral transform (PIT) to Monte Carlo samples drawn from the posterior distribution. Both the PIT-based credible intervals and the reported posterior standard deviations were estimated using 1000 Monte Carlo samples. The 95\% coverage plots show that the majority of the reconstruction error is contained within the corresponding uncertainty bounds. For velocity and pressure, these bounds correspond to ($\pm 2\sigma$) intervals, whereas for WSS magnitude they correspond to PIT-based 95\% credible intervals. Across all geometries, approximately 99\% of nodal values within the region of interest lie inside their nominal 95\% credible intervals. This indicates that the uncertainty estimates are slightly conservative, with the predicted intervals being wider than required for perfect calibration. Similar behavior was observed for quantities not shown here, including the $x$-velocity, $y$-velocity, and velocity magnitude. Elevated uncertainty is not observed in the centerline flow-rate and average-pressure near the inlet or outlet because these quantities were evaluated only after a small amount of distance away from the inlet and outlet boundary.


\begin{figure}
\centering
\begin{subfigure}[t]{.32\textwidth}
  \centering
  \includegraphics[width=0.99\linewidth]{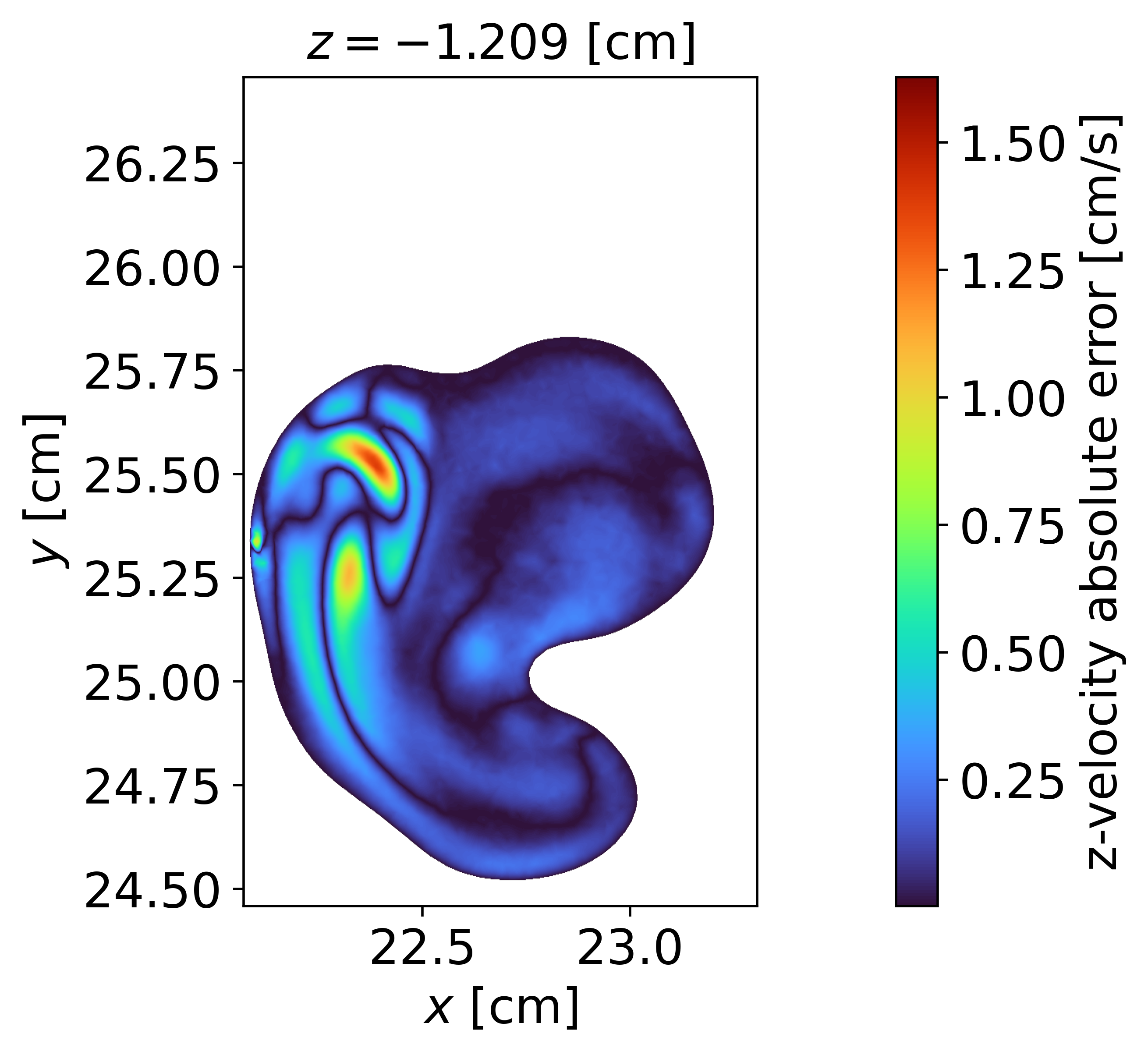}
  \caption{Error}
\end{subfigure}
\hfill
\begin{subfigure}[t]{.32\textwidth}
  \centering
  \includegraphics[width=0.99\linewidth]{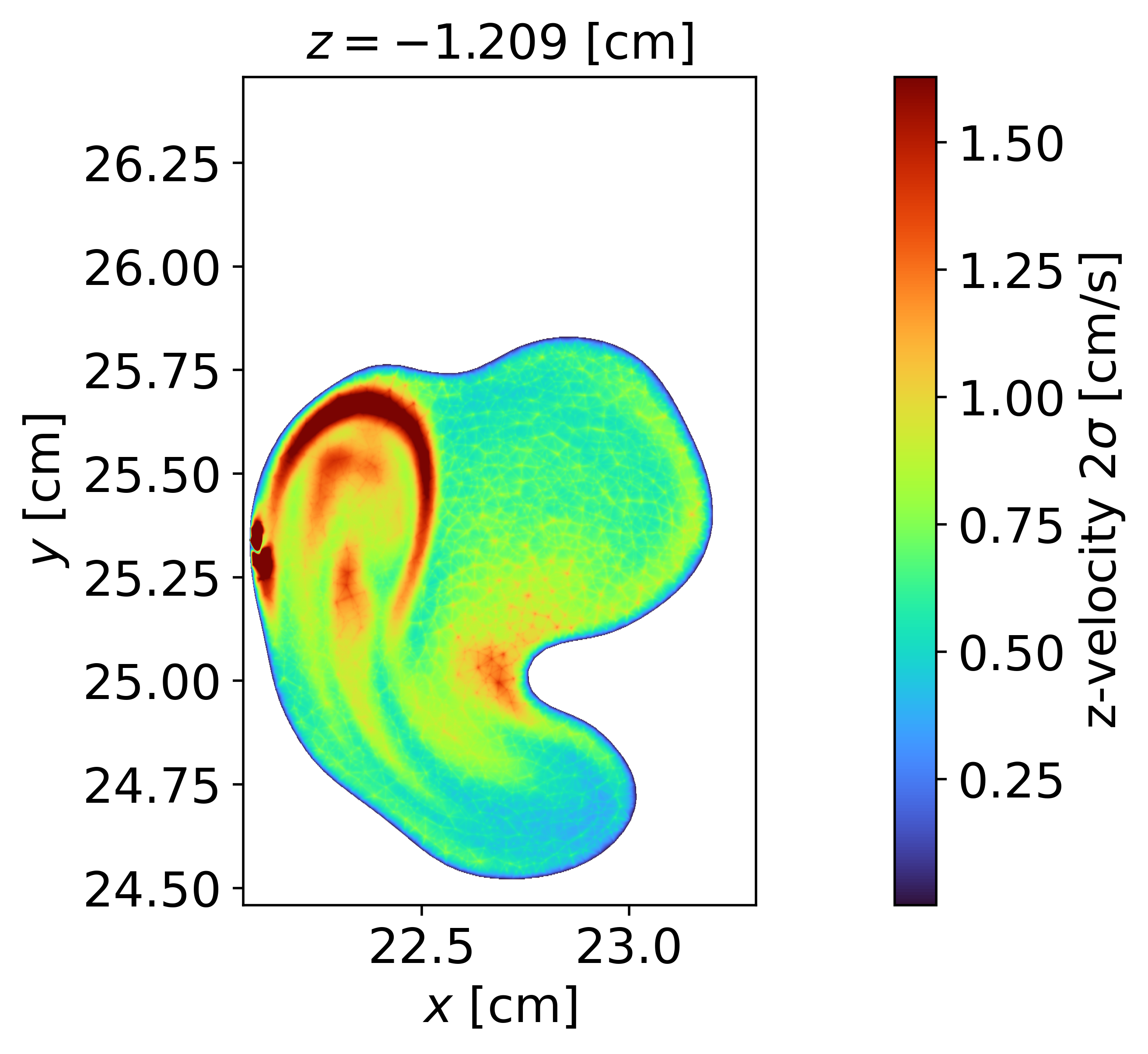}
  \caption{Predicted uncertainty.}
\end{subfigure}
\hfill
\begin{subfigure}[t]{.32\textwidth}
  \centering
  \includegraphics[width=0.99\linewidth]{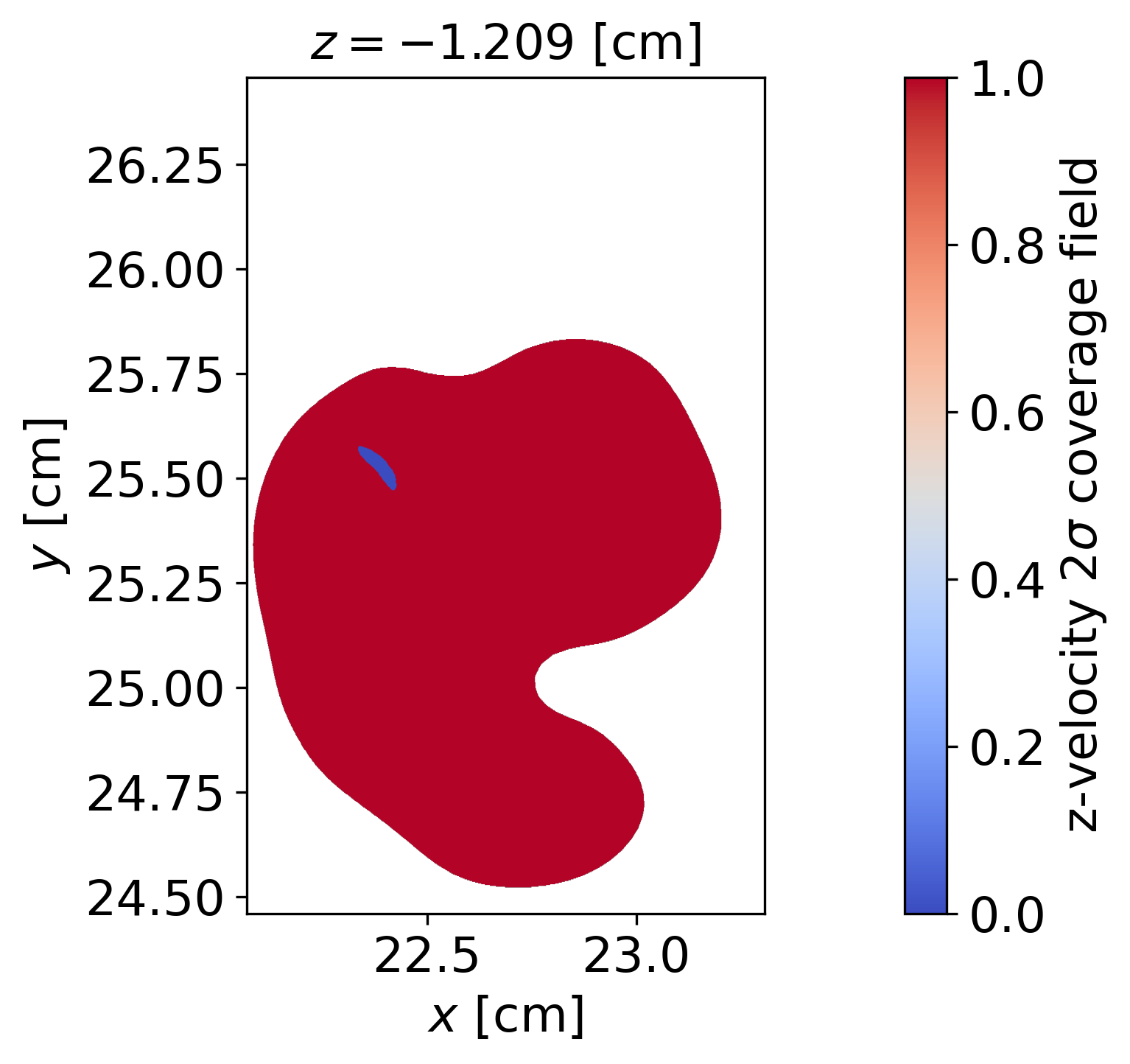}
  \caption{95\% coverage.}
\end{subfigure}

\vspace{0.5em}

\begin{subfigure}[t]{.32\textwidth}
  \centering
  \includegraphics[width=0.99\linewidth]{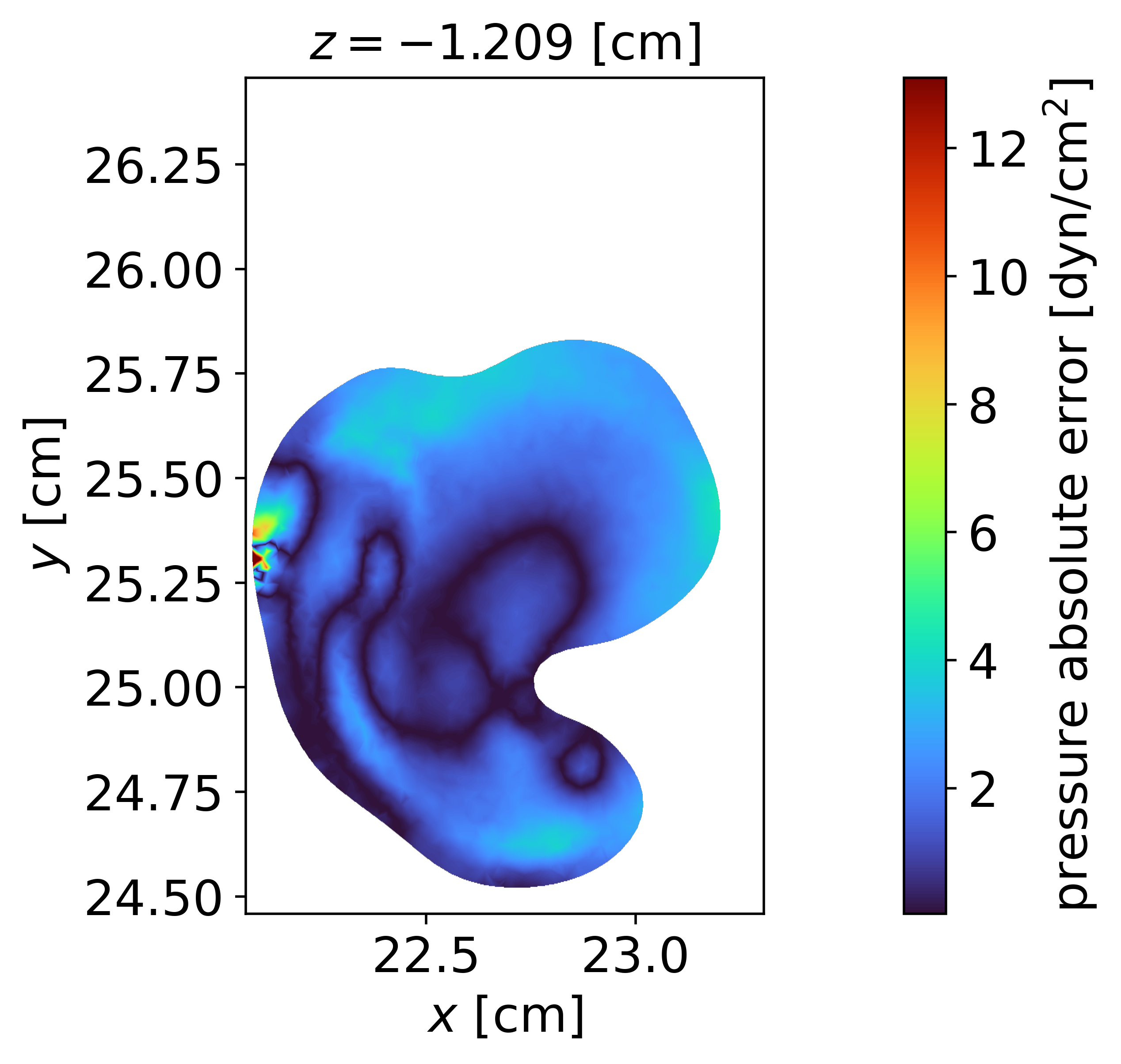}
  \caption{Error.}
\end{subfigure}
\hfill
\begin{subfigure}[t]{.32\textwidth}
  \centering
  \includegraphics[width=0.99\linewidth]{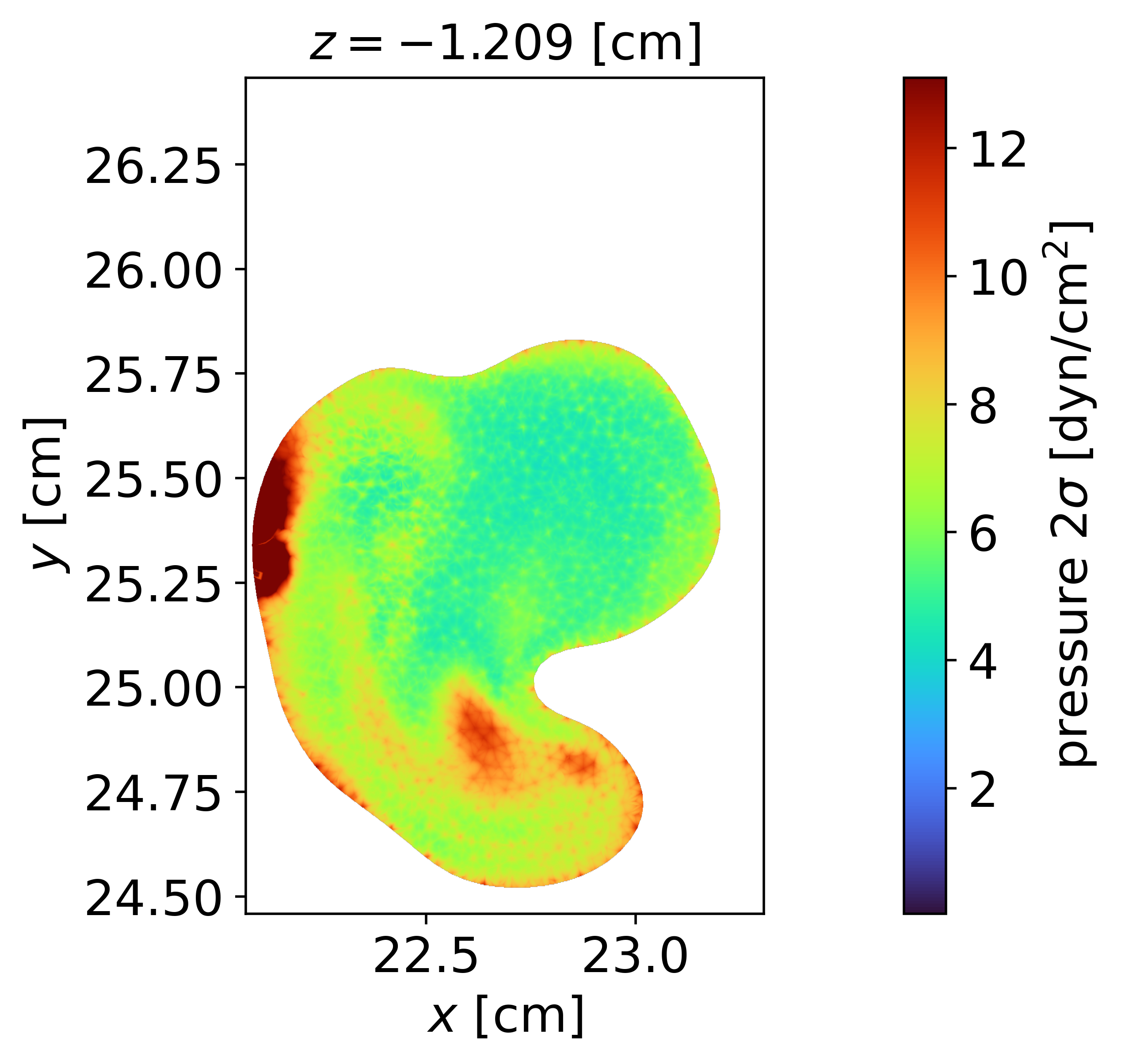}
  \caption{Predicted uncertainty.}
\end{subfigure}
\hfill
\begin{subfigure}[t]{.32\textwidth}
  \centering
  \includegraphics[width=0.99\linewidth]{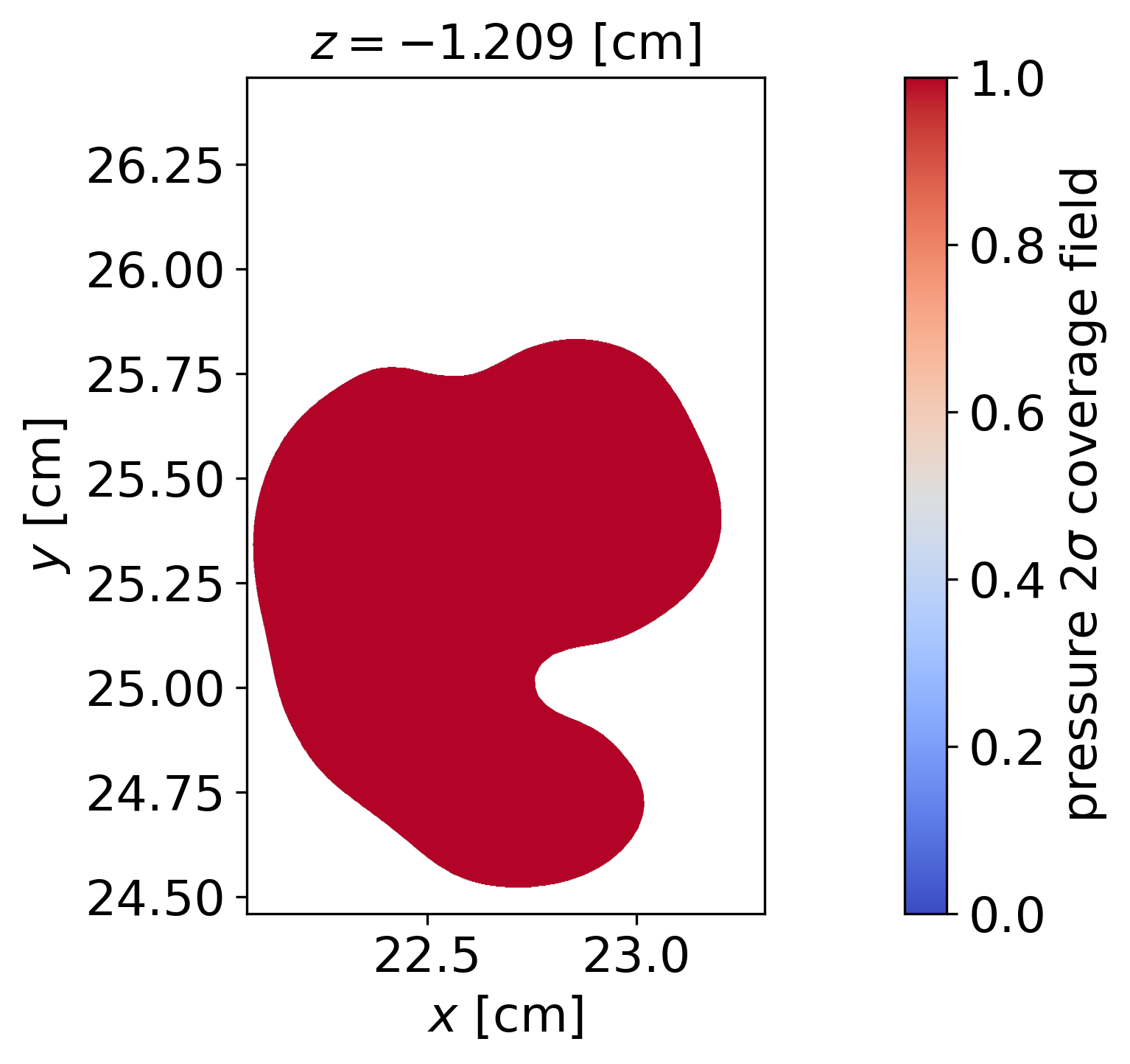}
  \caption{95\% Coverage.}
\end{subfigure}

\vspace{0.5em}

\begin{subfigure}[t]{.32\textwidth}
  \centering
  \includegraphics[width=0.99\linewidth]{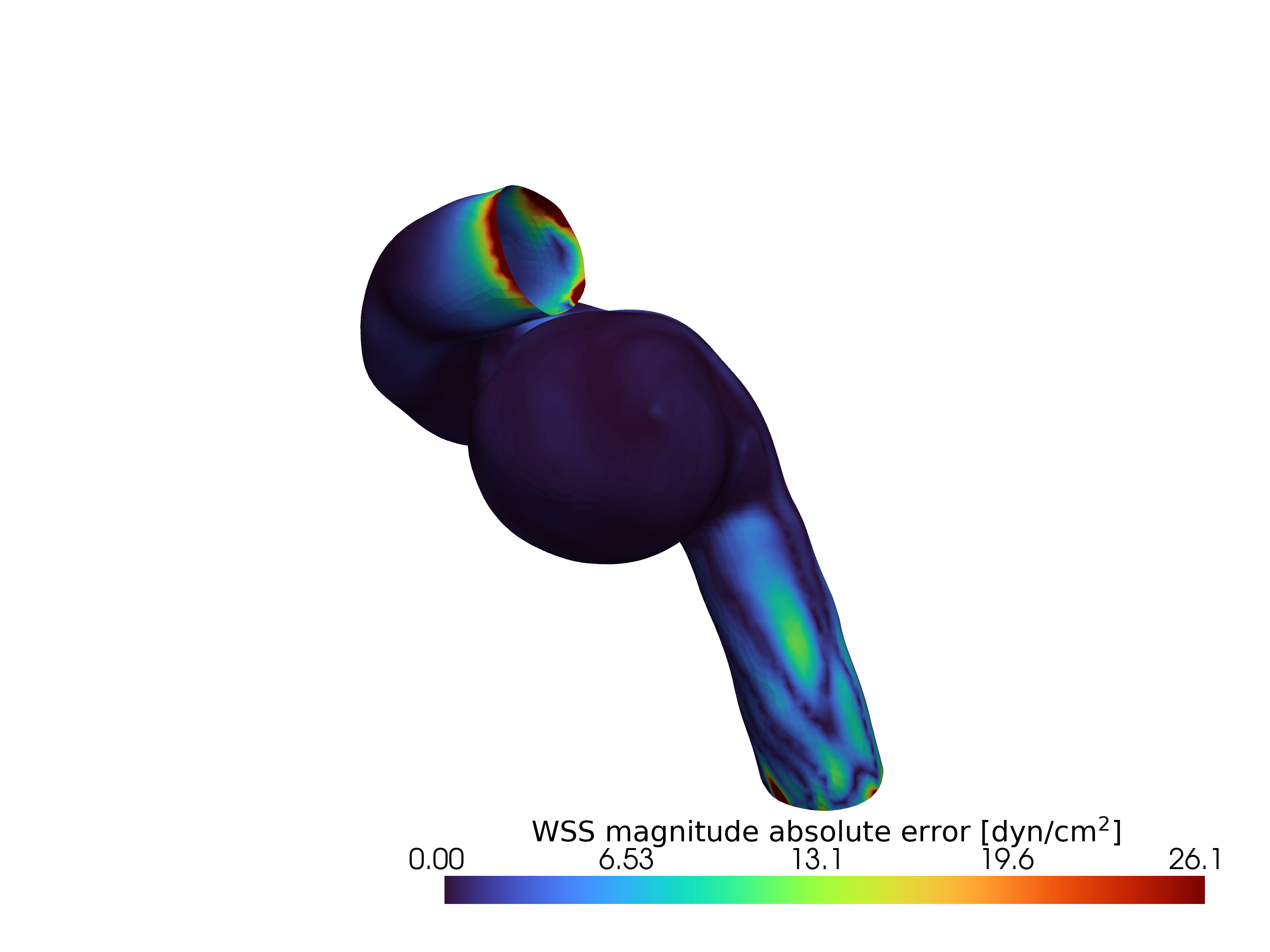}
  \caption{Error.}
\end{subfigure}
\hfill
\begin{subfigure}[t]{.32\textwidth}
  \centering
  \includegraphics[width=0.99\linewidth]{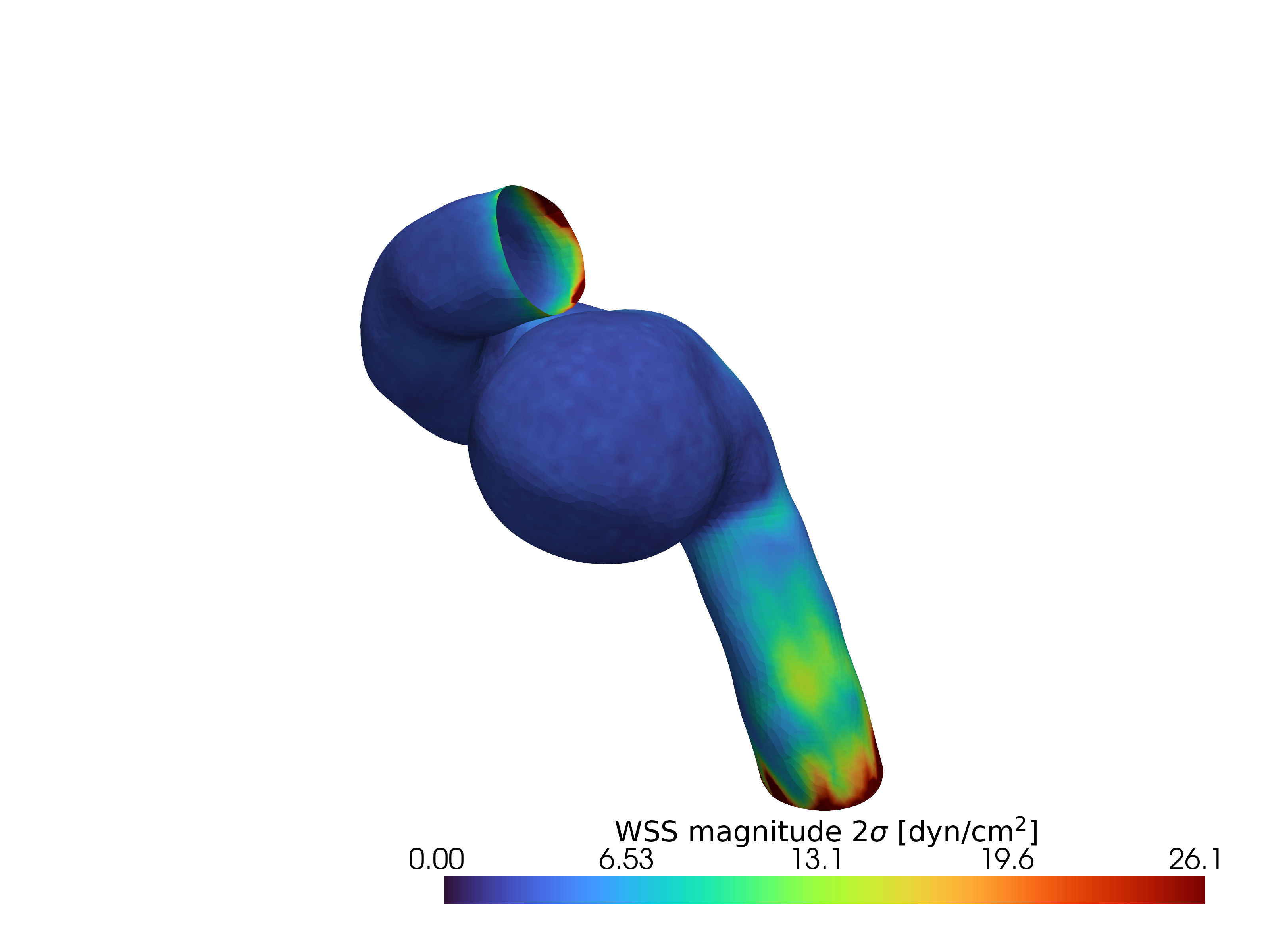}
  \caption{Predicted uncertainty}
\end{subfigure}
\hfill
\begin{subfigure}[t]{.32\textwidth}
  \centering
  \includegraphics[width=0.99\linewidth]{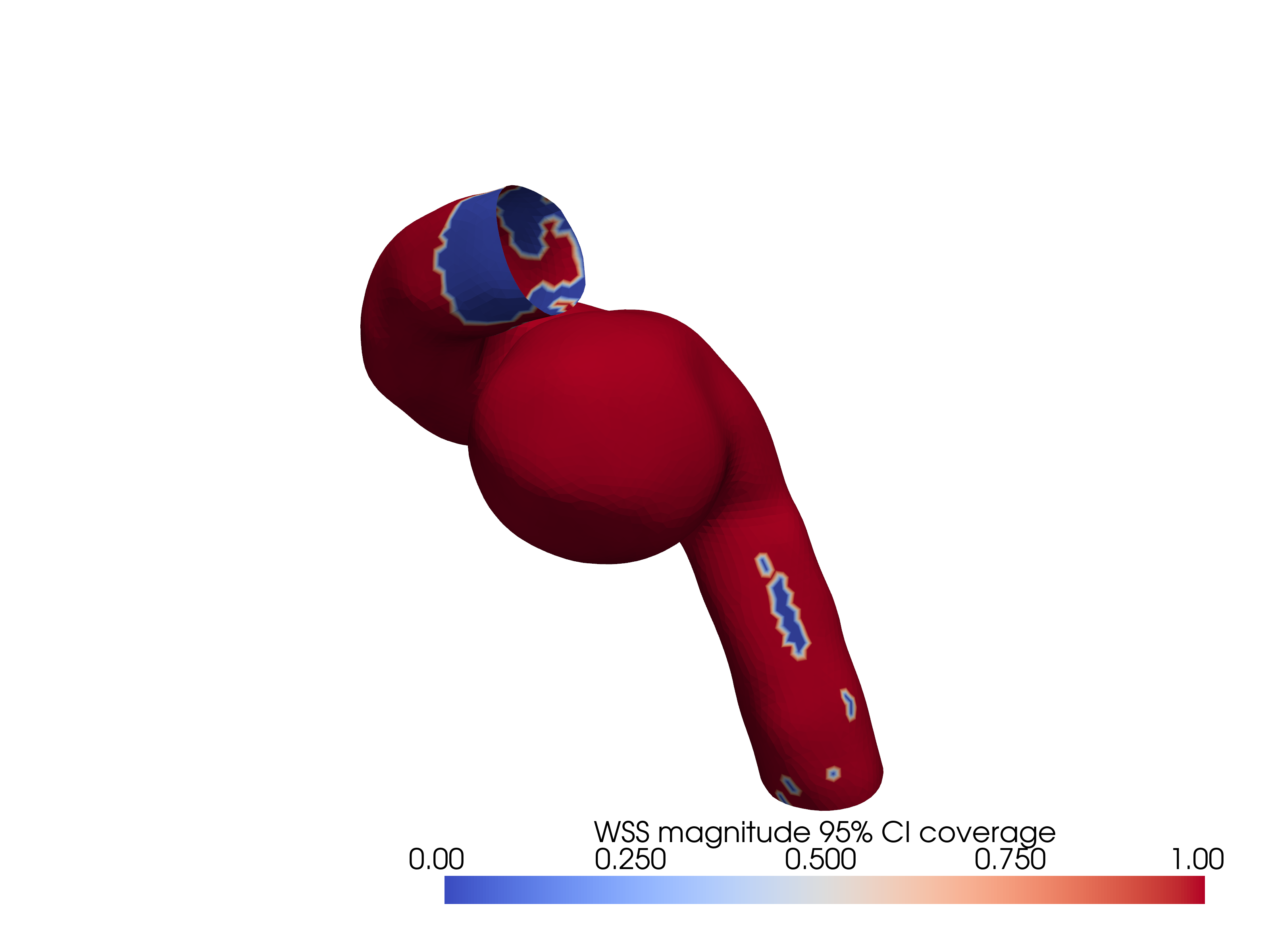}
  \caption{95\% Coverage.}
\end{subfigure}
\caption{Reconstruction errors, posterior uncertainties, and coverage fields for the z-velocity, pressure, and WSS magnitude in the cerebral aneurysm geometry, for $\Delta x = 0.075$ cm and $\mathrm{SNR}=5.0$.}
\label{fig:ca_uq}
\end{figure}

\FloatBarrier

\subsection{Comparison to baselines}

\begin{figure}[h!]
\center
\begin{subfigure}[t]{.49\textwidth}
  \centering
  \includegraphics[width=0.8\linewidth]{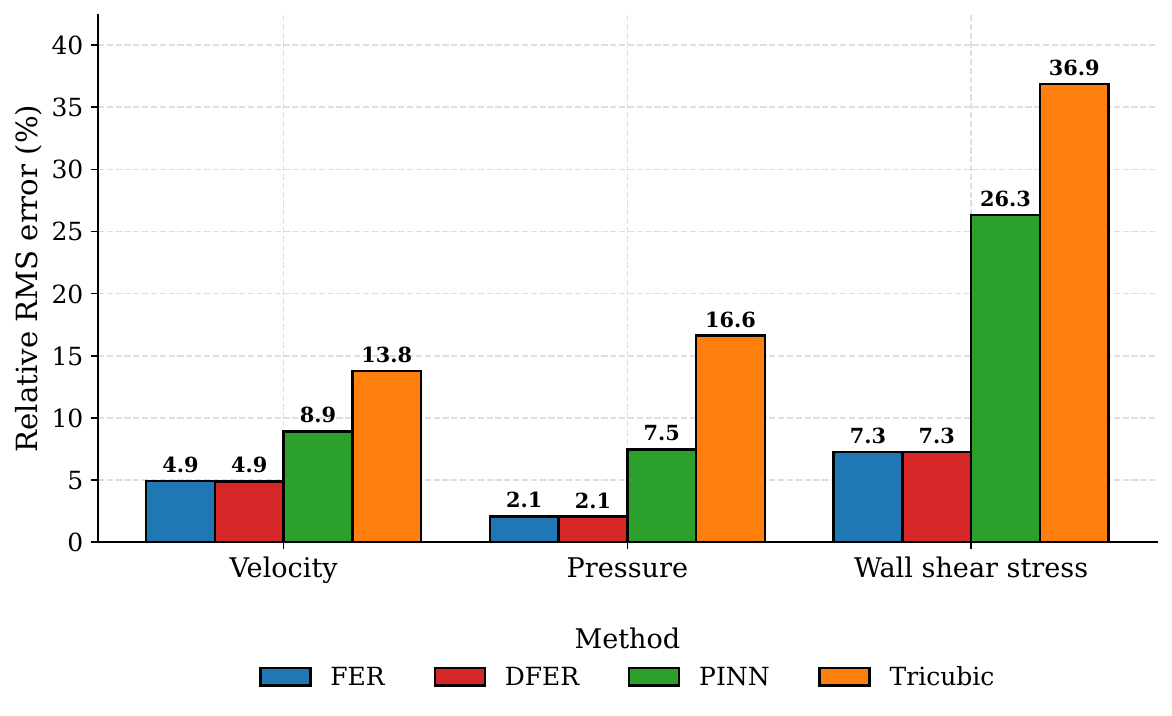}
\caption{Global errors for aortic aneurysm.}
\end{subfigure}
\hfill
\begin{subfigure}[t]{.49\textwidth}
  \centering
  \includegraphics[width=0.8\linewidth]{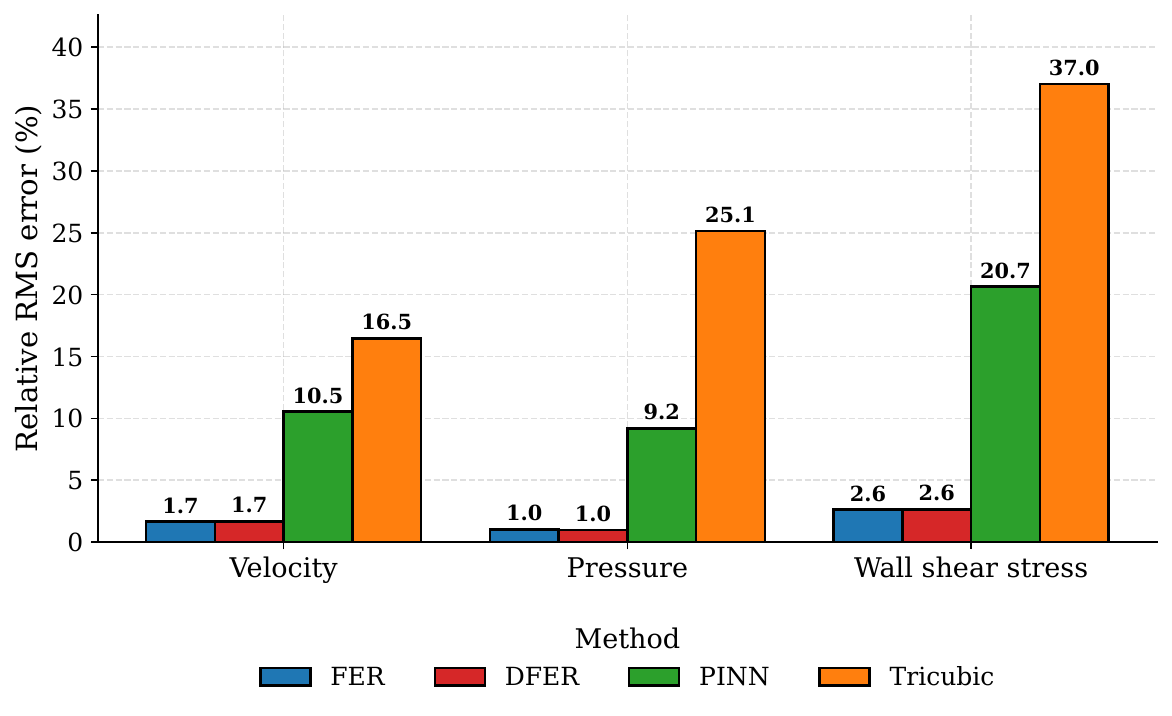}
\caption{ROI errors for aortic aneurysm.}
\end{subfigure}
\begin{subfigure}[t]{.49\textwidth}
  \centering
  \includegraphics[width=0.8\linewidth]{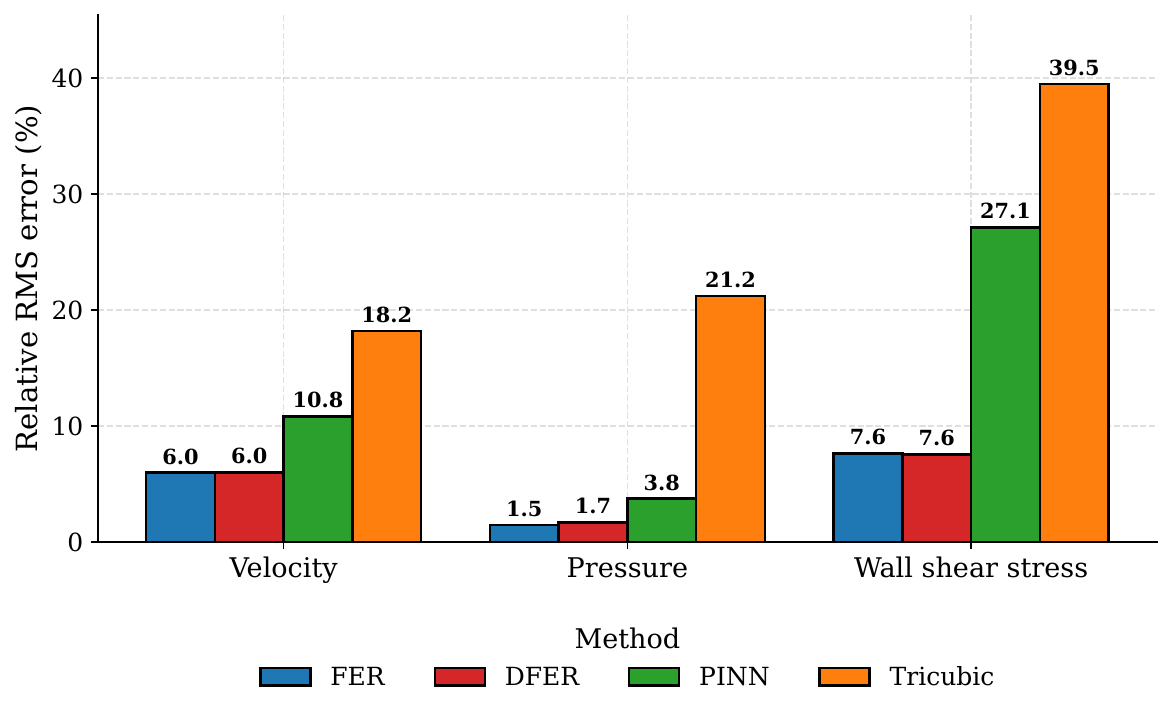}
\caption{Global errors for aortic coarctation.}
\end{subfigure}
\hfill
\begin{subfigure}[t]{.49\textwidth}
  \centering
  \includegraphics[width=0.8\linewidth]{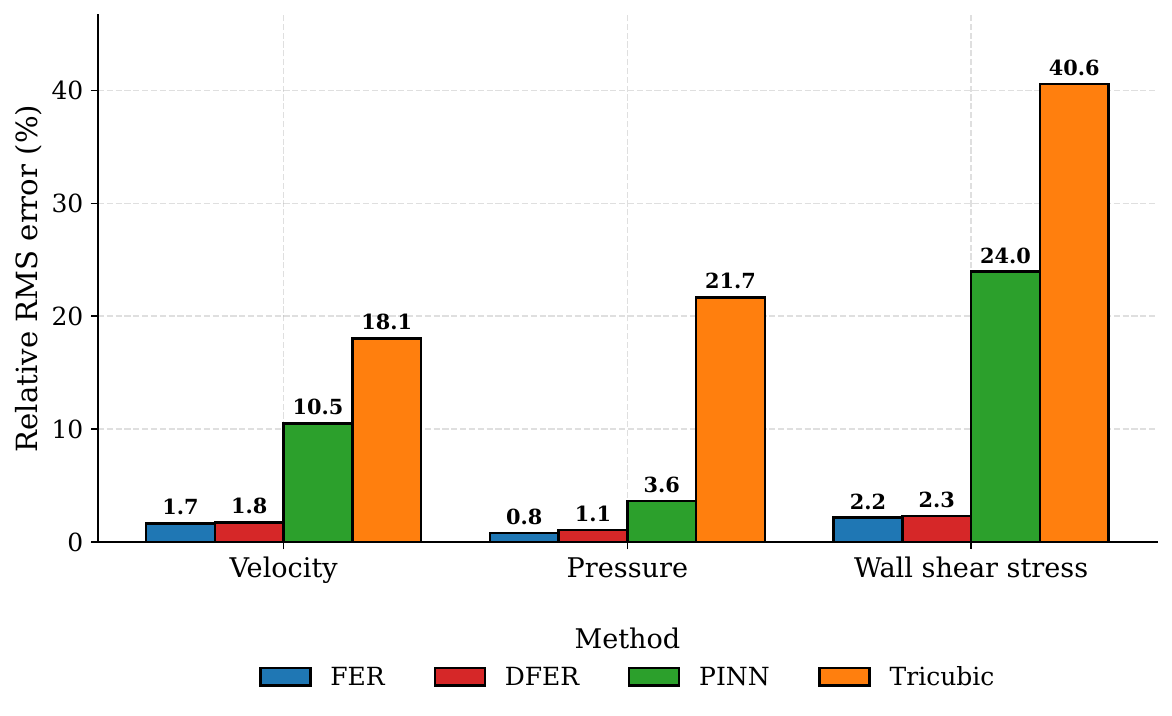}
\caption{ROI errors for aortic coarctation.}
\end{subfigure}
\begin{subfigure}[t]{.49\textwidth}
  \centering
  \includegraphics[width=0.8\linewidth]{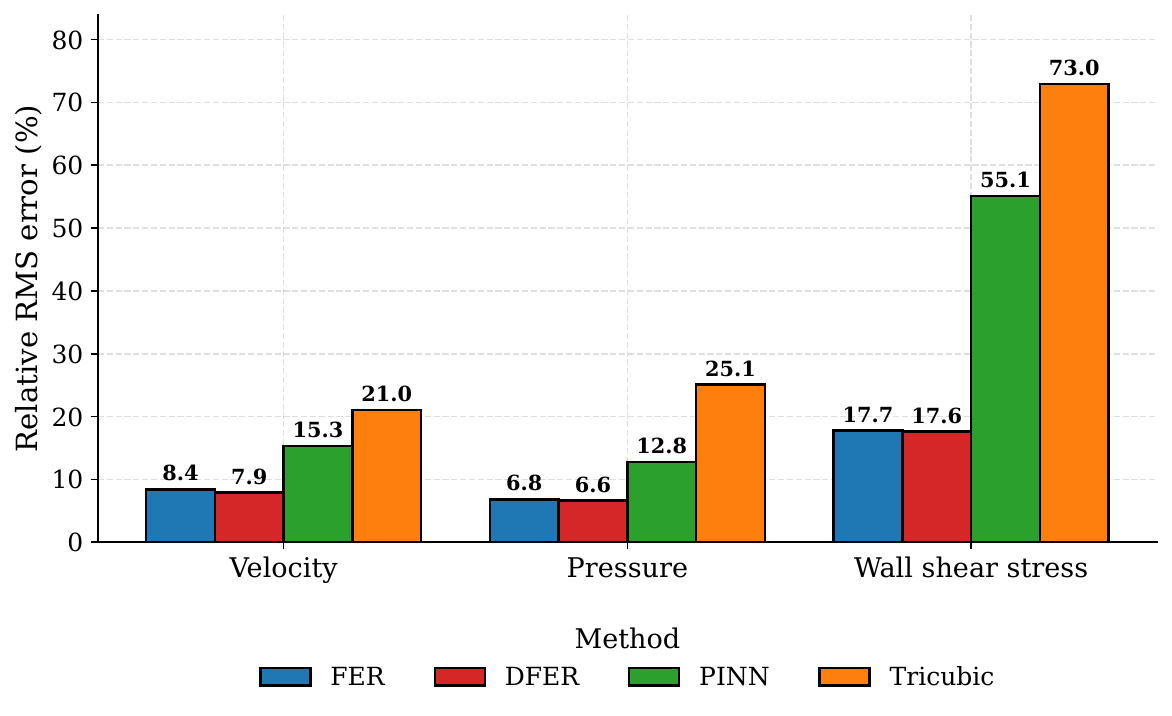}
\caption{Global errors for cerebral aneurysm.}
\end{subfigure}
\hfill
\begin{subfigure}[t]{.49\textwidth}
  \centering
  \includegraphics[width=0.8\linewidth]{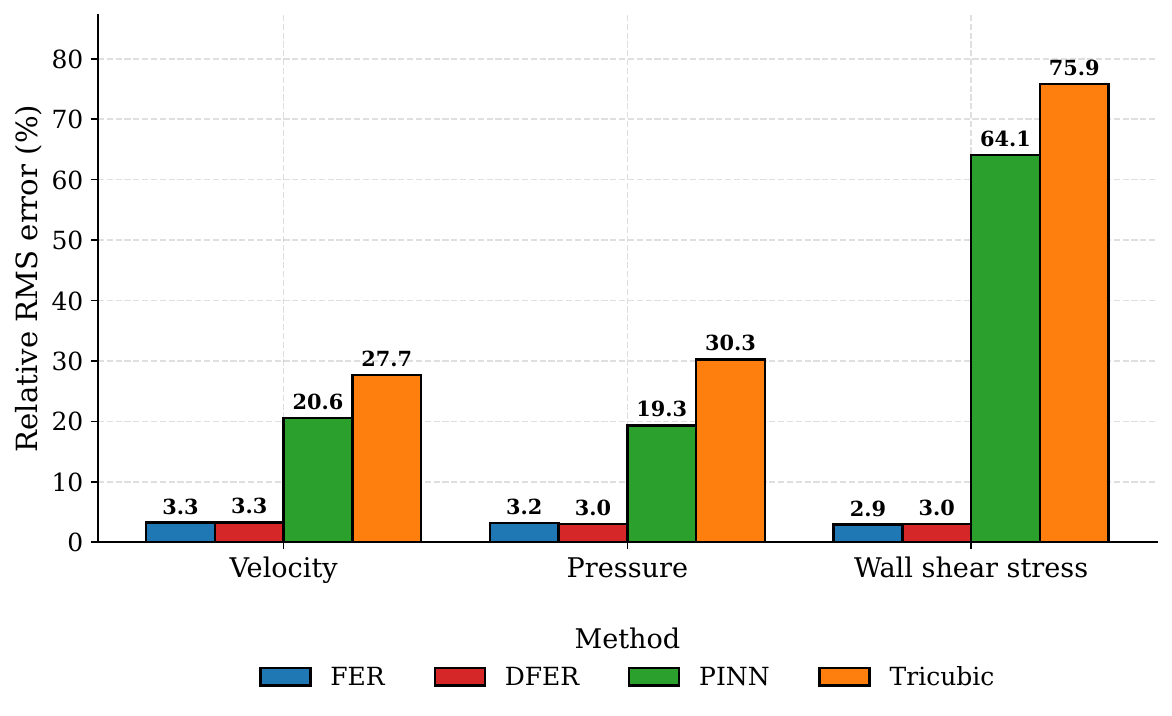}
\caption{ROI errors for cerebral aneurysm.}
\end{subfigure}
\caption{Relative RMS reconstruction errors (\%) for velocity, pressure, and WSS obtained with each method, averaged across all SNR and resolution settings. Errors are reported over the entire domain (left) and over the region of interest (right).}
\label{fig:baseline_comparison}
\end{figure}

Figure (\ref{fig:baseline_comparison}) shows the relative RMS reconstruction errors (\%) for velocity, pressure, and WSS, computed over the entire domain as well as the the region of interest, and averaged across all SNR and resolution settings. We refer to our proposed approach as FER and its weakly divergence-free variant as DFER (divergence-free finite element regression). Across all geometries and quantities of interest, FER and DFER consistently achieve the lowest reconstruction errors, while tricubic interpolation yields the largest errors and PINN performs in between these methods. This ranking remains unchanged when errors are restricted to the region of interest. The reconstruction accuracy of FER and DFER is nearly identical for all quantities and geometries. While PINN produces velocity and pressure reconstructions that are only moderately less accurate than FER and DFER, its WSS errors are substantially larger across all geometries. Reconstruction errors for all methods for all SNRs and resolutions are available in appendix \ref{sec:appendix}

We did not perform a systematic timing study, but our numerical experiments indicate that tricubic interpolation is the fastest method, requiring approximately one minute per reconstruction. FER and DFER require approximately 15 minutes, whereas PINN requires roughly one hour. These observations are consistent with the findings of \cite{Karnakov2023}, who reported that optimization of discretized objective functions is faster than PINNs on flow reconstruction problems.

\begin{figure}[h!]
\centering
\begin{subfigure}[t]{.31\textwidth}
  \centering
  \includegraphics[width=0.99\linewidth]{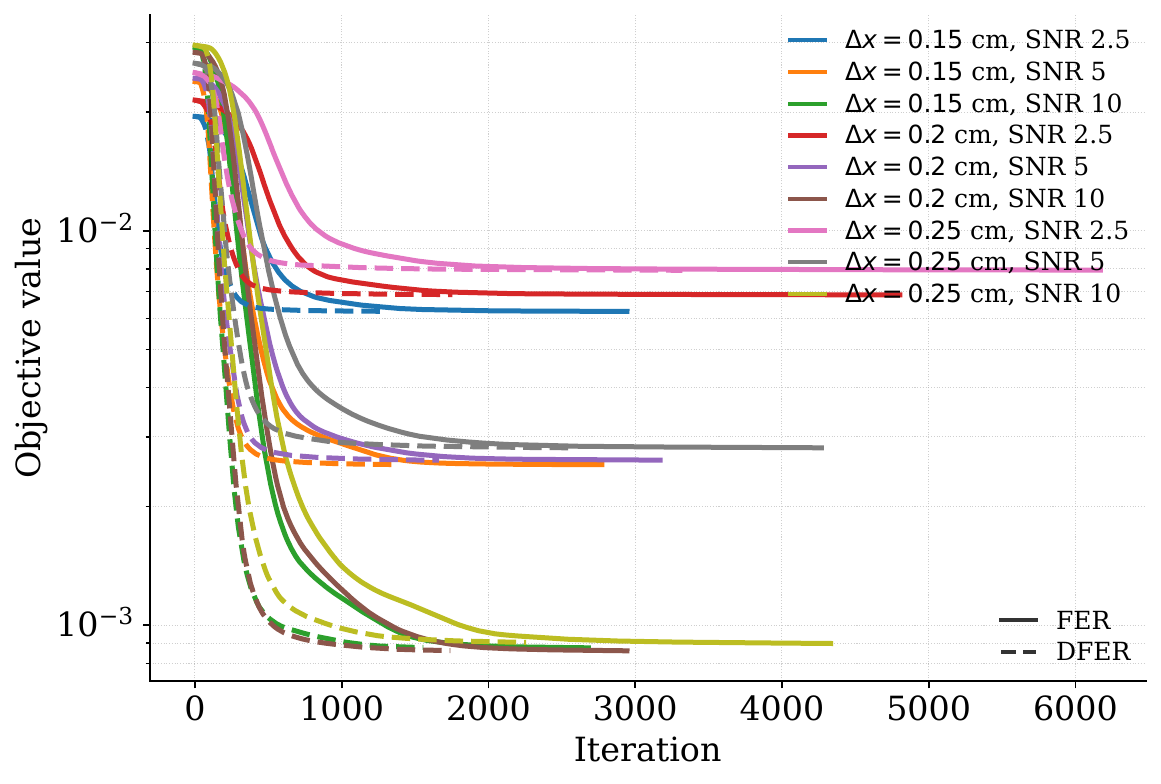}
  \caption{}
\end{subfigure}
\hfill
\begin{subfigure}[t]{.31\textwidth}
  \centering
  \includegraphics[width=0.99\linewidth]{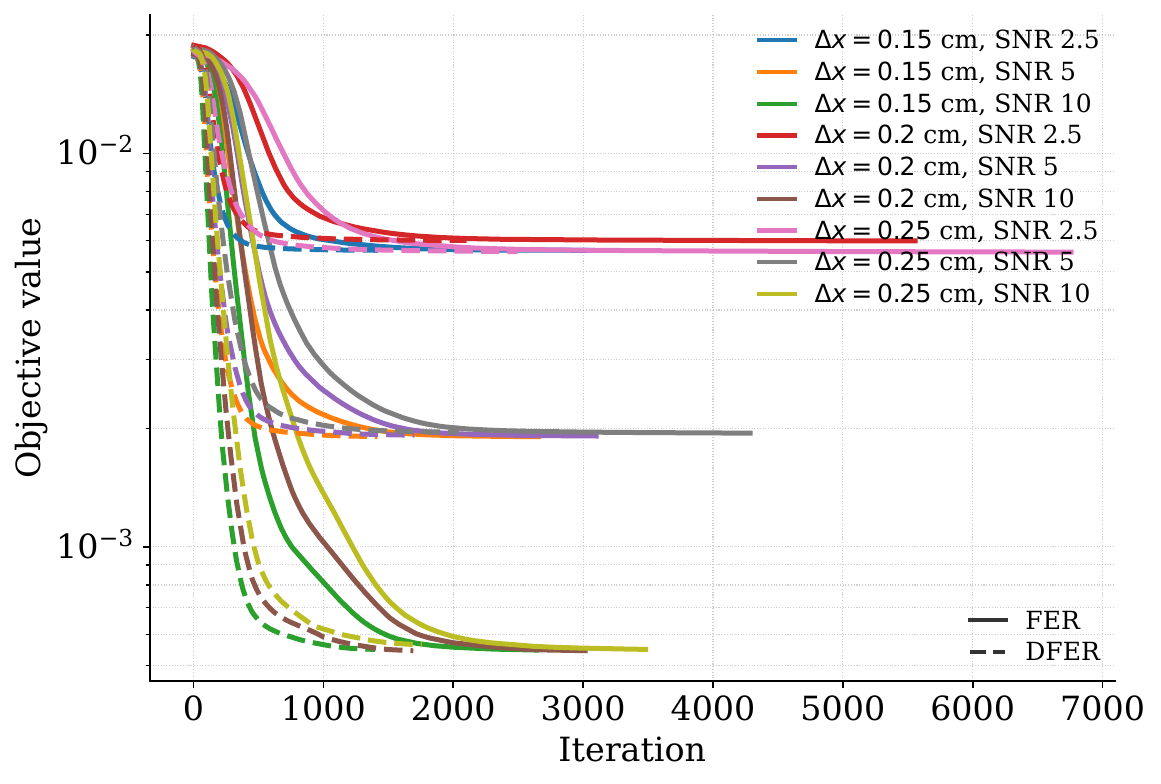}
  \caption{}
\end{subfigure}
\hfill
\begin{subfigure}[t]{.31\textwidth}
  \centering
  \includegraphics[width=0.99\linewidth]{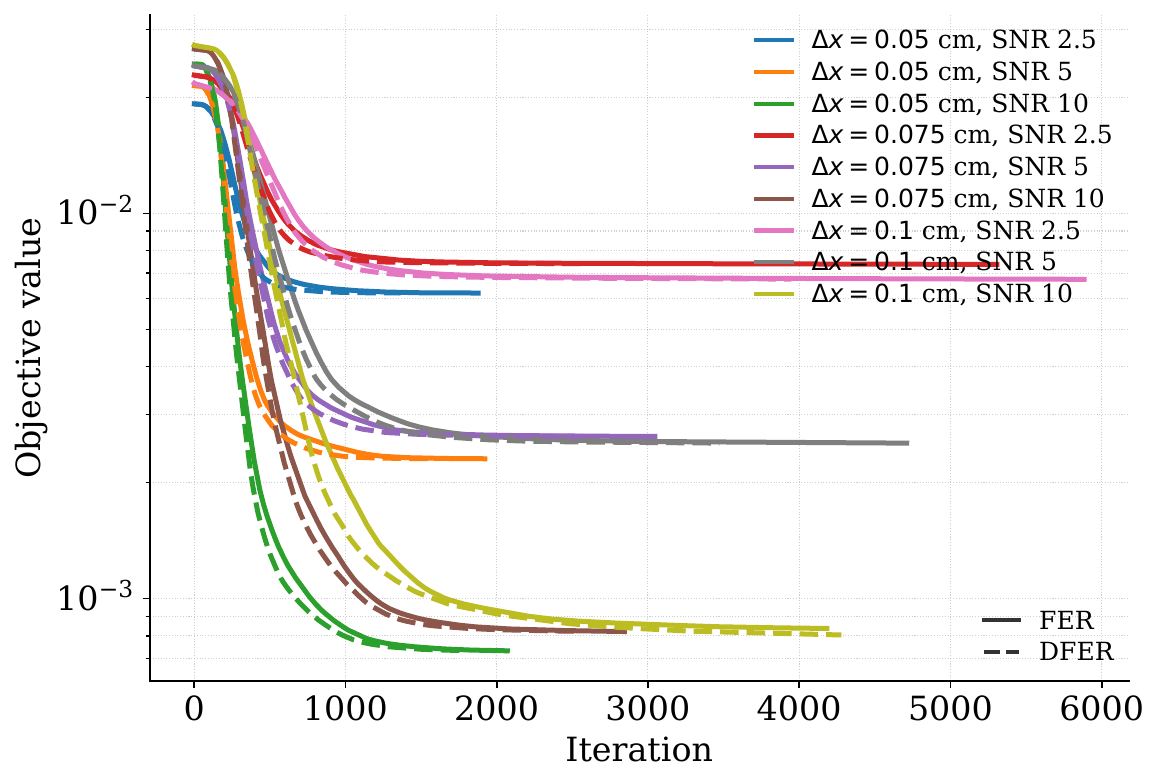}
  \caption{}
\end{subfigure}
\caption{Objective function value history for different geometries. Aortic aneurysm (left), aortic coarctation (middle), cerebral aneurysm (right).}
\label{fig:objective_iterations}
\end{figure}

We additionally observed that DFER converges in fewer iterations than FER for all geometries. Averaged over all SNR and resolution settings, the mean number of iterations required by FER and DFER was 3787 and 1949 for the aortic aneurysm, 3841 and 1780 for the aortic coarctation, and 3538 and 2678 for the cerebral aneurysm, respectively. Figure (\ref{fig:objective_iterations}) shows the objective value histories for FER and DFER across all SNR and resolution settings. Consistent with the iteration counts, DFER exhibits faster convergence while achieving reconstruction accuracies comparable to those of FER.

\FloatBarrier

\section{Discussion and Conclusion}

We have presented a Bayesian finite-element regression framework for flow reconstruction from noisy and under-resolved velocity images. The proposed method does not require external training data, forward and adjoint solves, or automatic differentiation. The method accurately reconstructed velocity, pressure, and wall shear stress fields while recovering clinically relevant flow features such as recirculation zones, high-speed jets, and localized pressure gradients, across a range of spatial resolutions, noise levels, and vascular geometries. Furthermore, the Laplace approximation provided conservative uncertainty estimates that were consistent with the reported reconstruction errors.

Our reconstruction and uncertainty estimates exhibited several trends across all geometries. Reported errors and estimated uncertainties for all quantities were largest near the inlet boundaries. This is expected because, in advection-dominated flows, information propagates predominantly downstream. As a result, distinct inlet profiles relax to a similar downstream flow profile, making the reconstruction of inlet conditions ill-posed. A possible remedy is to regularize the inlet velocity field by representing the velocity on each inlet cap using a truncated Laplace--Beltrami eigenfunction expansion with Dirichlet boundary conditions prescribed along the rim. Limiting the expansion to a small number of modes reduces the dimensionality of the inverse problem and suppresses nonphysical oscillatory inlet profiles. 

While the uncertainty estimates successfully identified regions of high reconstruction error, the coverage rates were higher than their nominal values, implying that the Laplace approximation was conservative. Conservative uncertainty estimates are preferable to over-confident ones, yet this observation suggests that the Laplace approximation does not fully capture the geometry of the true posterior distribution.

Although FER and DFER produced similar reconstruction accuracies, DFER converged in fewer number of iterations. This makes sense bacause enforcing weak incompressibility during optimization reduces the effective search space and improves the conditioning of the inverse problem. Consequently, DFER can accelerate the reconstruction procedure without compromising accuracy. In comparison with the PINN baseline, FER and DFER produced only moderately lower velocity and pressure errors but substantially lower wall shear stress errors. This is because PINN enforces no-slip boundary condition only via a penalty term, making velocity and velocity gradients less accurate close to wall. In contrast, FER enforces no-slip conditions exactly by imposing zero velocity at the nodes on the vessel walls. Moreover, PINNs are globally supported function approximators, meaning that changes in the weights of the network affect the prediction throughout the entire domain. FER, on the other hand, is a local approach where each degree of freedom influences only its neighbors, allowing it to describe localized flow features and near-wall gradients more effectively.

All experiments were performed using synthetic data generated from CFD simulations and therefore do not fully capture the measurement artifacts and uncertainties present in clinical PC-MRI acquisitions. In addition, the current formulation is restricted to steady flow and assumes that the vascular geometry is known a priori.

Future work will focus on extending the framework to unsteady and periodic flows by representing the solution with a Fourier series in time and jointly inferring the corresponding Fourier coefficients and finite element degrees of freedom. Another direction is the simultaneous reconstruction of vascular segmentation, velocity, and pressure fields by representing the vessel boundary through a signed distance function and enforcing out-of-vessel regions using Brinkman penalization. The development of a likelihood model directly based on the PC-MRI acquisition process is another direction which may provide a better description of measurement noise and imaging artifacts. Incorporating temporal variability will substantially increase the dimensionality of the inverse problem and may require strategies to improve computational tractability. Parallel and GPU implementations of this framework are promising directions of future work for accelerating reconstructions on larger problems. Finally, uncertainty quantification techniques that capture the posterior geometry more accurately than Laplace approximation are needed to give better calibrated reconstruction uncertainties.

\bibliographystyle{ieeetr}
\bibliography{biblio} 

\newpage

\section{Appendix}
\label{sec:appendix}

\subsection{Performance tables for the proposed approach and the baseline methods.}

The tables below summarize the relative (\%) reconstruction errors in velocity, pressure, and wall shear stress (WSS). For each (SNR, resolution) pair the top cell contains the global error over the entire region while the bottom cell contains the error over the region of interest.

\subsubsection{FER}

\vspace{0.5cm}

\noindent \textbf{Aortic aneurysm:}

\begin{table}[htbp]
\small
\centering
\renewcommand{\arraystretch}{1.3}
\begin{tabular}{|c||c|c|c||c|c|c||c|c|c|}
\hline
\multirow{2}{*}{\textbf{SNR}}
& \multicolumn{3}{c||}{\textbf{Velocity}}
& \multicolumn{3}{c||}{\textbf{Pressure}}
& \multicolumn{3}{c|}{\textbf{WSS}} \\ \cline{2-10}
& \textbf{1.5 mm} & \textbf{2.0 mm} & \textbf{2.5 mm}
& \textbf{1.5 mm} & \textbf{2.0 mm} & \textbf{2.5 mm}
& \textbf{1.5 mm} & \textbf{2.0 mm} & \textbf{2.5 mm} \\
\hhline{|=||=|=|=||=|=|=||=|=|=|}
\multirow{2}{*}{\textbf{2.5}}
& 5.06 & 8.75 & 10.90 & 1.64 & 3.06 & 5.19 & 7.36 & 11.51 & 14.83 \\ \cline{2-10}
& 2.14 & 2.75 & 4.03 & 1.17 & 1.25 & 4.02 & 3.52 & 4.64 & 6.36 \\
\hhline{|=||=|=|=||=|=|=||=|=|=|}
\multirow{2}{*}{\textbf{5.0}}
& 2.49 & 4.09 & 5.32 & 1.07 & 1.37 & 1.82 & 4.11 & 6.11 & 6.82 \\ \cline{2-10}
& 1.02 & 1.32 & 1.55 & 0.40 & 0.45 & 0.59 & 1.67 & 2.07 & 2.38 \\
\hhline{|=||=|=|=||=|=|=||=|=|=|}
\multirow{2}{*}{\textbf{10.0}}
& 1.59 & 2.22 & 3.99 & 1.13 & 0.83 & 2.60 & 3.15 & 3.04 & 8.40 \\ \cline{2-10}
& 0.68 & 0.68 & 0.73 & 0.60 & 0.27 & 0.62 & 1.06 & 1.11 & 0.94 \\
\hline
\end{tabular}
\end{table}

\noindent \textbf{Aortic coarctation:}

\begin{table}[htbp]
\small
\centering
\renewcommand{\arraystretch}{1.3}
\begin{tabular}{|c||c|c|c||c|c|c||c|c|c|}
\hline
\multirow{2}{*}{\textbf{SNR}}
& \multicolumn{3}{c||}{\textbf{Velocity}}
& \multicolumn{3}{c||}{\textbf{Pressure}}
& \multicolumn{3}{c|}{\textbf{WSS}} \\ \cline{2-10}
& \textbf{1.5 mm} & \textbf{2.0 mm} & \textbf{2.5 mm}
& \textbf{1.5 mm} & \textbf{2.0 mm} & \textbf{2.5 mm}
& \textbf{1.5 mm} & \textbf{2.0 mm} & \textbf{2.5 mm} \\
\hhline{|=||=|=|=||=|=|=||=|=|=|}
\multirow{2}{*}{\textbf{2.5}}
& 5.94 & 10.67 & 14.04 & 0.87 & 2.38 & 3.20 & 7.22 & 10.72 & 17.41 \\ \cline{2-10}
& 1.52 & 2.36 & 4.39 & 0.25 & 0.45 & 1.88 & 2.10 & 2.18 & 6.29 \\
\hhline{|=||=|=|=||=|=|=||=|=|=|}
\multirow{2}{*}{\textbf{5.0}}
& 3.44 & 4.39 & 6.59 & 1.50 & 0.81 & 1.57 & 5.70 & 5.61 & 7.15 \\ \cline{2-10}
& 1.04 & 1.01 & 2.11 & 1.25 & 0.16 & 1.52 & 1.43 & 0.98 & 3.00 \\
\hhline{|=||=|=|=||=|=|=||=|=|=|}
\multirow{2}{*}{\textbf{10.0}}
& 2.36 & 2.62 & 3.88 & 1.48 & 0.85 & 0.87 & 5.36 & 4.93 & 4.71 \\ \cline{2-10}
& 0.85 & 0.69 & 1.09 & 1.23 & 0.56 & 0.22 & 1.15 & 0.71 & 1.75 \\
\hline
\end{tabular}
\end{table}

\noindent \textbf{Cerebral aneurysm:}

\begin{table}[htbp]
\small
\centering
\renewcommand{\arraystretch}{1.3}
\begin{tabular}{|c||c|c|c||c|c|c||c|c|c|}
\hline
\multirow{2}{*}{\textbf{SNR}}
& \multicolumn{3}{c||}{\textbf{Velocity}}
& \multicolumn{3}{c||}{\textbf{Pressure}}
& \multicolumn{3}{c|}{\textbf{WSS}} \\ \cline{2-10}
& \textbf{0.5 mm} & \textbf{0.75 mm} & \textbf{1.0 mm}
& \textbf{0.5 mm} & \textbf{0.75 mm} & \textbf{1.0 mm}
& \textbf{0.5 mm} & \textbf{0.75 mm} & \textbf{1.0 mm} \\
\hhline{|=||=|=|=||=|=|=||=|=|=|}
\multirow{2}{*}{\textbf{2.5}}
& 4.72 & 11.97 & 17.19 & 3.58 & 8.55 & 16.09 & 8.66 & 23.69 & 39.34 \\ \cline{2-10}
& 1.65 & 4.15 & 7.34 & 1.85 & 5.26 & 6.69 & 1.63 & 4.28 & 5.46 \\
\hhline{|=||=|=|=||=|=|=||=|=|=|}
\multirow{2}{*}{\textbf{5.0}}
& 2.96 & 5.02 & 8.60 & 2.96 & 3.96 & 5.76 & 6.98 & 10.05 & 17.77 \\ \cline{2-10}
& 0.85 & 2.20 & 3.32 & 0.48 & 2.02 & 2.96 & 0.75 & 2.54 & 2.80 \\
\hline
\end{tabular}
\end{table}


\newpage
\subsubsection{DFER}

\vspace{0.5cm}

\noindent \textbf{Aortic aneurysm:}

\begin{table}[htbp]
\small
\centering
\renewcommand{\arraystretch}{1.3}
\begin{tabular}{|c||c|c|c||c|c|c||c|c|c|}
\hline
\multirow{2}{*}{\textbf{SNR}}
& \multicolumn{3}{c||}{\textbf{Velocity}}
& \multicolumn{3}{c||}{\textbf{Pressure}}
& \multicolumn{3}{c|}{\textbf{WSS}} \\ \cline{2-10}
& \textbf{1.5 mm} & \textbf{2.0 mm} & \textbf{2.5 mm}
& \textbf{1.5 mm} & \textbf{2.0 mm} & \textbf{2.5 mm}
& \textbf{1.5 mm} & \textbf{2.0 mm} & \textbf{2.5 mm} \\
\hhline{|=||=|=|=||=|=|=||=|=|=|}
\multirow{2}{*}{\textbf{2.5}}
& 5.07 & 8.13 & 11.19 & 1.57 & 2.61 & 5.68 & 7.36 & 11.03 & 16.01 \\ \cline{2-10}
& 2.09 & 3.04 & 4.05 & 0.93 & 1.13 & 4.02 & 3.34 & 4.84 & 6.44 \\
\hhline{|=||=|=|=||=|=|=||=|=|=|}
\multirow{2}{*}{\textbf{5.0}}
& 2.47 & 4.01 & 5.39 & 1.03 & 1.12 & 1.98 & 4.00 & 5.17 & 6.96 \\ \cline{2-10}
& 1.00 & 1.37 & 1.53 & 0.42 & 0.43 & 0.91 & 1.46 & 2.29 & 2.40 \\
\hhline{|=||=|=|=||=|=|=||=|=|=|}
\multirow{2}{*}{\textbf{10.0}}
& 1.63 & 2.24 & 3.86 & 1.06 & 0.86 & 2.66 & 3.22 & 3.15 & 8.53 \\ \cline{2-10}
& 0.61 & 0.72 & 0.73 & 0.48 & 0.28 & 0.27 & 0.90 & 1.27 & 0.88 \\
\hline
\end{tabular}
\end{table}

\noindent \textbf{Aortic coarctation:}

\begin{table}[htbp]
\small
\centering
\renewcommand{\arraystretch}{1.3}
\begin{tabular}{|c||c|c|c||c|c|c||c|c|c|}
\hline
\multirow{2}{*}{\textbf{SNR}}
& \multicolumn{3}{c||}{\textbf{Velocity}}
& \multicolumn{3}{c||}{\textbf{Pressure}}
& \multicolumn{3}{c|}{\textbf{WSS}} \\ \cline{2-10}
& \textbf{1.5 mm} & \textbf{2.0 mm} & \textbf{2.5 mm}
& \textbf{1.5 mm} & \textbf{2.0 mm} & \textbf{2.5 mm}
& \textbf{1.5 mm} & \textbf{2.0 mm} & \textbf{2.5 mm} \\
\hhline{|=||=|=|=||=|=|=||=|=|=|}
\multirow{2}{*}{\textbf{2.5}}
& 6.05 & 10.01 & 13.49 & 1.15 & 2.20 & 3.32 & 7.85 & 10.50 & 16.73 \\ \cline{2-10}
& 1.58 & 2.32 & 4.54 & 0.67 & 0.84 & 2.44 & 2.30 & 2.35 & 6.26 \\
\hhline{|=||=|=|=||=|=|=||=|=|=|}
\multirow{2}{*}{\textbf{5.0}}
& 3.54 & 5.43 & 6.55 & 1.59 & 1.83 & 1.08 & 5.92 & 5.94 & 7.14 \\ \cline{2-10}
& 1.09 & 1.10 & 2.02 & 1.31 & 0.18 & 0.74 & 1.56 & 0.97 & 3.06 \\
\hhline{|=||=|=|=||=|=|=||=|=|=|}
\multirow{2}{*}{\textbf{10.0}}
& 2.39 & 2.68 & 4.07 & 1.73 & 1.10 & 1.41 & 4.01 & 5.17 & 4.84 \\ \cline{2-10}
& 1.02 & 0.79 & 1.36 & 1.53 & 0.89 & 0.99 & 1.45 & 0.90 & 2.18 \\
\hline
\end{tabular}
\end{table}

\noindent \textbf{Cerebral aneurysm:}

\begin{table}[htbp]
\small
\centering
\renewcommand{\arraystretch}{1.3}
\begin{tabular}{|c||c|c|c||c|c|c||c|c|c|}
\hline
\multirow{2}{*}{\textbf{SNR}}
& \multicolumn{3}{c||}{\textbf{Velocity}}
& \multicolumn{3}{c||}{\textbf{Pressure}}
& \multicolumn{3}{c|}{\textbf{WSS}} \\ \cline{2-10}
& \textbf{0.5 mm} & \textbf{0.75 mm} & \textbf{1.0 mm}
& \textbf{0.5 mm} & \textbf{0.75 mm} & \textbf{1.0 mm}
& \textbf{0.5 mm} & \textbf{0.75 mm} & \textbf{1.0 mm} \\
\hhline{|=||=|=|=||=|=|=||=|=|=|}
\multirow{2}{*}{\textbf{2.5}}
& 4.76 & 9.69 & 16.24 & 3.75 & 6.92 & 15.00 & 8.82 & 17.43 & 40.84 \\ \cline{2-10}
& 1.70 & 4.40 & 7.19 & 1.91 & 4.93 & 6.16 & 1.69 & 5.00 & 5.32 \\
\hhline{|=||=|=|=||=|=|=||=|=|=|}
\multirow{2}{*}{\textbf{5.0}}
& 2.99 & 5.02 & 8.70 & 3.17 & 4.11 & 6.76 & 7.15 & 10.34 & 21.01 \\ \cline{2-10}
& 0.84 & 2.17 & 3.44 & 0.44 & 1.95 & 2.81 & 0.74 & 2.53 & 2.72 \\
\hline
\end{tabular}
\end{table}


\newpage
\subsubsection{PINN}

\vspace{0.5cm}

\noindent \textbf{Aortic aneurysm:}

\begin{table}[htbp]
\small
\centering
\renewcommand{\arraystretch}{1.3}
\begin{tabular}{|c||c|c|c||c|c|c||c|c|c|}
\hline
\multirow{2}{*}{\textbf{SNR}}
& \multicolumn{3}{c||}{\textbf{Velocity}}
& \multicolumn{3}{c||}{\textbf{Pressure}}
& \multicolumn{3}{c|}{\textbf{WSS}} \\ \cline{2-10}
& \textbf{1.5 mm} & \textbf{2.0 mm} & \textbf{2.5 mm}
& \textbf{1.5 mm} & \textbf{2.0 mm} & \textbf{2.5 mm}
& \textbf{1.5 mm} & \textbf{2.0 mm} & \textbf{2.5 mm} \\
\hhline{|=||=|=|=||=|=|=||=|=|=|}
\multirow{2}{*}{\textbf{2.5}}
& 10.06 & 10.71 & 11.99 & 9.78 & 8.78 & 9.96 & 29.68 & 29.84 & 35.06 \\ \cline{2-10}
& 11.61 & 12.40 & 14.05 & 13.55 & 9.43 & 14.86 & 23.61 & 22.42 & 26.56 \\
\hhline{|=||=|=|=||=|=|=||=|=|=|}
\multirow{2}{*}{\textbf{5.0}}
& 7.92 & 8.61 & 8.32 & 5.73 & 6.59 & 6.32 & 24.50 & 24.63 & 24.91 \\ \cline{2-10}
& 9.34 & 10.44 & 10.08 & 7.26 & 7.01 & 8.75 & 18.52 & 19.56 & 20.09 \\
\hhline{|=||=|=|=||=|=|=||=|=|=|}
\multirow{2}{*}{\textbf{10.0}}
& 7.36 & 7.22 & 8.04 & 6.76 & 4.66 & 8.61 & 22.66 & 21.72 & 24.07 \\ \cline{2-10}
& 8.64 & 9.04 & 9.31 & 6.50 & 5.78 & 9.53 & 18.30 & 18.28 & 18.58 \\
\hline
\end{tabular}
\end{table}

\noindent \textbf{Aortic coarctation:}

\begin{table}[htbp]
\small
\centering
\renewcommand{\arraystretch}{1.3}
\begin{tabular}{|c||c|c|c||c|c|c||c|c|c|}
\hline
\multirow{2}{*}{\textbf{SNR}}
& \multicolumn{3}{c||}{\textbf{Velocity}}
& \multicolumn{3}{c||}{\textbf{Pressure}}
& \multicolumn{3}{c|}{\textbf{WSS}} \\ \cline{2-10}
& \textbf{1.5 mm} & \textbf{2.0 mm} & \textbf{2.5 mm}
& \textbf{1.5 mm} & \textbf{2.0 mm} & \textbf{2.5 mm}
& \textbf{1.5 mm} & \textbf{2.0 mm} & \textbf{2.5 mm} \\
\hhline{|=||=|=|=||=|=|=||=|=|=|}
\multirow{2}{*}{\textbf{2.5}}
& 11.03 & 12.32 & 15.20 & 4.00 & 4.41 & 4.51 & 27.58 & 31.32 & 35.44 \\ \cline{2-10}
& 10.44 & 11.50 & 14.19 & 3.39 & 4.54 & 5.21 & 23.93 & 25.77 & 32.39 \\
\hhline{|=||=|=|=||=|=|=||=|=|=|}
\multirow{2}{*}{\textbf{5.0}}
& 9.76 & 10.32 & 10.80 & 3.55 & 4.01 & 3.36 & 25.10 & 26.28 & 26.34 \\ \cline{2-10}
& 9.57 & 9.97 & 10.81 & 3.05 & 3.67 & 3.61 & 22.29 & 22.66 & 24.15 \\
\hhline{|=||=|=|=||=|=|=||=|=|=|}
\multirow{2}{*}{\textbf{10.0}}
& 8.57 & 9.44 & 10.17 & 3.14 & 3.46 & 3.39 & 22.53 & 24.17 & 25.40 \\ \cline{2-10}
& 8.61 & 9.38 & 10.28 & 2.71 & 3.11 & 3.43 & 20.43 & 21.12 & 22.82 \\
\hline
\end{tabular}
\end{table}

\noindent \textbf{Cerebral aneurysm:}

\begin{table}[htbp]
\small
\centering
\renewcommand{\arraystretch}{1.3}
\begin{tabular}{|c||c|c|c||c|c|c||c|c|c|}
\hline
\multirow{2}{*}{\textbf{SNR}}
& \multicolumn{3}{c||}{\textbf{Velocity}}
& \multicolumn{3}{c||}{\textbf{Pressure}}
& \multicolumn{3}{c|}{\textbf{WSS}} \\ \cline{2-10}
& \textbf{0.5 mm} & \textbf{0.75 mm} & \textbf{1.0 mm}
& \textbf{0.5 mm} & \textbf{0.75 mm} & \textbf{1.0 mm}
& \textbf{0.5 mm} & \textbf{0.75 mm} & \textbf{1.0 mm} \\
\hhline{|=||=|=|=||=|=|=||=|=|=|}
\multirow{2}{*}{\textbf{2.5}}
& 10.61 & 20.74 & 31.24 & 11.01 & 17.50 & 22.97 & 39.77 & 82.15 & 99.33 \\ \cline{2-10}
& 14.38 & 27.53 & 43.06 & 10.89 & 29.29 & 39.69 & 37.53 & 108.63 & 110.56 \\
\hhline{|=||=|=|=||=|=|=||=|=|=|}
\multirow{2}{*}{\textbf{5.0}}
& 5.07 & 9.54 & 14.52 & 5.96 & 8.04 & 11.19 & 20.34 & 38.35 & 50.57 \\ \cline{2-10}
& 6.35 & 12.18 & 20.06 & 6.58 & 11.95 & 17.53 & 20.11 & 48.96 & 58.92 \\
\hline
\end{tabular}
\end{table}


\newpage
\subsubsection{Tricubic interpolation}

\vspace{0.5cm}

\noindent \textbf{Aortic aneurysm:}

\begin{table}[htbp]
\small
\centering
\renewcommand{\arraystretch}{1.3}
\begin{tabular}{|c||c|c|c||c|c|c||c|c|c|}
\hline
\multirow{2}{*}{\textbf{SNR}}
& \multicolumn{3}{c||}{\textbf{Velocity}}
& \multicolumn{3}{c||}{\textbf{Pressure}}
& \multicolumn{3}{c|}{\textbf{WSS}} \\ \cline{2-10}
& \textbf{1.5 mm} & \textbf{2.0 mm} & \textbf{2.5 mm}
& \textbf{1.5 mm} & \textbf{2.0 mm} & \textbf{2.5 mm}
& \textbf{1.5 mm} & \textbf{2.0 mm} & \textbf{2.5 mm} \\
\hhline{|=||=|=|=||=|=|=||=|=|=|}
\multirow{2}{*}{\textbf{2.5}}
& 14.38 & 17.22 & 19.29 & 12.98 & 17.82 & 21.16 & 41.89 & 44.46 & 47.28 \\ \cline{2-10}
& 17.56 & 20.89 & 24.33 & 22.85 & 27.03 & 35.74 & 43.29 & 44.67 & 47.46 \\
\hhline{|=||=|=|=||=|=|=||=|=|=|}
\multirow{2}{*}{\textbf{5.0}}
& 11.06 & 13.59 & 15.60 & 12.70 & 16.64 & 20.05 & 31.36 & 35.99 & 39.25 \\ \cline{2-10}
& 13.08 & 16.00 & 18.99 & 19.28 & 23.57 & 31.38 & 32.53 & 35.50 & 39.88 \\
\hhline{|=||=|=|=||=|=|=||=|=|=|}
\multirow{2}{*}{\textbf{10.0}}
& 8.60 & 11.00 & 13.13 & 13.46 & 15.99 & 18.87 & 25.16 & 31.55 & 35.03 \\ \cline{2-10}
& 9.69 & 12.37 & 15.24 & 17.82 & 21.13 & 27.41 & 25.04 & 29.76 & 35.21 \\
\hline
\end{tabular}
\end{table}

\noindent \textbf{Aortic coarctation:}

\begin{table}[htbp]
\small
\centering
\renewcommand{\arraystretch}{1.3}
\begin{tabular}{|c||c|c|c||c|c|c||c|c|c|}
\hline
\multirow{2}{*}{\textbf{SNR}}
& \multicolumn{3}{c||}{\textbf{Velocity}}
& \multicolumn{3}{c||}{\textbf{Pressure}}
& \multicolumn{3}{c|}{\textbf{WSS}} \\ \cline{2-10}
& \textbf{1.5 mm} & \textbf{2.0 mm} & \textbf{2.5 mm}
& \textbf{1.5 mm} & \textbf{2.0 mm} & \textbf{2.5 mm}
& \textbf{1.5 mm} & \textbf{2.0 mm} & \textbf{2.5 mm} \\
\hhline{|=||=|=|=||=|=|=||=|=|=|}
\multirow{2}{*}{\textbf{2.5}}
& 18.68 & 22.40 & 26.04 & 19.61 & 25.94 & 26.79 & 40.55 & 45.95 & 49.04 \\ \cline{2-10}
& 17.83 & 21.64 & 25.24 & 21.05 & 27.09 & 28.79 & 39.28 & 45.62 & 48.17 \\
\hhline{|=||=|=|=||=|=|=||=|=|=|}
\multirow{2}{*}{\textbf{5.0}}
& 14.30 & 17.61 & 21.09 & 17.98 & 21.96 & 23.58 & 33.41 & 38.76 & 42.89 \\ \cline{2-10}
& 13.91 & 17.39 & 21.03 & 18.11 & 22.16 & 24.16 & 34.59 & 40.10 & 44.57 \\
\hhline{|=||=|=|=||=|=|=||=|=|=|}
\multirow{2}{*}{\textbf{10.0}}
& 11.18 & 14.46 & 18.11 & 15.98 & 18.53 & 20.72 & 29.96 & 35.18 & 39.81 \\ \cline{2-10}
& 11.41 & 15.06 & 19.00 & 15.14 & 18.13 & 20.50 & 32.50 & 37.44 & 43.03 \\
\hline
\end{tabular}
\end{table}

\noindent \textbf{Cerebral aneurysm:}

\begin{table}[htbp]
\small
\centering
\renewcommand{\arraystretch}{1.3}
\begin{tabular}{|c||c|c|c||c|c|c||c|c|c|}
\hline
\multirow{2}{*}{\textbf{SNR}}
& \multicolumn{3}{c||}{\textbf{Velocity}}
& \multicolumn{3}{c||}{\textbf{Pressure}}
& \multicolumn{3}{c|}{\textbf{WSS}} \\ \cline{2-10}
& \textbf{0.5 mm} & \textbf{0.75 mm} & \textbf{1.0 mm}
& \textbf{0.5 mm} & \textbf{0.75 mm} & \textbf{1.0 mm}
& \textbf{0.5 mm} & \textbf{0.75 mm} & \textbf{1.0 mm} \\
\hhline{|=||=|=|=||=|=|=||=|=|=|}
\multirow{2}{*}{\textbf{2.5}}
& 18.76 & 24.03 & 28.33 & 27.65 & 26.77 & 36.87 & 76.78 & 78.34 & 92.33 \\ \cline{2-10}
& 24.64 & 30.96 & 38.29 & 26.09 & 30.72 & 48.43 & 80.34 & 87.66 & 87.50 \\
\hhline{|=||=|=|=||=|=|=||=|=|=|}
\multirow{2}{*}{\textbf{5.0}}
& 13.85 & 18.69 & 22.55 & 16.79 & 17.55 & 24.79 & 56.54 & 59.82 & 73.96 \\ \cline{2-10}
& 18.09 & 24.00 & 30.02 & 20.10 & 24.50 & 31.77 & 59.89 & 67.91 & 71.92 \\
\hline
\end{tabular}
\end{table}
 
\end{document}